\documentclass[apj,natbib209]{aastex}
\newcommand{\hers}{{\it Herschel}}

\newcommand{\apex}{{\it APEX}}
\newcommand{\spitz}{{\it  Spitzer}}
\newcommand{\iso}{{\it ISO}}
\newcommand{\iras}{{\it IRAS}}
\newcommand{\jcmt}{{\it JCMT}}

\newcommand{\planck}{{\it Planck}}
\newcommand{\iram}{{\it IRAM}}

\newcommand{\lsun}{$L_\odot$}
\newcommand{\msun}{$M_\odot$}
\newcommand{\zsun}{$Z_\odot$}
\newcommand{\mic}{$\mu$m}
\newcommand{\sfr}{$\rm M_\odot\, yr^{-1}$}

\def\revised{}

\shorttitle{The Dwarf Galaxy Survey}
\shortauthors{Madden et al.}

\begin{document}


\title{An Overview of the Dwarf Galaxy Survey}

 
  \author{S.~C.~Madden
  		\altaffilmark{1}, 
		A.~R\'emy-Ruyer
		\altaffilmark{1}, 
		M.~Galametz
		\altaffilmark{2}, 
		D.~Cormier
		\altaffilmark{1}, 
		V.~Lebouteiller
		\altaffilmark{1}, 
		F.~Galliano
		\altaffilmark{1},  
		S.~Hony
		\altaffilmark{1}, 
		G.~J.~Bendo
		\altaffilmark{3}, 
		M.~W.~L.~Smith
		\altaffilmark{4}, 
		M.~Pohlen
		\altaffilmark{4}, 
		H.~Roussel
		\altaffilmark{5}, 
		M.~Sauvage
		\altaffilmark{1},
		R.~Wu
		\altaffilmark{1},
		E.~Sturm
		\altaffilmark{6}, 
		A.~Poglitsch
		\altaffilmark{6}, 
		A.~Contursi
		\altaffilmark{6}, 
		V.~Doublier
		\altaffilmark{6}, 
			M.~Baes
		\altaffilmark{7},
			M.~J.~Barlow
		\altaffilmark{8},
			A.~Boselli
		\altaffilmark{9},
			M.~Boquien
		\altaffilmark{9},
				L.~R.~Carlson
			\altaffilmark{10},
			L.~Ciesla
		\altaffilmark{9},
			A.~Cooray
		\altaffilmark{11}
		\altaffilmark{12},
			L.~Cortese
		\altaffilmark{13},
				I.~De~Looze
		\altaffilmark{7}, 
					J.~A.~Irwin
			\altaffilmark{14},
			K.~Isaak
		\altaffilmark{15},
			J.~Kamenetzky
		\altaffilmark{16},
			O.~\L.~Karczewski
		\altaffilmark{8},
			N. Lu
			\altaffilmark{19},
				J.~A.~MacHattie
				\altaffilmark{14}		   
			B.~O'Halloran
		\altaffilmark{17},
			T.~J.~Parkin
		\altaffilmark{18},
			N.~Rangwala
			\altaffilmark{16},
			M.~R.~P.~Schirm
		\altaffilmark{18},
			B.~Schulz
		\altaffilmark{19},
			L. Spinoglio
		\altaffilmark{20},
			 M.~Vaccari
			\altaffilmark{21}
			\altaffilmark{22},
			C.~D.~Wilson
		\altaffilmark{17}~and
			H.~Wozniak
			\altaffilmark{23}~			
			}
			
    
\altaffiltext{1} {Laboratoire AIM, CEA, Universit\'{e} Paris VII, IRFU/Service d'Astrophysique, Bat. 709, 91191 Gif-sur-Yvette, France}
\altaffiltext{2}{Institute of Astronomy, University of Cambridge, Madingley Road, Cambridge CB3 0HA, UK}
\altaffiltext{3}{UK ALMA Regional Centre Node, Jodrell Bank Centre for Astrophysics, School of Physics \& Astronomy, University of Manchester, Oxford Road, Manchester M13 9PL, UK}
\altaffiltext{4}{ School of Physics \& Astronomy, Cardiff University, The Parade, Cardiff, CF24 3AA, UK}
\altaffiltext{5}{Institut d'Astrophysique de Paris, UMR7095 CNRS, Universit\'{e} Pierre \& Marie Curie, 98 bis Boulevard Arago, 75014 Paris, France}
\altaffiltext{6}{Max-Planck-Institute for Extraterrestrial Physics (MPE), Giessenbachstrasse 1, 85748 Garching, Germany}
\altaffiltext{7}{Sterrenkundig Observatorium, Universiteit Gent, Krijgslaan 281 S9, B-9000 Gent, Belgium}
\altaffiltext{8}{Department of Physics and  Astronomy, University College London, Gower St, London WC1E 6BT, UK}
\altaffiltext{9}{Laboratoire d'Astrophysique de Marseille - LAM, Universit\'e d'Aix-Marseille \& CNRS, UMR7326, 38 rue F. Joliot-Curie, 13388 Marseille Cedex 13, France}
\altaffiltext{10}{Leiden Observatory, Leiden University, Leiden, The Netherlands }
\altaffiltext{11}{Center for Astrophysics and Space Astronomy, University of Colorado, 1255 38th Street, Boulder, CO, 80303, USA}
\altaffiltext{12}{Division of Physics, Astronomy and Mathematics, California Institute of Technology, Pasadena, CA 91125 USA}
\altaffiltext{13}{European Southern Observatory, Karl Schwarzschild Str. 2, 85748 Garching bei M\"unchen, Germany}
\altaffiltext{14}{Department of Physics, Engineering Physics \& Astronomy, Queen's University, Kingston, ON K7L 3N6, Canada}
\altaffiltext{15}{ESA Research and Scientific Support Department, ESTEC/SRE-SA, Keplerlaan 1, 2201 AZ Noordwijk, The Netherlands}
\altaffiltext{16}{Department of Physics and Astronomy, University of California, Irvine, CA 92697 USA}
\altaffiltext{17}{Astrophysics Group, Imperial College, Blackett Laboratory, Prince Consort Road, London SW7 2AZ, UK}
\altaffiltext{18}{Department of Physics and Astronomy, McMaster University, Hamilton, ON L8S 4M1 Canada}
\altaffiltext{19}{Infrared Processing and Analysis Center, Mail Code 100-22, California Institute of Technology, Pasadena, CA 91125, USA}
\altaffiltext{20}{Istituto di Astrofisica e Planetologia Spaziali, INAF-IAPS, Via Fosso del Cavaliere 100, I-00133 Roma, Italy}
\altaffiltext{21}{Astrophysics Group, Physics Department, University of the Western Cape, Private Bag X17, 7535, Bellville, Cape Town, South Africa}
\altaffiltext{22}{Dipartimento di Astronomia, Universit\`a di Padova, vicolo Osservatorio, 3, 35122 Padova, Italy}
\altaffiltext{22}{Observatoire astronomique de Strasbourg, Universit\'e de Strasbourg, CNRS, UMR 7550, 11 rue de lÕUniversit\'e , F-67000 Strasbourg, France}
 

\begin{abstract}
The Dwarf Galaxy Survey (DGS) program is studying low-metallicity galaxies using 230h of far-infrared (FIR) and submillimetre (submm) photometric and spectroscopic observations of the {\it Herschel Space Observatory} and draws to this a rich database of a wide range of wavelengths tracing the dust, gas and stars.  This sample of 50 galaxies includes the largest metallicity range achievable in the local Universe including the lowest metallicity (Z) galaxies, 1/50 \zsun, and spans 4 orders of magnitude in star formation rates.  The survey is designed to get a handle on the physics of the  interstellar medium (ISM) of low metallicity dwarf galaxies, especially on their dust and gas properties and the ISM heating and cooling processes. The DGS produces PACS and SPIRE maps of low-metallicity galaxies observed at 70, 100, 160, 250, 350, and 500 \mic\ with the highest sensitivity achievable to date in the FIR and submm. The FIR fine-structure lines,  [CII]~158\mic, [OI]~63\mic, [OI]~145\mic, [OIII]~88\mic, [NIII]~57\mic\ and [NII]~122 and 205 \mic\ have also been observed with the aim of studying the gas cooling in the neutral and ionized phases.  The SPIRE FTS observations include many CO lines (J=4-3 to J=13-12), [NII] 205 \mic\ and [CI] lines at 370 and 609 \mic. This paper describes the sample selection and global properties of the galaxies, the observing strategy as well as the vast ancillary database available to complement the \hers\ observations. The scientific potential of the full DGS survey is described with some example results included.
\end{abstract}





\section{Introduction}
One of the keystones in understanding how galaxies form and evolve is establishing the role of dwarf galaxies. Hierarchical models put dwarf galaxies forward as crucial building blocks of much larger galaxies, yet there is indication that many dwarf galaxies in our local Universe have stellar populations much younger than their larger cousins \citep[e.g.][]{maiolino08}. A major source of this ambiguity may arise from the fact that accurate measurements to conclude on the properties of the Interstellar Medium (ISM) and star formation histories of local Universe low metallicity dwarf galaxies have been lacking, thus preventing a consistent picture to explain the interplay between the evolution of the metals and the process of star formation. How genuinely young galaxies harboring no old stellar population are related, if at all, to chemically young galaxies, is not at all clear.  The scenery is quickly evolving, given the explosion of new wavelength coverage as well as spatial resolution and sensitivity provided by the current far-infrared (FIR)  and submillimetre (submm) {\it Herschel Space Observatory}, which has made it possible to study in detail the vast variefof dwarf galaxies in our local Universe.
These convenient laboratories allow us to study star formation and ISM properties in conditions that may be representative of early Universe environments, such as primordial galaxies where initial bursts of star formation may be taking place. 

Before \hers, much of what we have gleaned of the details of dust and gas properties and the processes of dust recycling, heating, and cooling in galaxies has been limited mostly to Galactic studies. 
\iso\ and \spitz\ space missions have proven successful in demonstrating the importance of mid-infrared (MIR) to FIR observations to study the ionized and photodissociated gas emission measures and the unusual dust properties in the low luminosity, low metallicity galaxies \citep[e.g.][]{engelbracht05, madden06,wu06, engelbracht08}. Systematic statistical surveys, that include submm wavelengths, spanning a wide variety of galactic physical parameter space such as metallicity, gas mass fraction, star formation activity, infrared (IR) luminosity, can  be used to analyse in detail the physical processes at play in the low metallicity gas and dust over galaxy-wide scales. Only with such studies can we understand the impact of galactic properties on processes controlling the evolution of galaxies. To this end, the \hers\ Dwarf Galaxy Survey (HDGS; P.I. Madden; 144.7h) is a compilation of a SPIRE SAG2 guaranteed time (GT) key program plus SPIRE GT2 observations, using the PACS and SPIRE instruments onboard the {\it Herschel Space Observatory} to obtain 55 to 500 \mic\  photometry and spectroscopy of 48 dwarf galaxies in our local Universe, chosen to cover a broad range of physical conditions.  Our larger \hers\ sample also includes 85.5h of complementary spectroscopy of dwarf galaxies, including pointings in 2 more galaxies, the Small Magellanic Cloud (SMC) and Large magellanic Cloud (LMC), carried out with the PACS GT key program and GT2 program, SHINING, which is also surveying starbursts and AGNs (P.I. Sturm). The total observing time of our \hers\ PACS GT plus SPIRE GT surveys dedicated to 50 dwarf galaxies from our HDGS and SHINING programs, is 230h and is hereafter together called the DGS (Dwarf Galaxy Survey).

The DGS {\revised sample} has also been observed with MIR photometry and spectroscopy with \spitz\ \citep[i.e.][]{bendo12}. Thus these data, together with the ancillary database being assembled, create a rich legacy for the community to perform dust and gas analyses in unprecedented detail in low metallicity galaxies from ultraviolet (UV) to millimetre (mm) wavelengths.  The properties of the DGS makes it ideally suited for such studies (see section \ref{source_selection}). Total infrared luminosity ({\it L$_{TIR}$}) values of the survey range from 6.6$\times$10$^7$\lsun\ for UGC4483, to our most luminous target, Haro 11, {\it L$_{TIR}$} = 1.6$\times$10$^{11}$ \lsun. 
Typical star formation rates range over more than four orders of magnitude ($\sim$ 0.0002 to 28 \msun yr$^{-1}$), with many of the dwarf galaxies containing super star clusters. Their metallicity properties span the largest range achievable in the local Universe, from a nearly solar value in He~2-10 to extremely low metallicity (ELM) environments with metallicity values, 12+log(O/H) $<$ 7.6 ($\sim$ 1/20 \zsun\footnote{throughout we assume solar O/H =4.90$\times$10$^{-4}$ \cite{asplund09}; i.e. 12+log(O/H)$_{\odot}$= 8.7}) and ranging as low as $\sim$ 1/40 to 1/50 \zsun\ in galaxies such as I~Zw~18 and SBS~0335-052 - among the most metal-poor in the local Universe.  
Furthermore, these objects have wide variations of gas mass fractions {\revised relative to stellar mass} (0.001 to 0.7), dust-to-metal mass ratios (0.07 to 0.33), and IR luminosities-to-gas mass ratios (0.4 to 20).

The DGS obtains detailed spatial information on the well-studied nearby star-forming dwarf galaxies of the Local Group. The well-resolved objects serve as anchors between the detailed local physical processes and the larger statistical sample. Size scales of a few pc to 100 pc allow for  the detailed multiphase study of low metallicity ISM, where stellar winds and supernovae can have a violent impact on the ISM, controlling the evolution of the dust and gas properties in dwarf galaxies.
The DGS includes PACS spectroscopic maps in several bright star forming regions in the LMC (1/2 \zsun) and the SMC (1/5 \zsun) which complement the study of HERITAGE - the  \hers\ FIR and submm photometric survey of the Magellanic Clouds \citep{meixner13}.

The DGS, along with ancillary data existing and continuing to be collected, provides the community with an unprecedented database of high impact legacy value and will lead to many future follow-up studies. To date, the DGS observations have been the objective of several case studies already presented or in preparation: NGC~6822 \citep{galametz10, carlson13}, 
NGC~1705 \citep{ohalloran10}, NGC~4214 \citep{cormier10}, Haro~11 \citep{cormier12}, LMC/N11B \citep{lebouteiller12}, NGC~4449 \citep{karczewski13}. 
The results of the full photometric and spectrometric DGS will be presented in \cite{remy13} and \cite{cormier13}, respectively.
The utility of the FIR fine-structure lines as star formation indicators and the dependence on metallicity will be presented in \cite{delooze13}.
The interpretation of this full database will lay the groundwork for understanding star formation activity under low metallicity ISM conditions, thus providing benchmarks against which primordial galaxies can be compared. Some of the outstanding questions which can be addressed with this database include the following:

\noindent {\it The global properties and diagnostics of low metallicity galaxies}
\begin{itemize}
\item What properties of the galaxy as a whole, and local processes within it, control the observed Spectral Energy Distribution (SED) and its evolution? How can we disentangle the effects of galaxy metallicity, star formation properties, ISM morphology and dust properties (dust size distribution, composition, etc.), etc on the observed SED?
\item In which galactic environments does the submm excess\footnote{The submm excess is emission enhanced beyond that explicable by current dust models and normally appears near or longward of $\lambda$ $\geq$ 500 \mic\ in low metallicity environments (see Section \ref{sciencegoals}).} manifest and what is the nature of this submm excess?
\item How can we most accurately trace the {\it total} molecular gas reservoirs which are fueling the prominent star formation activity in low metallicity dwarf galaxies? Is the warm molecular gas a significant component in star forming dwarf galaxies? Reliable CO-to-H$_{2}$ conversion factors for low metallicity ISM conditions are urgently needed. 
\item How can we calibrate the technique of using observed dust masses to get at the gas reservoirs?
\item Is the 158 \mic\ [CII] line a valid star formation rate tracer in local galaxies, and can it be used to quantify star formation in high-{\it z} galaxies?
\end{itemize}

\noindent {\it The impact of star formation on the ISM}
\begin{itemize}
\item What are the effects of the ensemble of star formation sites on full galaxy scales and the processes controlling the feedback between the surrounding metal-poor ISM and star formation under conditions approaching those of the earliest, most metal-poor galaxies? 
\item What is the impact of the super star clusters on the nature of the surrounding dust and gas? How much star formation activity is actually enshrouded and optically thick out to and beyond MIR wavelengths even in these dust-poor environments?
\item What is the effect of metallicity (direct or indirect) on the morphology and filling factors of the different gas phases?
\end{itemize}

\noindent {\it The interplay between the gas and dust in galaxies}
\begin{itemize}
\item What are the consequences of the dust properties on the gas heating and cooling processes in low metallicity environments and how does this compare with  metal-rich ISM?
\item What is the relationship between metals in the gas phase and those in the dust as a function of the metallicity of a galaxy? How do the dust-to-gas mass ratios (D/G) evolve as a function of galaxy metallicity?
\item What is the impact of dust abundance and composition and gas reservoirs on the evolution of the ISM and vice versa?      
\end{itemize}

The improved angular resolution and sensitivity of \hers\ in the 60-160 \mic\ range compared to \iso\ and \spitz, is allowing us to probe local regions {\it within} galaxies as well as statistically-important global galaxy properties. Thus we can study the effect of varying spatial scales on the interpretation of the observations
and relate the observed variations in the dust SEDs and gas properties to the actual physical phenomena occurring within the galaxy \citep[e.g.][]{bendo12, parkin12, foyle12, aniano12, galametz12, fritz12, smith12, xilouris12, mentuch12}. The added benefit of the combination of the \hers\ spectroscopy and photometry will make it possible to address the long-standing issues on the total gas and dust reservoirs in low metallicity galaxies, with particular attention to the enigma of quantifying the molecular gas reservoir.
This will lead to the construction of a rich interpretative framework for unresolved, more distant galaxies in the early Universe. 

In this paper we present the overview of the scientific goals in section \ref{sciencegoals}; the source selection and properties are described section \ref{source_selection}; the observing strategy and data treatment of the DGS and ancillary data of the survey are described in section \ref{observing_section}; selected science results are presented in section \ref{science_section}; future products to come out of the survey are mentioned in section \ref{products_section}; and a final summary is given in section \ref{summary_section}. 
 
 \section{Scientific Goals}
\label{sciencegoals}
 \subsection{The dust properties of dwarf galaxies
 }
One avenue to understand the star formation activity and the evolution of dwarf galaxies is by studying the dust emission properties. In addition to being the culprit in obscuring the starlight in galaxies, potentially biasing our view of stellar activity, dust plays a more positive role as a major actor in the thermal equilibrium of galaxies.  Dust absorbs starlight of a wide range of energies, emitting in the MIR to submm/mm wavelength regime. In spite of their metal-poor ISM, dwarf galaxies have been known, even since \iras\ observations, to possess non-negligible amounts of dust with some low metallicity galaxies globally harboring warmer dust on average than their dust-rich counterparts. 
The more active blue compact dwarf galaxies (BCDs), for example, revealed higher \iras\ 60/100 \mic\ flux ratios (i.e. warmer dust) compared to those of disk-type metal-rich galaxies while the elevated 25/12 \mic\ fluxes highlighted the presence of warm stochastically-heated MIR-emitting dust and the dearth of the smallest grains/large molecules, Polycyclic Aromatic Hydrocarbons (PAHs) \citep[e.g.][]{helou86, hunter89, sauvage90, melisse94b}. The dramatic improvement in sensitivity of \spitz\  accentuated these findings of galaxy-wide warmer dust in star forming dwarf galaxies \citep[e.g.][]{jackson06, rosenberg06, cannon06b, walter07}, with an extreme case being one of the lowest metallicity galaxies known in the Universe, SBS0335-052 \citep{houck04}. {\revised \iso\ and \spitz\ spectroscopy confirmed the relatively weak PAH bands in dwarf galaxies \citep[e.g.][]{madden06, ohalloran06, wu06, hunter07, smith07, hunt10}} and are sometimes completely absent as metallicities drop to a threshold level of about 12+log(O/H)=8.2 - about 1/5 to 1/3 \zsun   \citep{engelbracht05}. 

While observations have been modeled with sufficient MIR to submm coverage in only a limited number of local Universe metal-poor galaxies with \iras, \iso, \spitz\ and ground-based telescopes, their SEDs were already exhibiting evidence for dust properties with other notable differences from those of our Galaxy as well as other metal-rich starburst galaxies.  Excess emission in the submm/mm wavelength range (Figure~\ref{SED}) over the expected Rayleigh-Jeans behavior was noted in the COBE observations of our Galaxy \citep{reach95} and has been detected in low-metallicity dwarf galaxies using ground-based telescopes observing 850/870 \mic\ with SCUBA on the \jcmt\ and with LABOCA on \apex\ \citep[e.g.][]{galliano03, galliano05, bendo06, galametz09, zhu09, galametz11}. In the LMC, where \hers\ brings 10 pc resolution at 500 \mic, \cite{galliano11} have highlighted locations of submm excess ranging from 15\% to 40\% compared to that expected from dust SED models. The intensity of the excess is anti-correlated with the dust mass surface density. Possible explanations put forth to explain the submm excess include: 1) very cold dust component \citep[e.g.][]{galliano05, galametz11};  2) unusual dust emissivity properties \citep[e.g.][]{lisenfeld02, meny07, galliano11, paradis11};  3) anomalous spinning dust which has explained excess emission appearing at radio wavelengths \citep[e.g.][]{draine98a, hoang11, ysard11} and 4) magnetic dipole emission from magnetic grains \citep{draine12}.  No single explanation put forth to date as the origin of the submm excess is completely satisfactory for all cases observed. Thus, the origin of the submm excess remains enigmatic.
With more sensitive submm wavelength coverage, \hers\ is confirming the flatter submm slope indicative of the submm excess present in some dwarf galaxies in the DGS \citep[e.g.][]{remy13}.

The global MIR SEDs of dwarf galaxies, often dominated by prominent ionized gas lines, weak PAH emission, resemble HII regions \citep[e.g.][]{madden06}, or even, to some extent, ULIRGs (Figure~\ref{SEDs_galaxies}).
Galaxy-wide, star forming dwarf galaxies often have a larger proportion of warm dust than spiral galaxies, with steeply rising MIR to FIR slopes and overall SEDs peaking at wavelengths shortward of  $\sim 60$ \mic\ (Figure~\ref{SEDs_galaxies}) \citep[e.g.][]{galametz09, galametz11}. Starbursting dwarf galaxies have a higher relative abundance of small grains, compared to our Galaxy \citep[e.g.][]{lisenfeld02, galliano03, galliano05}. In one of the most metal-poor galaxies detected to date, SBS 0335-052 (1/40 \zsun), the MIR to FIR SED is unlike that of any other galaxy observed so far, with silicate absorption bands superimposed on a featureless continuum peaking at an unusually short wavelengths, $\sim 30$ \mic\ \citep{thuan99, houck04}.   The smallest grains in galaxies, the most numerous in low metallicity galaxies being the very small grains, not the PAHs which are abundant in disk galaxies, are normally playing the most important role in gas heating, commonly via the photoelectric effect {\citep{bakes94}. Thus how the dust size distribution and abundance varies within galaxies is critical to understand the heating and cooling processes.  Given these observed dust properties, the heating and cooling processes will, undoubtedly, be altered in dust-poor galaxies relative to dust-rich galaxies. The structure of the ISM in terms of relative filling factors of star clusters, atomic, ionized and molecular gas looks to be  different from those in our Galaxy or other metal-rich starburst galaxies: the ionized regions and the galaxy-wide hard radiation field, play an important role in shaping the observed SED and the gas properties \citep{lebouteiller12, cormier12}.  Simply assuming Galactic dust properties in the analyses of SEDs of low metallicity galaxies may lead to erroneous results in the determination of dust mass, extinction properties, gas mass, star formation properties, etc.

\begin{figure}
plotone{f1.eps}
\caption{Comparison of a dwarf galaxy with the SEDs of 4 other galaxies: star-forming dwarf galaxy, NGC~1705 \citep{ohalloran10} with a spiral galaxy, NCG~6946 \citep{aniano12}, a Ultra Luminous InfraRed Galaxy (ULIRG), Arp~220, a metal-rich starburst, M82 \citep{roussel10} and a dwarf elliptical, NGC~205 \citep{delooze12}. The dwarf galaxy, NGC~1705, is one of the faintest example here with a SED characteriztic of star-forming low metallicity dwarf galaxies. The vertical bars indicate the \hers\ PACS and SPIRE channels. The SED models are derived using the Galliano et al. "AC" (amorphous carbon) model \citep{galliano11}. 
}
\label{SEDs_galaxies}
\end{figure}


The choice of dust composition to use when interpreting the observations complicates the problem. For example, assuming the often-used graphite dust analog, which, along with a silicate and PAH components, well-describes the dust composition of the Galaxy \citep[e.g.][]{zubko04, draine07} can lead to inexplicably large dust masses in low metallicity galaxies such as the LMC, as shown by \cite{galliano11}. While either amorphous carbon or graphite and silicate and PAHs in the SED model can successfully fit the observations, the resulting larger dust mass using graphite, is sometimes difficult to reconcile with the metallicity, the observed gas mass and the expected D/G (Section~\ref{section_D2G}). Posessing higher emissivity properties, an amorphous carbon component \citep{zubko96} can result in a lower dust mass that agrees better in some cases with the measured gas mass and the expected D/G. A recent reevaluation of the optical properties of (hydrogenated) amorphous carbon may lead to yet a different view of the variations in the observed dust properties in galaxies \citep{jones12a, jones12b, jones12c}.

Many of the conclusions on the dust properties of low metallicity ISM rest on the need for higher spatial resolution and longer wavelength coverage into the submm, as provided by \hers. While \cite{draine07b} concluded that there is little effect overall on the dust masses of galaxies with or without submm observations, \cite{galametz11} found that including submm constraints in the SED modelling process can affect the dust masses estimated. To complicate interpretation, the single temperature modified black body, often used to describe the dust FIR-submm properties, can lead to different dust masses, for example, compared to more realistic dust SED models \citep[e.g.][]{galliano11, dale12, galametz12}.  This is a consequence of the fact that 70-500 \mic\ emission from galaxies can originate from different heating sources and a single modified blackbody fitting these data can overestimate the average dust temperature and thus underestimate the dust mass within the galaxies \citep[e.g.][]{bendo10, boquien11, bendo12, dale12}. \hers\ is well-positioned to lend clarity to these issues.

\subsection{FIR Fine-Structure lines as gas coolants in the interstellar medium}
 
The Kennicutt-Schmidt law demonstrates the empirical relationship between total gas column density and star formation rate surface density. The observed slope of $\sim$1.4 indicates that the efficiency in forming stars increases with the gas surface density \citep{kennicutt98}. Evidence is growing for {\it molecular} gas surface density, in particular, as the regulator for star formation in galaxies of the local Universe as well as at high-redshift, and not necessarily the {\it total} gas density  \citep[e.g.][]{kennicutt07, bigiel08, genzel12}. While mostly more massive galaxies have been used to derive this relationship, does the same recipe hold for star formation in the wide range of low mass dwarf galaxies, or, for that matter, high-{\it z} low metallicity dwarf galaxies?  

We take for granted that CO (J=1-0) is the best tracer of molecular gas in local and distant galaxies, considering the observational difficulties in detecting directly the most abundant molecule, H$_2$. While HI halos are often very extended in dwarf galaxies, CO has been notoriously difficult to detect. Traversing from CO (J=1-0) observations to the total H$_{2}$ gas mass in dwarf galaxies is a perplexing issue, one that has been hampering studies of low metallicity ISM for decades. This CO (J=1-0) deficit is often difficult to reconcile with the molecular gas necessary to fuel star formation and valiant effort has been invested in  determining a recipe for the CO (J=1-0)-to-H$_{2}$ conversion factor in low metallicity galaxies \citep[e.g.][and references within]{leroy11, schruba12, bigiel11}. On the other hand, surprisingly prominent FIR fine structure cooling lines, such as the 158 \mic\ [CII] line, often interpreted as the best tracer of photodissociation regions (PDRs) around molecular clouds illuminated by massive stars, have been detected in low metallicity dwarf galaxies with the Kuiper Airborne Observatory (KAO) and the ISO LWS \citep{stacey91, poglitsch95, madden97, madden00, bergvall00, hunter01, vermeij02, brauher08, israel11}.  Thus, exceptionally bright 158 \mic\ [CII] with exceptionally faint CO (J=1-0) renders the ISM of low metallicity galaxies and the conditions for star formation rather elusive.

L$_{[CII]}$ has been hailed as an important tracer of star formation activity \citep{stacey91, boselli02, delooze11}. For example, values of L$_{[CII]}$/L$_{CO(J=1-0)}$ of more active normal starburst galaxies have ratios of $\sim$ 4 000 while more quiescent spiral galaxies show a factor of 2 lower \citep{stacey91}. In dwarf galaxies, however, values of L$_{[CII]}$/L$_{CO(J=1-0)}$ can range from about 4 000 to as high as $\sim$ 80 000 \citep{madden00} - much higher than the most active metal-rich galaxies (Figure~\ref{cii_co}).  What is the explanation behind such strikingly high L$_{[CII]}$/L$_{CO(J=1-0)}$?

Toward regions of the Local Group galaxy, IC10, for example, [CII] excitation from collisions with H atoms could not explain the surprisingly high observed [CII] intensity. Assuming the [CII] excitation originated in PDRs, comparison of the observed HI column density required an additional excitation source, namely H$_{2}$ molecules, to reconcile the cooling rate with the photoelectric UV heating rate.
This lead to the suggestion that up to 100 times more molecular hydrogen gas may be present, compared to the H$_2$ inferred from the relatively faint CO (J=1-0) lines \citep{madden97}. These early KAO detections of the 158 $\mu$m [CII] line in low metallicity galaxies, namely the LMC \citep{poglitsch95, israel96, israel11} and IC10 \citep{madden97} were the first observational evidence for the presence of a reservoir of H$_{2}$ gas not traced by CO (J=1-0), but residing in the C$^{+}$-emitting region - the "CO-dark" molecular gas. 

The presence of the CO-dark molecular gas is regulated by conditions present in low metallicity environments. Molecules which are not effective at self-shielding, such as CO, will be photo-dissociated before H$_2$ in the PDRs. The combined effect of the strong FUV field from the nearby young star clusters and the decreased dust abundance is to shift the atomic-molecular gas transition deeper into the cloud, reducing the relative size of the CO core and increasing the volume of the PDR. In this scenario, a potentially significant reservoir of H$_2$ can be present in C$^{+}$- emitting PDR, outside the smaller CO core. This component, not traced by CO nor HI, has also been uncovered in our Galaxy via $\gamma$ - ray observations \citep{grenier05} and more recently with Planck \citep{planck_galaxy_dg, paradis12}. This effect of the CO-dark molecular gas in the PDR envelope is shown theoretically to be controlled strongly by {\it A$_{V}$} and should increase in importance as the metallicity decreases \citep{wolfire10, krumholz11, glover12}. More recent detailed studies comparing the dust and observed gas properties in the LMC and other nearby galaxies with \hers\ and {\it Planck} have likewise highlighted the presence of this otherwise undetected gas phase (Section~\ref{section_D2G}), which could also be in the form of optically thick HI.

 \begin{figure}
\includegraphics[width=15cm]{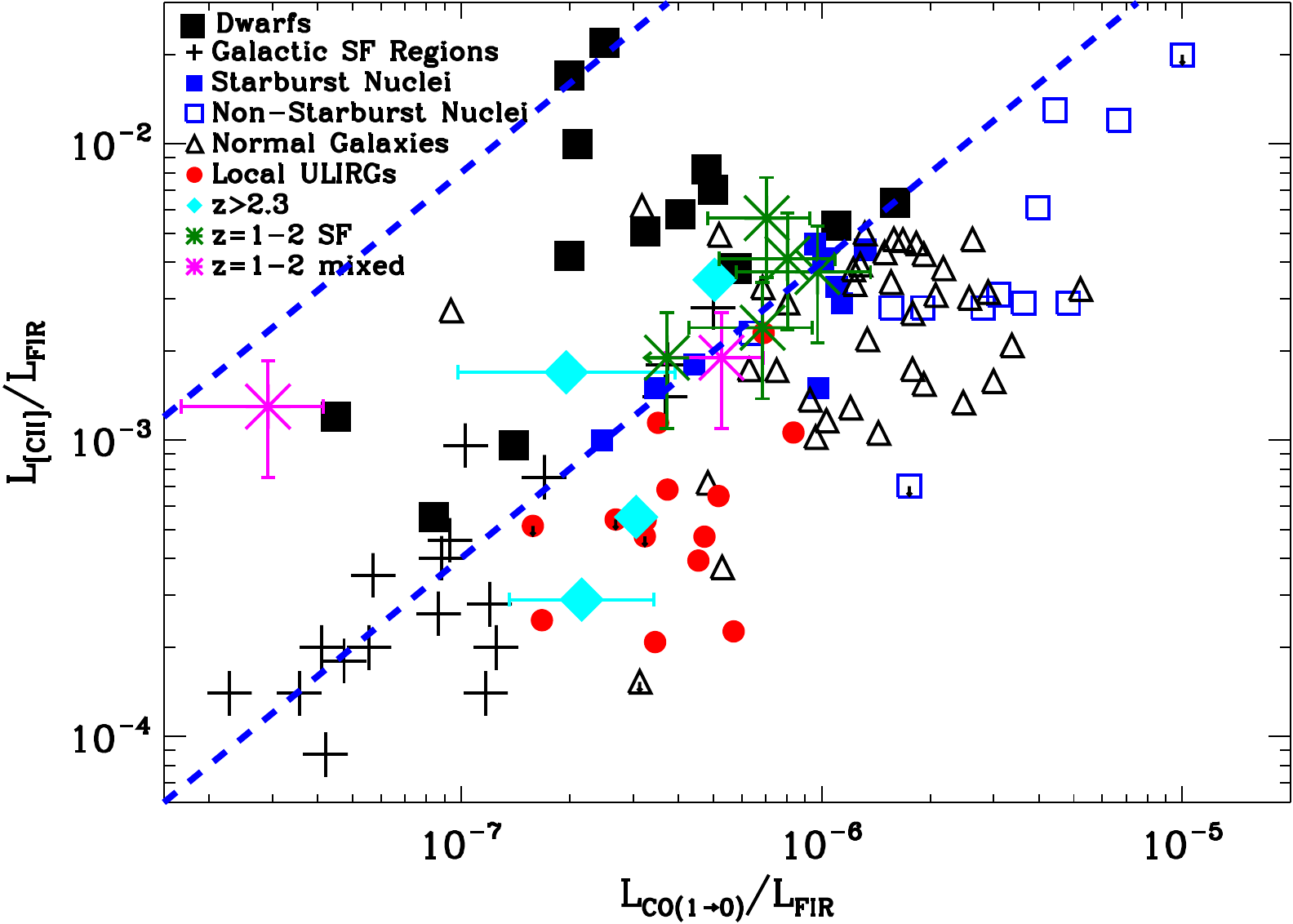}
\caption{L$_{[CII]}$/L$_{FIR}$ vs L$_{CO}$/L$_{FIR}$ for local galaxies, and high-{\it z} galaxies \citep{haileydunsheath10, stacey10} and low metallicity dwarf galaxies \citep{madden00}.  Lines of constant L$_{[CII]}$/L$_{CO}$ are shown (blue, dashed) for 4,000 (local starburst and high-{\it z} galaxies) and 80,000, the maximum end of dwarf galaxies. These observations are from the \iso/LWS and the KAO. \hers\ observations of dwarf galaxies will populate this plot further to the upper left region.  
}
\label{cii_co}
  \end{figure}

In addition to the widely-used [CII] line, the DGS spectroscopy programme traces the other key FIR ISM cooling lines such as [OIII], [OI], [NIII] and [NII] (Table~\ref{FIR_lines}). These important cooling lines are diagnostics to probe the FUV flux, the gas density (n) and temperature (T) and the filling factor of the ionized gas and photodissociation regions \citep[e.g.][]{wolfire90,kaufman06}. While the new NASA-DFG airborne facility, SOFIA, will continue observing the FIR fine-structure lines beyond {\it Herschel's} lifetime, it must contend with the residual effects of the Earth's atmosphere. Hence, \hers\ allows us the best sensitivity and spatial resolution to probe a variety of emission lines from these intrinsically faint dwarf galaxies.

Since the ionization potential of C$^0$ is 11.3 eV (Table~\ref{FIR_lines}) - less than that required to ionize atomic hydrogen - [CII] can be found in the ionized as well as the neutral phases.   One way to determine how much of {\revised the observed [CII] may be excited by collisions with electrons in low density ionized regions}, for example, is to use the [NII] lines at 122 and 205 \mic\ \citep[e.g.][]{oberst11}. Since N$^0$ requires an ionization  potential of 14.5 eV,  these lines arise unambiguously from a completely ionized phase. Their relatively low critical densities ({\it n$_{crit}$}) for collisions with electrons (48 to 200 cm$^{-3}$) make them useful to associate any observed [CII] emission possibly arising from the diffuse ionized gas, and thus, provide a means to extract the PDR-only origin component of the [CII] emission.  The [NII] 122 \mic\ line is detected with PACS only in the brightest regions  of the DGS galaxies. The PACS [NII] 205 \mic\ line was not observable in the dwarf galaxies: leakage from other orders of the grating renders the flux calibration of the [NII] 205 \mic\ line unreliable and the sensitivity of the instrument beyond 190 \mic\ is degraded. The [NII] 205 \mic\ line is, however, covered by the SPIRE FTS which has observed 4 DGS galaxies. The [OIII] 88 \mic\ line has a similarly low {\it n$_{crit}$} for collisions with electrons (Table~\ref{FIR_lines}) and can also be associated with the low density ionized medium. It has the highest ionization energy of these FIR lines and traces the hard photons traversing the diffuse ionized medium.
Thus, with \hers, along with the existing ancillary data, we can more accurately model the diffuse and dense atomic, ionized and molecular phases of the low-metallicity ISM. 

 
The sensitivity provided by \hers, along with the wide range of MIR lines that have been observed for the target galaxies with the \spitz\ IRS spectrograph (Table ~\ref{ancillary}), provides $\sim$ 20 MIR and FIR fine structure lines in a wide variety of low metallicity galaxies in the DGS. Detailed studies of the photodissociated and photoionized gas can characterize the physical properties of the ISM as well as the properties of the energetic sources in low metallicity environments (see \cite{cormier12} for an example of such complex extragalactic modeling). Additionally, these observations will be able to shed light on quantifying the molecular gas reservoir in dwarf galaxies. 
 
\begin{deluxetable}{lccccl}
\tabletypesize{\scriptsize}
\tablecolumns{6}
\tablecaption{PACS FIR emission lines for the DGS: their properties and diagnostic capabilities \label{FIR_lines}}
\tablehead{
\colhead{Species}            &     \colhead{$\lambda$} & \colhead{Transition} & \colhead{Excitation$^{a}$} & \colhead{{\it n$_{crit}$}} & \colhead{Comments}  \\
\colhead{} 	&	\colhead{(\mic)} 	& \colhead{}	&	\colhead{Potential (eV)}	&  \colhead{cm$^{-3}$} & \colhead{} \\
}
\startdata
{[}OI]	& 63.1 & $^3P_2- ^3P_1$	&-	&4.7$\times$10$^5$$^b$ 	& Emission from neutral regions only   \\
&	&	&	&	& With [CII] and L$_{FIR}$ - constrains incident FUV flux. \\
{[}OI]	& 145.5  	& $^3P_1 - ^3P_0$ 	&-	&9.5$\times$10$^4$$^b$ 	& Ratio of both [OI] lines - measures local T $\sim$ 300K \\
{[}OIII]	& 88.3  	& $^3P_1 - ^3P_0$  	& 35.1	& 510$^c$	&Ionized regions only - radiation dominated by OB stars. \\
&	&	&	&	& With optical [OIII] lines (e.g. 495.9 and 500.7 nm) - \\
&	&	&	&	& probes T and n in hot optically thin plasmas. \\
{[}CII]	& 157.9  	& $^2P_{3/2} - ^2P_{1/2}$	& 11.3 & 2.8$\times$10$^3$$^b$,  50$^c$	& 
Most important coolant of the warm, neutral ISM. \\
&	&	&	&	& Found in both neutral and ionized regions \\
{[}NII]	& 121.7 	& $^3P_{2} - ^3P_{1}$ & 14.5	&310$^c$	&Excited by collisions with electrons. \\
&	&	&	&	& Found only in ionized gas  \\
{[}NII]	& 203.9 	& $^3P_{1} - ^3P_{0}$ & 14.5 	&48$^c$ & Ratio of [NII] lines - excellent probe of low n$_e$ \\
{[}NIII] 	& 57.3  	& $^3P_{3/2} - ^3P_{1/2}$	& 29.6 	& 3.0$\times$10$^3$$^c$	& Ratio with [NII] can trace radiation field hardness \\
\enddata
\scriptsize{
$^a$ Excitation Potential: Here we refer to the energy needed to remove electrons to reach this state. \\
$^b$ Critical density for collisions with hydrogen atoms (kinetic T = 300K) \\
$^c$ Critical density for collisions with electrons  \\
}
\end{deluxetable}

\subsection{Possible Warm Gas Reservoir in Dwarf Galaxies?}
MIR H$_{2}$ lines, possible tracers or warm gas,  can sometimes be present in Spitzer IRS spectra of dwarf galaxies, most often at low levels \citep{wu06, hunt10}, but are not normally thought to account for a substantial reservoir of the molecular gas in dwarf galaxies because most of this gas lies at lower T ($<$100K). 

The SPIRE FTS, covering 194 to 670 \mic, provides simultaneous coverage of CO rotational lines ranging from J=4-3 through J=13-12, a range within which the spectral line energy distribution (SLED) of CO lines has been found to peak for AGNs,  star forming galaxies and for high-{\it z} QSOs unveiling, often, a previously unknown warm molecular component \citep[e.g.][]{panuzzo10, rangwala11, kamenetzky12, spinoglio12}.   The warm molecular gas component of the relatively bright, well-studied nearby dwarf galaxies, NGC~4214, IC10, He2-10 and  30 Doradus, the most massive star forming region in our neighboring LMC, are observed with the SPIRE FTS. 30Doradus, He2-10 and IC10 have already been detected from the ground in CO transitions as high as J=7-6 \citep{bayet04, bayet06, pineda12}. 
 While CO J=1-0 can be excited at low densities (10$^{2}$ to 10$^{3}$ cm$^{-3}$) and low temperatures ($\sim 10K - 50 K$), higher-J CO lines effectively trace the denser and warmer gas connected to ongoing star-formation. The high J ground-based CO lines (up to J= 7-6) have been touted to be the best available SFR indicators accessible with ground-based instruments \citep{bayet09}. 
The FTS observations, aided by PDR and chemical models, 
will lead to the determination of the physical parameters (including density, temperature, filling factor and mass) of the CO-emitting molecular reservoir in low metallicity galaxies. This warm component is essential to help provide a complete inventory of all of the gas in these star-forming low metallicity galaxies, much of it possibly residing in the warmest reservoirs more reliably traced by the high-J CO lines from the FTS observations.


\subsection{Dust to Gas Mass ratios (D/G)}
\label{section_D2G}
Limits in sensitivity, spatial resolution and wavelength coverage in the past have left an ambiguity in the behavior of the D/G at low metallicities \citep[e.g.][]{galametz11}, The DGS is finally extending studies on D/G in galaxies to the lowest metallicities in the local Universe \citep{remy13}.

While the proportion of heavy elements in the gas and dust scales with the metallicity in the disk of our Galaxy \citep{dwek98}, it is not completely clear that this also holds for low metallicity dwarf galaxies. For a sample of low-metallicity galaxies, \cite{lisenfeld98} and \cite{issa90} found that the dependence of metallicity on D/G can deviate from the expected linear relationship which implies that the fraction of metals in the dust phase varies over the history of a galaxy. These studies, however, were performed for dust masses determined from \iras\ observations, therefore not tracing cooler dust beyond 100 \mic. \citet{james02}, who did a similar study using some ground-based 850 \mic\ observations, include a few low metallicity galaxies in their large sample and find a flatter trend of the D/G at lower metallicities that needs to be tested for a larger range of metallicities, particularly for lower metallicities.   In a study conducted before \hers,  \cite{galametz11} demonstrated that submm wavelength observations were indispensable to constrain the total dust masses, leaving some ambiguity in the measured D/G even without considering a submm excess. On the other hand, \cite{draine07b} concluded that the D/G of the SINGS sample of galaxies was well determined with \spitz\ observations up to 160 $\mu$m with an additional cirrus component to explain the 850 \mic\ observations of the metal-rich galaxies in the sample. More recently, \cite{dale12} results extending the SINGS sample to \hers\ wavelengths with the KINGFISH survey \citep{kennicutt11}, note a submm excess present in their low Z galaxies.  \hers\ is beginning to uncover more examples of possible candidates to study the submm excess as shown, for example, in the dwarf galaxies in the \hers\ Virgo Cluster Survey \citep{grossi10}. Likewise, a metallicity dependence of the SPIRE band ratios in the \hers\ Reference Survey \citep{boselli12} suggests the presence of a submm excess. More complete wavelength sampling by \hers\ will help resolve the submm behavior and refine our extragalactic dust SED models. 

\citet{bot04} modelled the dust in the diffuse ISM of the SMC (Z = 1$/$10), 
and concluded that the D/G is 30 times lower than that of the Galaxy - 3 times lower than {\revised the value a linear scaling of} the metallicity would suggest.
The SMC result supports the idea of depletion of heavy elements in the dust, such as through supernova events. The large sample of DGS galaxies surveyed from MIR to the submm, having widely varying metallicities, allows us to rigorously test various chemical evolution models \citep[e.g.][]{edmunds01, dwek98, galliano08a}.  Tighter constraints were put on the metallicity dependence of the D/G by \cite{galametz11} but the behavior below {\it Z}=12+log(O/H) $\sim$ 8 has been ambiguous due to lack of low metallicity observations. Investigating the behavior of D/G with metallicity in the DGS will help to decipher the origin of the dust: for example, from supernovae versus evolved low mass stars. 


One popular way to get at possible CO-dark molecular gas is via the dust mass.  Using the dust mass determined from observations, along with an assumed D/G (most often used as a linear or near-linear relation with metallicity), can give the total gas mass which can be compared to that observed in HI and CO. This has been shown to reveal gas reservoirs not emitting in CO or HI \citep[e.g.][and references within]{leroy11}. Measuring dust mass and using an expected D/G has also been used to determine the CO-to-H$_{2}$ conversion factor for high-{\it z} galaxies \citep{magdis12, magnelli12}. Such methods to get at any possible dark gas, however, require confidence in the dust mass determination as well as understanding metallicity variations and the D/G variations throughout a galaxy. 

Possible uncertainties in the total dust mass, as well as the total molecular gas reservoir, again highlights the difficulty in understanding the ISM of low metallicity galaxies and the local conditions for star formation. With the new \hers\ data and self consistent modeling of the gas and dust together we can begin to unravel the mysteries of the ISM physics in dwarf galaxies.


\section{Source Selection}
\label{source_selection}
Until very recently, we have had information only on a small number of relatively well-known dwarf galaxies with metallicities mostly ranging from 1/2 to 1/10 \zsun, with the exceptions of I~Zw~18 and SBS~0335-052, which are 1/50 and 1/40 \zsun. Deep optical emission line and photometric observations from the Hamburg/SAO Survey and the First and Second Byurakan Surveys \citep[e.g.][]{izotov91,ugryumov03}  have unveiled a treasure trove of low metallicity galaxies, with metallicity values ranging as low as 1/50 \zsun, including a large number of  ELM galaxies. We have chosen a broad sample of 50 galaxies ranging from very low metallicity ($\sim$1/50 \zsun) to moderate metallicity ($\sim$ 1/3 \zsun) with $\sim$ 1/3 of the sample targeting ELMs. The first panel in Figure~\ref{Histo_SAG2} shows the distribution of the metallicity values of our sample. Beyond the Local Group sources, our DGS sample was selected from several surveys including the Hamburg/SAO Survey and the First and Second Byurakan Surveys picking up emission line galaxies. Our target sources come from follow-up observations which specifically identified the metal-deficient Blue Compact Dwarf galaxies (BCDs) from the survey \citep[e.g.][] {izotov91,ugryumov03,popescu00}, as well as the sample of \citet{izotov99} and \citet{hopkins02}, which contain some of the better-studied BCDs. The sources range from the nearest neighbor galaxies, the Large and Small Magellanic Clouds (LMC and SMC; 50 kpc) and NGC~6822 (500 kpc) to the more distant Haro 11 (92 Mpc) and HS0052+2536 (191 Mpc).

The design of the DGS is an attempt to obtain a statistically significant sample of about 9 galaxies in 7 bins over a wide metallicity range: 12+log(O/H) ranging from 8.4 (1/3 \zsun), in bins of 12+log(O/H)=0.2 (a factor of 1.4 difference) to provide an accuracy of $\sim$ 30$\%$. As can be seen from Figure~\ref{Histo_SAG2}, it was not possible to have a uniform sample in terms of metallicity, due to lack of availability or detectability of such extreme low metallicity sources in the local Universe. We also did not insist on populating our highest metallicity bins, since metal-rich galaxies are plentiful in other surveys, such as the \hers\ Reference Survey \citep{boselli10, ciesla12}, \hers\ Virgo Cluster Survey \citep{davies12} and the KINGFISH survey \citep{kennicutt11}, for example. 

 
 \begin{figure}
     \centering
     \vspace{-0.8cm}
\includegraphics[width=6.7cm]{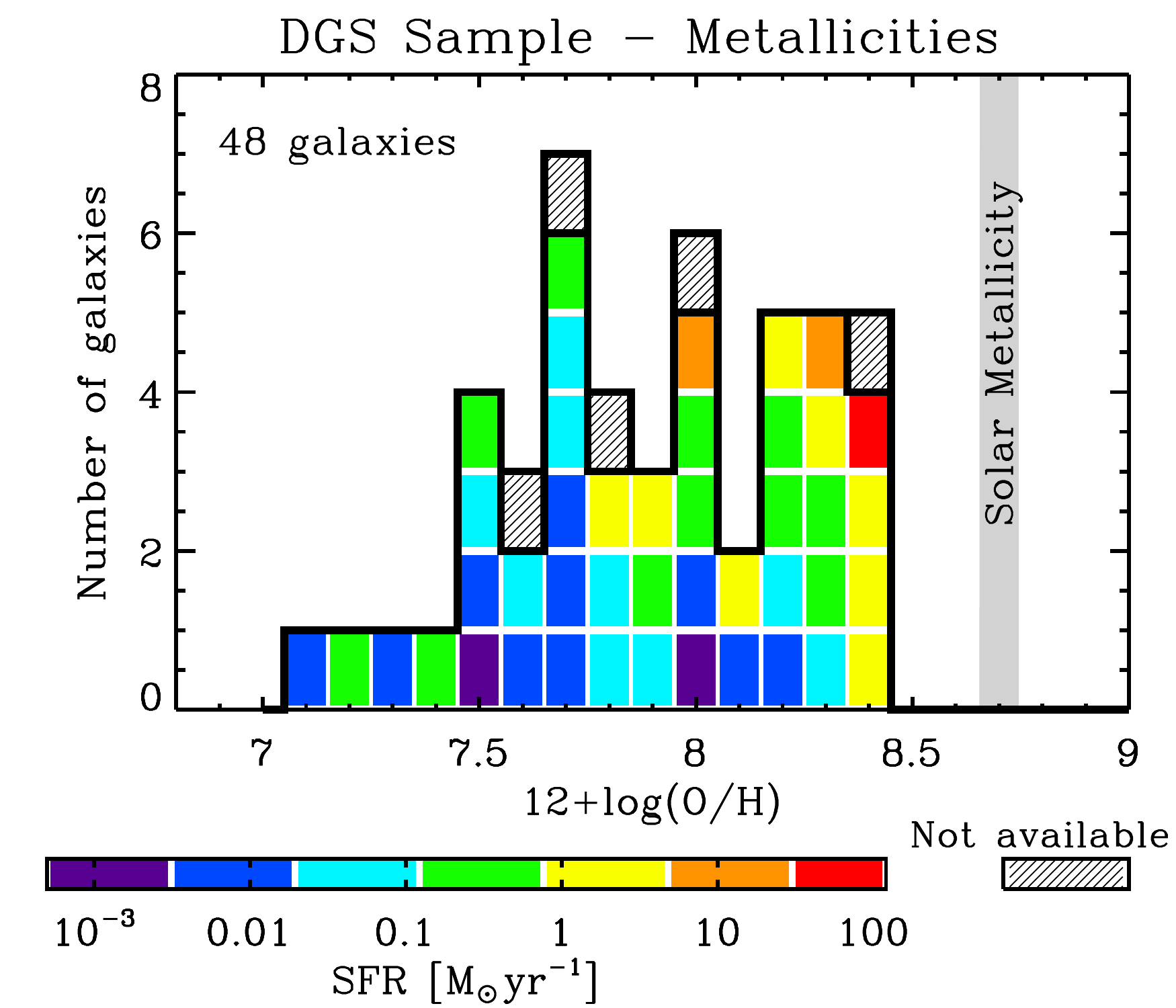}
\vspace{1cm}
\hspace{1cm}
\includegraphics[width=6.7cm]{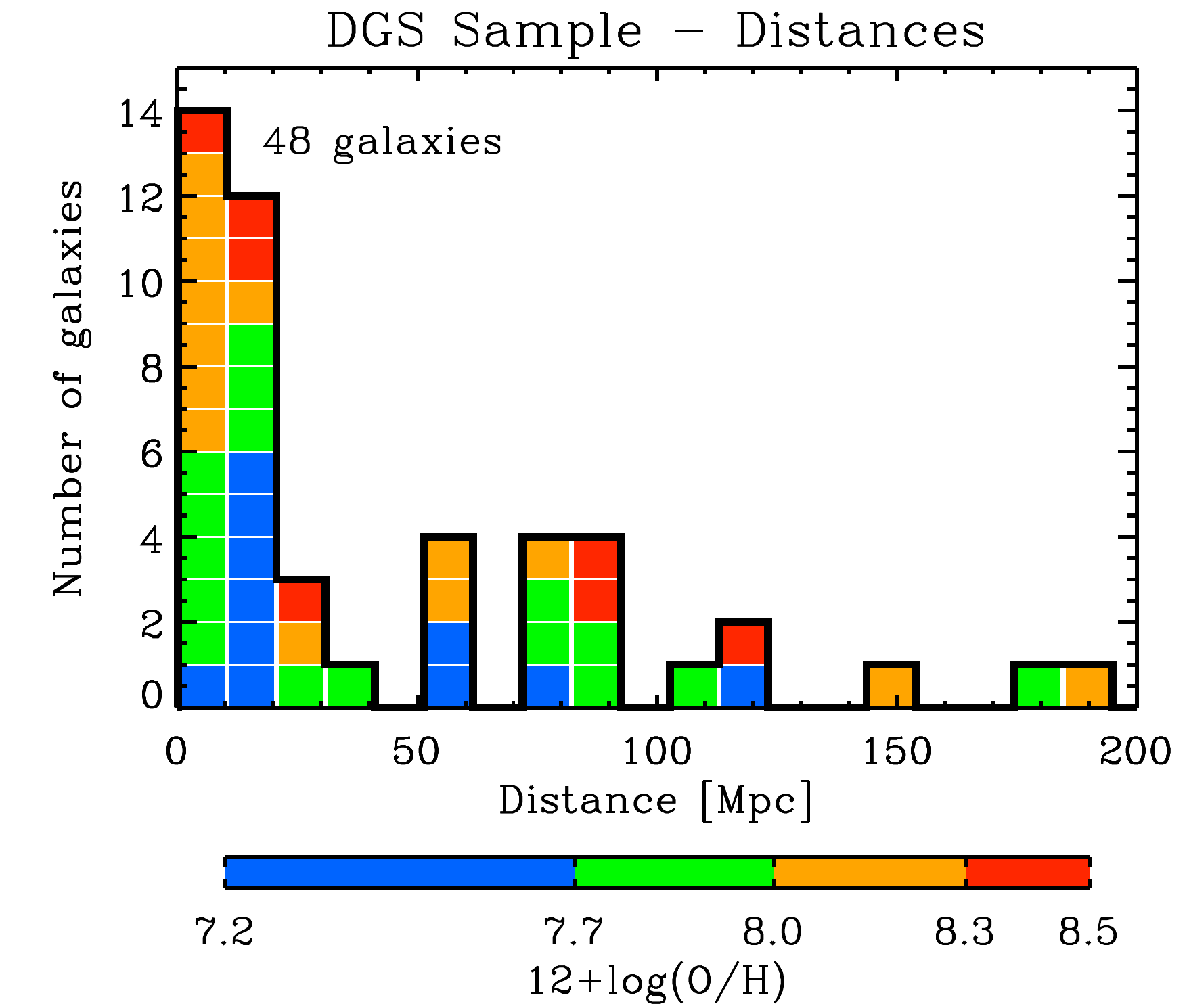}
\includegraphics[width=6.7cm]{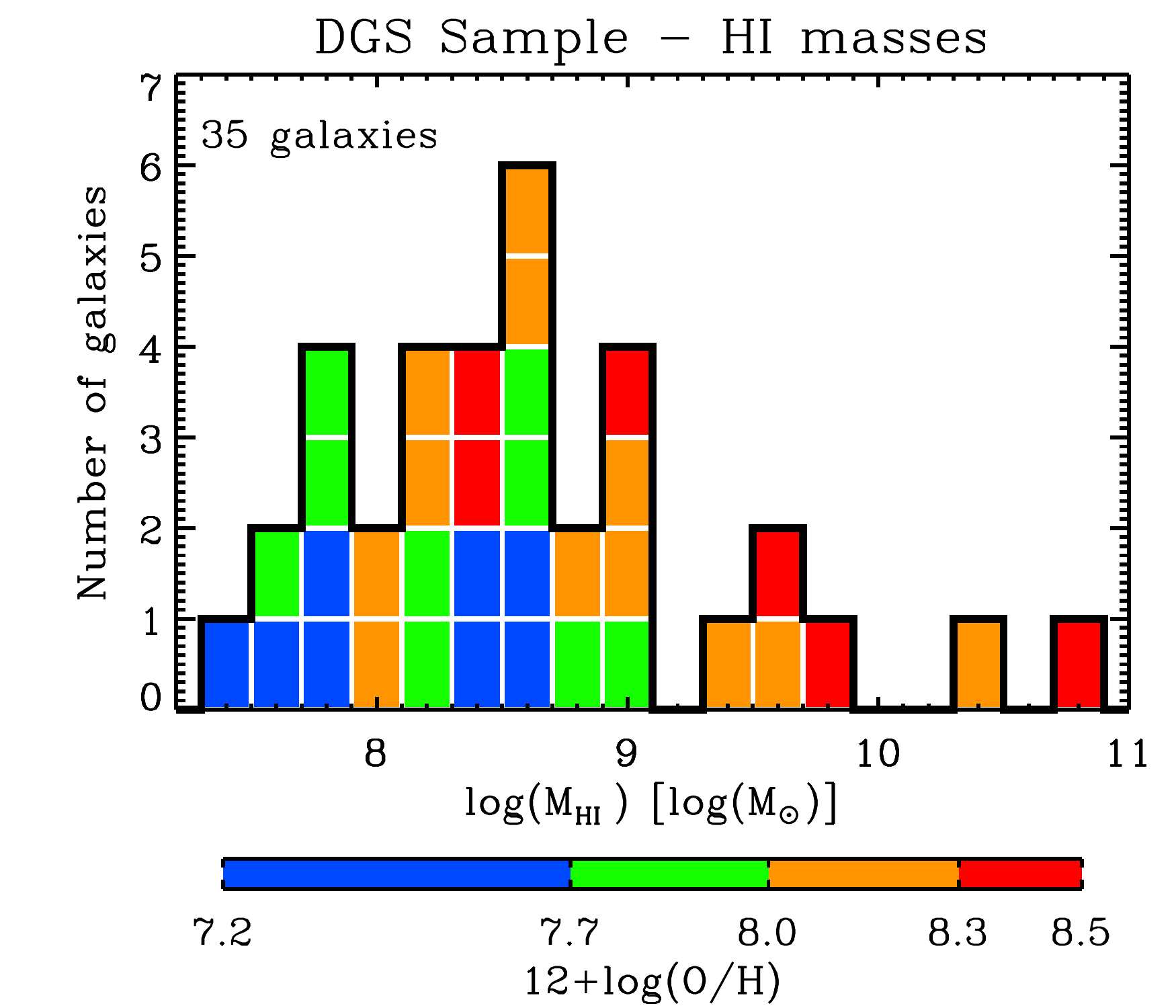}
\vspace{1cm}
\hspace{1cm}
\includegraphics[width=6.7cm]{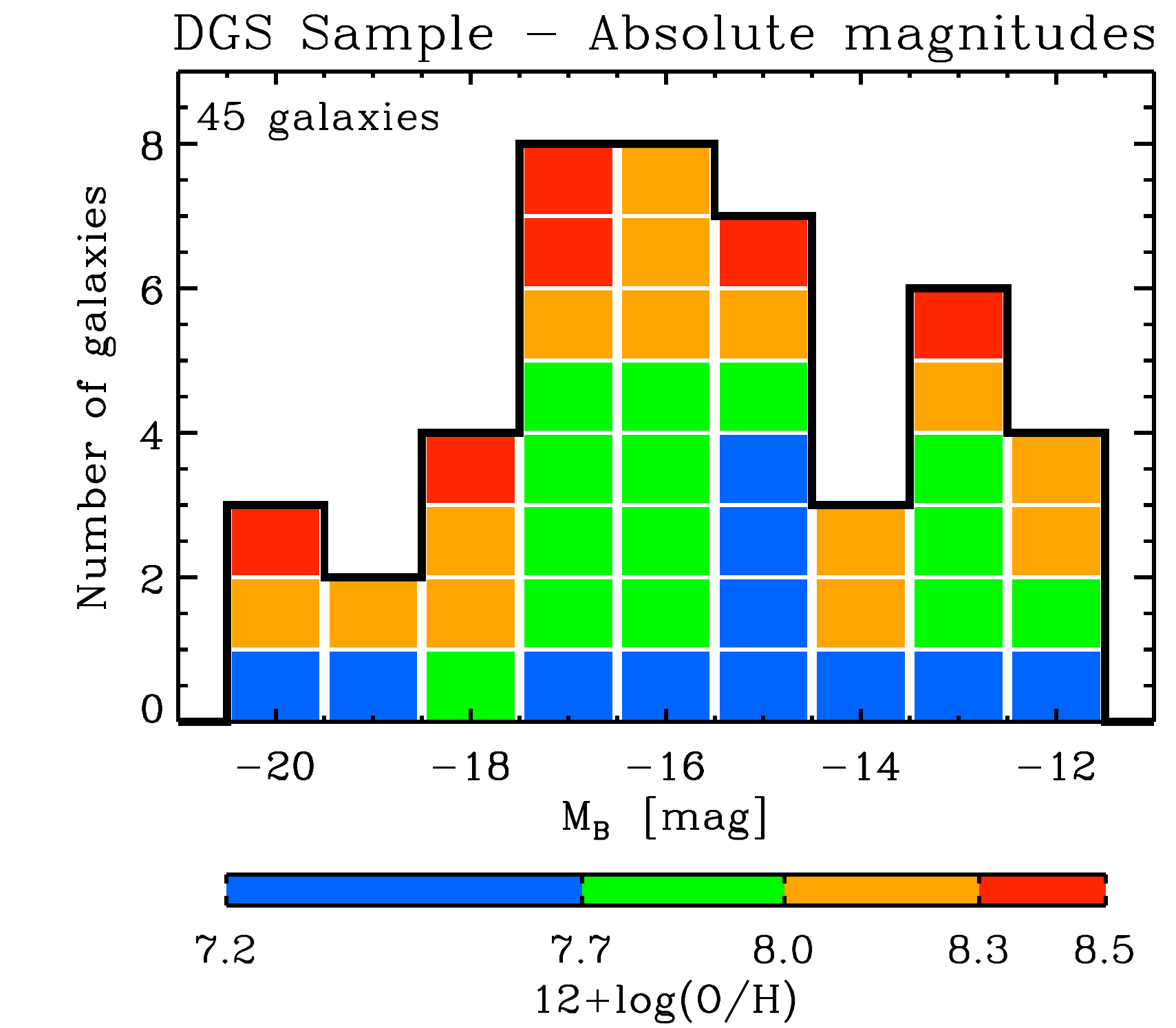}
\includegraphics[width=6.7cm]{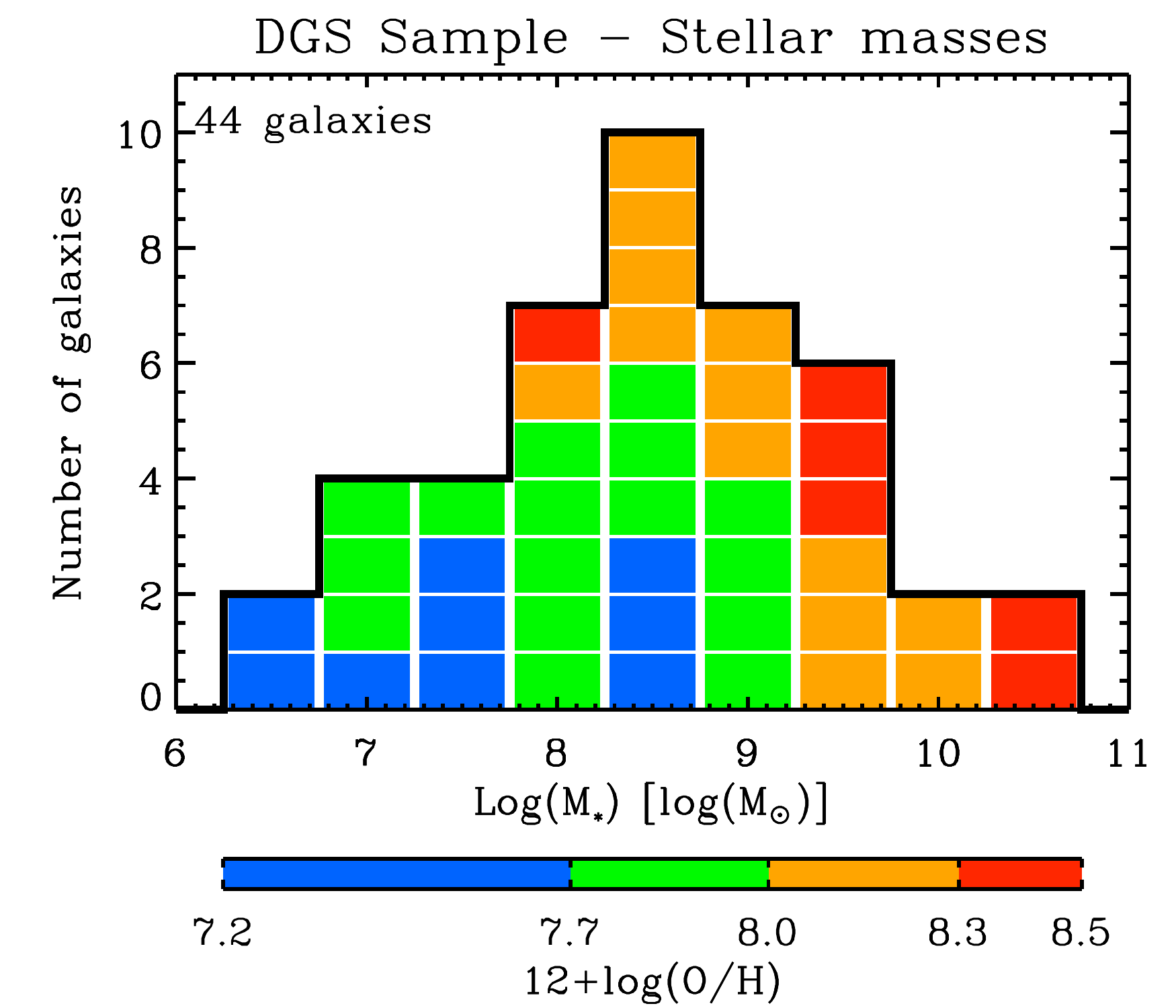}
\vspace{1 cm}
 \vspace{-1.0 cm}
\caption{\footnotesize{Properties of the DGS sample : Distribution of metallicity and SFR properties {\it (top left)} computed from {\it L$_{TIR}$} as in \cite{dale02}, if \iras\ or \spitz\ data is available. Otherwise SFR was determined from H$_{\alpha}$ or H$_{\beta}$ \citep{kennicutt98}. When data were not available they are labeled 'not available'. Histograms of distances {\it (top right)}, HI masses {\it(middle left)}, absolute B magnitude {\it(middle right)} and stellar masses, {\it M$_*$} {\it(bottom)}, from the formula of \cite{eskew12} using 3.6 and 4.5 \mic. The color scale represents the range of star formation rate for the top panel and metallicity for the others.}}
\label{Histo_SAG2}
  \end{figure}

The target list is presented in Table~\ref{sample} along with galaxy characteristics that include:
Galaxy name (column 1), RA and DEC (columns 2 and 3) and distance (column 4).  Any distance quoted here not referenced to the literature, has been determined using the velocities given in NED and the \cite{mould00} model, assuming H$_0$ = 70 km s$^{-1}$ Mpc$^{-1}$. Metallicities (in terms of oxygen abundances) are in column 5; M$_B$ (column 6), galaxy size in terms of D$_{25}$ (column 7), {\it M$_*$}, stellar mass, from the formula of \cite{eskew12} using 3.6 and 4.5 \mic\ (column 8);  M$_{HI}$, mass of HI (column 9);  {\it L$_{TIR}$} determined from Spitzer bands \citep{dale02} (column 10) and star formation rate (SFR) using the IR-based calibration of \cite{kennicutt98} (column 11). 

{\revised Determination of absolute metallicity values can vary, depending on the calibration method used \citep[e.g.][and references within]{kewley08}.  In order to ensure homogeneity in the calibration of the metallicity values in the DGS sample, metallicities were recomputed from optical line intensities following the methods of \cite{izotov06} (I06) and \cite{pilyugin05} (PT05) and compared. The method of I06 uses the direct measurement of the electron temperature T$_{e}$[OIII] from the [OIII]$\lambda$4363 line.
\cite{pilyugin05} is an empirical calibration of the metallicity value using the R$_{23}$ ratio determined from strong emission lines: R$_{23}$ = ([OII]$\lambda$3727+[OIII]$\lambda\lambda$4959,5007)/H$\beta$. To break the degeneracy between the upper and lower branches of the calibration \citep{kobulnicky04}, we use the [NII]$\lambda6584$/[OII]$\lambda$3727 ratio as advised in \cite{kewley08}. The mean of the variations between the PT05 and I06 values for this sample is about 0.1 dex (see Appendix A for comparison and discussion). For the presentation of the DGS, we choose the PT05 values for consistency (Table~\ref{sample}). Uncertainties on the metallicity values are computed by propagating the errors in the observed lines.
 }

 The wide range in the distribution of the DGS sample in metallicity, SFR, distance, M$_{HI}$, M$_{B}$, and M$_*$ are presented as histograms in Figure~\ref{Histo_SAG2}.
 
\section{Observation strategy and data reduction}
\label{observing_section}
The DGS observed {\it both} the dust and gas components in the target sources with a total observing time of  230h. With \hers\ the FIR PACS bands of 70, 100 and 160 \mic\ and the submm bands of SPIRE at 250, 350 and 500 \mic\ are mapped. The fine-structure lines of [CII] 158 \mic,  [OIII] 88 \mic\ , [OI] 63, [OI] 145 \mic\ and in few cases [NIII] 57 \mic,  [NII] 122 \mic\ and [NII] 205 \mic\ have been observed. SPIRE FTS submm observations have been obtained for 4 DGS sources.

The survey contains  37 compact galaxies with sizes ranging from 0.08' to 1.2' and 11 well-studied extended galaxies (Table~\ref{sample}), with the most extended ones being NGC~625, NGC~2366, NGC~4214, NGC~4449, NGC~4861 and the Local Group galaxies, IC~10, NGC~6822 and the Magellanic Clouds.  All of these galaxies have also been observed with \spitz\ photometry bands and the \spitz\ IRS (Figure ~\ref{detection_histo}). 

Since the low surface brightness dust in diffuse metal-poor regions is important in studying the dust processing and may be associated with the extended HI structures in dwarf galaxies, we planned observations to completely cover the star forming regions and where possible much of the HI structures. A nominal size of at least 2 $\times$ D$_{25}$ was considered for most photometry maps. 

\begin{figure}
     \centering
\includegraphics[width=16cm]{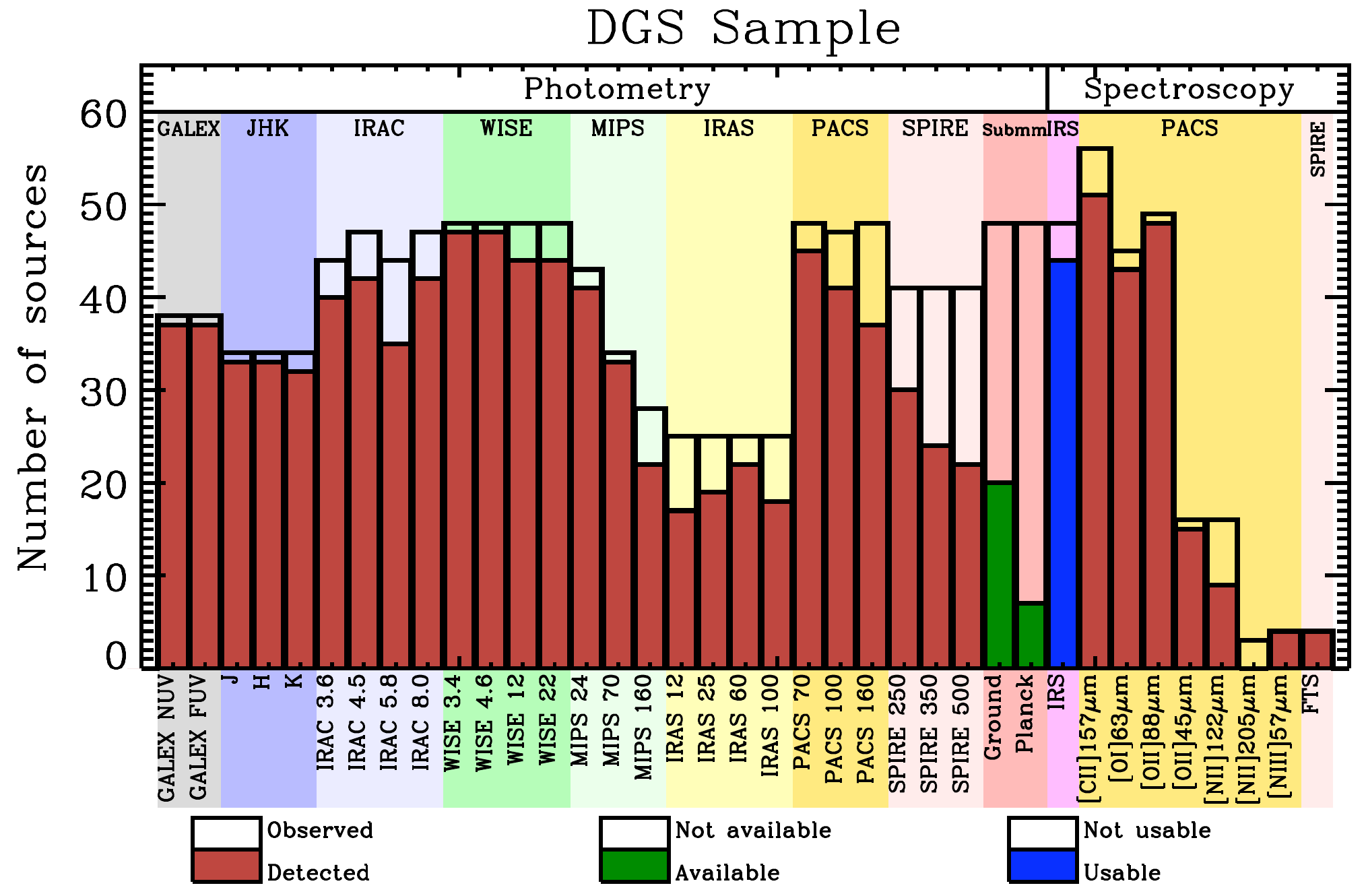}
\caption{Histogram of the number of observed and detected \hers\ sources in the DGS program along with the accompanying ancillary data that has been collected. Note that PACS spectroscopy includes 7 individual LMC and SMC sources, thus accounting for the larger number (55 sources) in the histogram while the accompanying LMC and SMC \hers\ photometry data is in the HERITAGE program \cite{meixner13}.}
\label{detection_histo}
  \end{figure}


\subsection{PACS imaging}
 As 2 of the 3 PACS bands \citep{poglitsch10} can be observed simultaneously (either 70~\mic\ and 160 \mic\ or 100 \mic\ and 160 \mic), 2 sets of observations were required to obtain all of the 3 PACS bands. Mapping the largest galaxies of the Local Group with the 3 PACS bands required $\sim$10h of observing time per galaxy. For each band, we obtained two sets of maps with scanning angles separated by 90 degrees to reduce the possibility of striping in the maps. All of the sources were observed in the PACS scan-map mode with map sizes varying from 4'x4' to 30'x30' using medium scan speed.
  
The PACS data were reduced using the \hers\ Interactive Processing Environment (HIPE) version 7.0. The details of the data reduction and the various map reconstruction techniques and tests are described in \cite{remy13}.

Three different map-making methods, PhotProject, MADmap, and Scanamorphos \citep{roussel12},
were performed on all of the DGS PACS data and resulting aperture photometry values were compared to determine the preferred method. The PACS bolometer thermal drifts, in the form of 1/f noise, leave imprints on the maps which manifest in large scale spatial structures. These are removed in PhotProject via high-pass filtering, having the tendency to suppress extended, large scale emission. The MADmap method creates maximum likelihood maps from the time ordered data, assuming the detector noise is Gaussian, which is not the case for the 1/f noise. Extended features tend to be reproduced somewhat better than in PhotProject maps. Scanamorphos takes advantage of the redundancy in the maps and accounts for the correlated and uncorrelated, non-thermal noise. From Level 1 data treatment in HIPE, the data were exported into Scanamorphos and maps reconstructed using the default options.  After detailed tests comparing all methods 
\cite{remy13},
PhotProject was in most cases found to be more adapted for point sources while Scanamorphos was chosen for the extended galaxies.

The FWHM beam sizes for PACS bands are 5.2", 7.7" and 12" for the 70, 100 and 160 \mic\ observations while final pixel sizes are 2", 2" and 4", respectively. The sensitivities we achieve in the PACS observations are normally a 1 $\sigma$ value ranging $\sim$ 1.2 - 8.4 MJy~sr$^{-1}$ for 70 and 100 \mic\  and 0.6 - 5.6 MJy~sr$^{-1}$ for 160 \mic\  (Table~\ref{rms_photom}). Calibration uncertainties are 3, 3, and 5\% for the 70, 100 and 160 \mic\ PACS bands. We obtain a S/N of at least 10 for the brightest galaxies or the brightest regions in galaxies, and 3 to 5 $\sigma$ detections typically for the fainter (lowest metallicity) galaxies.

\begin{deluxetable}{lcccccccccc}
\tabletypesize{\tiny}
\rotate
\tablewidth{0pt}
\tablecolumns{11}
\tablecaption{ Characteristics of the DGS sample \label{sample}} 
\tablehead
{ 
\colhead{Source} & \colhead{RA} & \colhead{DEC} & \colhead{Distance}  & \colhead{12 + log(O/H)}  & \colhead{{\it M$_B$}}  & \colhead{Size ({\it D$_{25}$})} & \colhead{{\it M$_{\star}$}} & \colhead{{\it M$_{HI}$}}  & \colhead{Log({\it L$_{TIR}$})}  & \colhead{{\it SFR}} \\
\colhead{}		& \colhead{}		&\colhead{}		& \colhead{Mpc (ref)}&		\colhead{(ref)}       & \colhead{(ref)}      &  \colhead{arcmin $\times$ arcmin} & \colhead{10$^{8}$ \msun} & \colhead{10$^8$ \msun (ref)} & \colhead{Log(\lsun)}&  \colhead{\sfr} 
}
 \startdata
Haro 11   	  	  	&0h36m52.7s		&-33d33m17.0s  	&92.1	(1) 		&8.36  $\pm$    0.01		(1)		&-20.55 	(1)	&0.5'x0.4'		&339.0 & 2.5 	(1) &	11.22	&	28.6  \\
Haro 2  	  	  	&10h32m32.0s 		&+54d24m02.0s 	&21.7 	(2)		&8.23  $\pm$    0.03	         (2) 		&-18.28 	(1)	&1.2'x0.8' 		& 20.8 &4.8			(2)	&	9.72	&	0.90	\\
Haro 3 	  	  	&10h45m22.4s 		&+55d57m37.0s 	&19.3 	(3)		&8.28  $\pm$    0.01		(3) 		&-18.29	(1) 	&1.2'x0.8'		& 18.8&11.3		(3)	&	9.67	&	0.80 \\
He 2-10 			&08h36m15.1s 		& -26d24m34.0s 	&8.7	 	(21)		&8.43  $\pm$    0.01		(4)		&-17.25 	(1) 	&1.9'x1.4'		& 18.2 &3.1			(4)	&	9.66	&	0.79 \\
HS 0017+1055  	&00h20m21.4s		&11d12m21.0s 	&79.1 	(3)		&7.63  $\pm$    0.10$^a$	(5)  		&-16.23	(2)	& - 			& 2.0 &$\leq$ 1.74	(5)	&	-	&	-	 \\
HS 0052+2536  	&00h54m56.4s		&+25d53m08.0s 	&191.0	(3)		&8.07  $\pm$    0.20$^a$ 	(5) 	&-19.56	(2)	& -			& 43.6 &$\leq$ 482	(6)	&	-	&	-	\\
HS 0822+3542 		&08h25m55.5s 		&+35d32m32.0s	&11.0	 	(4)		&7.32  $\pm$    0.03		(6) 		&-12.4$^d$ (3) 	&0.27'x0.12'	&0.03 & 0.77		(7)	&	$\leq$ 7.04	&	$\leq$ 0.002	\\
HS 1222+3741	  	&12h24m36.7s 	& +37d24m37.0s	&181.7 	(3)		&7.79  $\pm$    0.01		(7) 		&-17.9	(4)	&0.12'x0.11'	&14.6 &-				&	-	&	1.507	 \\
HS 1236+3937  	&12h39m20.2s 		&+39d21m05.0s 	&86.3 	(3)		&7.72  $\pm$    0.1$^c$		(8) 		&-16.0	(4)	&0.34'x0.09'	&12.8 &-				&	-	&	0.087	 \\
HS 1304+3529  	&13h06m24.2s 		&+35d13m43.0s 	&78.7 	(3)		&7.93  $\pm$    0.1$^c$		(8) 	&-17.73	(4)	&0.28'x0.21'	& 4.2 &-				&	-	&	0.50  \\
HS 1319+3224  	&13h21m19.7s 		&+32d08m25.0s	&86.3 	(3)		&7.81  $\pm$    0.1$^c$		(8) 		&-15.68	(4)	&0.14'x0.11'	&0.9 &-				&	-	&	-	\\
HS 1330+3651  	&13h33m08.3s 		&+36d36m33.0s 	&79.7 	(3)		&7.98  $\pm$    0.1$^c$		(8) 		&-17.81	(5)	&0.29'x0.19'	&6.3 &-				&	-	&	0.382 \\
HS 1442+4250  	&14h44m12.8s 		&+42d37m44.0s 	&14.4	(3)		&7.60  $\pm$    0.01		(9) 		&-15.15	(4)	&1.13'x0.26'	& - &3.10			(5)	&	$\leq$ 7.59	& $\leq$0.01 \\
HS 2352+2733  	&23h54m56.7s		&+27d49m59.0s 	&116.7 	(3)		&8.40  $\pm$    0.20$^a$	(5) 		&-17.05	(2)	& -			&1.8 &$\leq$ 230.0	(6)	&	-	&	-	 \\
I Zw 18  			&09h34m02.0s 		&+55d14m28.0s	&18.2 	(5)		&7.14  $\pm$    0.01		(10) 		&-15.22	(1)	&0.3'x0.3'		&0.2 &1.0		(8)	&	$\leq$ 7.74	&	$\leq$ 0.01	\\
II Zw 40  			&05h55m42.6s	 	&+03d23m32.0s	&12.1 	(20)		& 8.23  $\pm$    0.01		(11)		&-14.93	(1) 	&0.56'x0.22'	& 4.7 &5.7			(9)	&	9.4	&	0.43	\\
IC 10  	 		&00h20m17.3s		&+59d18m14.0s	&0.7 		(6)		&8.17  $\pm$    0.03 		(12) 		&-12.43	(1)	&6.8'x5.9'		& - &1.10		(10)	&	7.7$^e$	&	0.01 \\
Mrk 1089 			&05h01m37.7s 		&-04d15m28.0s 	&56.6 	(3)		&8.30/8.10  $\pm$    0.08$^b$ 	(13)	&-20.46	(1)	&0.61'x0.23'	&119.0 &275.0		(11)	&	7.1$^e$ 	&	0.002	\\
Mrk 1450  		&11h38m35.7s 		&+57d52m27.0s 	&19.8 	(3)		&7.84  $\pm$    0.01		(14) 		&-13.86$^d$ (3)&0.40'x0.31'	&0.86 &0.43		(12)	&	8.44	&	0.047	\\
Mrk 153  			&10h49m05.0s 		&+52d20m08.0s 	&40.3 	(3)		&7.86  $\pm$    0.04 		(15)		&-18.03	(1) 	&0.54'x0.30'	& 7.0 &     $\leq$6.93	(13)	&	8.97	&	0.16	 \\
Mrk 209  			&12h26m15.9s 		&+48d29m37.0s 	&5.8 		(7)		&7.74  $\pm$    0.01		(16) 		&-13.67	(1)	&0.8'x0.6'		&0.18 &0.74	(13)	&	7.49	&	0.005	\\
Mrk 930  	  		&23h31m58.3s 		&+28h56m50.0s	&77.8 	(3)		&8.03  $\pm$    0.01		(17) 		& - 			&0.43'x0.43'	&44.2 &31.9	(14)	&	10.26	&	3.12	\\
NGC 1140  		&02h54m33.6s		&-10d01m40.0s 	&20.0 	(8)		&8.38  $\pm$    0.01	 	(3) 		&-18.55	(1)	&1.7'x0.9'		&23.4 &35.0	(24)	&	9.52	&	0.57	\\
NGC 1569  		&04h30m49.0s		&+64d50m53s 		&3.1 		(9)		&8.02  $\pm$    0.02		(18) 		&-15.75	(1)	&3.6'x1.8'		&5.4 &1.8	(15)	&	9.04	&	0.19	\\
NGC 1705  		&04h54m13.5s		&-53d21m40.0s 	&5.1 		(10)		&8.27  $\pm$    0.11		(19) 		&-15.73	(1)	&1.9'x1.4'		&1.2 &0.9	(16)	&	7.82	&	0.01	\\
NGC 2366  		&07h28m54.6s		&+69d12m57.0s 	&3.2 		(11)		&7.70  $\pm$    0.01 		(20)		&-16.18	(1)	&8.1'x3.3'		&1.2&7.4	(17)	&	8.13	&	0.02	\\
NGC 4214  		&12h15m39.2s		&+36d19m37.0s 	&2.9 		(12)		&8.26  $\pm$    0.01		(4) 		&-17.11	(1)	&8.5'x6.6'		& 5.0 &4.08	(15)	&	8.73	&	0.09	\\
NGC 4449  		&12h28m11.1s		&+44d05m37.0s 	&4.2 		(13)		&8.20  $\pm$    0.11		(21)		&-17.94	(1)	&6.2'x4.4'		&14.9 &11.0	(15)	&	9.33	&	0.36	\\
NGC 4861  		&12h59m02.3s		&+34d51m34.0s 	&7.5	 	(14)		&7.89  $\pm$    0.01		(16) 		&-16.84	(1)	&4.0'x1.5'		& 1.8&4.79	(18)	&	8.49	&	0.05	\\
NGC 5253  		&13h39m55.9s		&-31d38m24.0s 	&4.0 		(12)		&8.25  $\pm$    0.02 		(4)		&-16.87	(1)	&5.0'x1.9'		& 6.7 &1.59	(19)	&	9.14	&	0.24	\\
NGC 625  		&01h35m04.6s		&-41d26m10.0s 	&3.9		(15)		&8.22  $\pm$    0.02		(22)		&-16.20	(1)	&5.8'x1.9'		& 3.0 &1.1	(29)	&	8.41	&	0.04	\\
NGC 6822  		&19h44m57.7s		&-14d48m12.0s 	&0.5 		(16)		&7.96  $\pm$    0.01		(23)		&-13.19	(1)	&15.5'x13.5'	&0.80 &1.34	(20)	&	7.6	&	0.007	\\
Pox 186  			&13h25m48.6s 		&-11d36m38.0s 	&18.3 	(3)		&7.70  $\pm$    0.01		(24)		&-12.17$^d$ (3)&-			&0.1 &$\leq$ 0.02	(21)	&	- &	0.04	\\
SBS 0335-052 		&03h37m44.0s		&-05d02m40.0s 	&56.0 	(3)		&7.25  $\pm$    0.01 		(17)		&-16.02	(6)	&0.23'x0.20'	& 4.3 &4.6	(22) 		&	$\leq$ 9.23	& $\leq$ 0.29 \\
SBS 1159+545 		&12h02m02.4s 		&+54d15m50.0s 	&57.0 	(3)		&7.44  $\pm$    0.01		(14)		&-15.08	(7)	&0.20'x0.10'	&0.5&$\leq$ 0.63	(5)	&	-	&	0.17 	\\
SBS 1211+540	  	&12h14m02.5s 		&+53d45m17.0s 	&19.3 	(3)		&7.58  $\pm$    0.01		(14)		&-13.43	(8) 	&0.26'x0.17'	& 0.1 &0.56	(12)	&	-	&	0.02	\\
SBS 1249+493 	&12h51m52.5s 		&+49d03m28.0s 	&110.8 	(3)		&7.68  $\pm$    0.02		(25)		& -			&0.10'x0.10'	&4.4 &10.0	(23)	&	-	&	-	\\
SBS 1415+437	   	&14h17m01.4s 		&+43d30m05.0s	&13.6	(17)		& 7.55  $\pm$    0.01		(26)		&-14.99$^d$ (3)&0.75'x0.15'	& - &4.37	(12)	&	$\leq$ 7.78	&	$\leq$ 0.01	\\
SBS 1533+574  	&15h34m13.8s 		&+57d17m06.0s	&54.2 	(3)		&8.05  $\pm$    0.01		(16)		&-16.69$^d$ (3)&0.67'x0.67'	&11.0 &30.0	(14)	&	$\leq$ 6.1$^e$ &	$\leq$0.0002	\\
Tol 0618-402  		&06h20m02.5s 		&-40d18m09.0s 	&150.8 	(3)		&8.09  $\pm$    0.01		(27) 		&-			&0.4'x0.3'		&109.0 &-			&	$\leq$ 9.91	&	$\leq$ 1.40	\\
Tol 1214-277  		&12h17m17.1s 		&-28d02m33.0s        &120.5      (3)            &7.52  $\pm$    0.01         (4)           &-17.00    (9)     & 0.08'x0.05'    &2.0 &$\leq$3.22      (5)   &	$\leq$9.3$^f$	&	$\leq$ 0.36	\\
 UGC 4483  		&08h37m03.0s 		&+69d46m31.0s 	&3.2 		(11)		&7.46  $\pm$    0.02		(28) 		&-12.53	(1)	&1.4'x0.7'		&0.03 &0.32	(28)	&	6.61	&	0.0007	\\
UGCA 20  	 	&01h43m14.7s		&+19d58m32.0s 	&11.0 		(18)		&7.50  $\pm$    0.02		(29)		&-14.43	(2)	&3.1'x0.8'		&0.18 &0.20	(25)	&	-	&	0.01 \\
UM 133  			&01h44m41.3s		&+04d53m26.0s 	&22.7 	(3)		&7.82  $\pm$    0.01		(4) 		&-15.72$^d$ (3)&1.0'x1.0'		&0.81 &4.0	(26)	&	-	&	0.02 \\
UM 311  			&01h15m34.40s	&-00d51m46.0s 	&23.5 	(3)		&8.38/8.36 $\pm$    0.01$^b$	(17,30,31) 		&-19.22	(10)	&0.11'x0.10'	& - &620.0	(27)	&	9.73	&	0.94	\\
UM 448  			&11h42m12.4s 		&+00d20m03.0s 	&87.8 	(3)		&8.32 $\pm$    0.01		(17) 		&-20.02	(1)	&0.4'x0.4'		&241.0 &60.0	(24)	&	10.86	&	12.36	\\
UM 461  			&11h51m33.3s 		&-02d22m22.0s 	&13.2 	(3)		&7.73 $\pm$    0.01		(15)		&-13.73	(6)	&0.30'x0.22'	&0.26 &1.34	(28)	&	7.8	&	0.01	\\
VII Zw 403  		&11h27m59.9s 		&+78d59m39.0s 	&4.5		(19)		&7.66 $\pm$    0.01		(16)		&-13.77	(1) 	&1.4'x0.8'		&0.10 &0.69	(2)	&	7.23	&	0.003 \\
 \enddata
 
\end{deluxetable} 
\clearpage

\begin{center}
Notes to Table~\ref{sample}
\end{center}

{\tiny
\noindent 
Positions are from the NASA/IPAC Extragalactic Database (NED). Distances are from the literature. Metallicity values in 12 + log(O/H) are determined as in \cite{pilyugin05} (PT05) using the R$_{23}$ ratio except for where noted. See text for further details on methods. Absolute B magnitude, M$_B$, is derived from the corresponding apparent magnitudes from NED. D$_{25}$ is the galaxy diameter from NED. M$_{\star}$ is stellar mass converted from 3.6 and 4.5 \mic\ following the method of \cite{eskew12}.  M$_{HI}$ is from the literature. Total infrared luminosities ({\it L$_{TIR}$}) have been computed using the formula of \cite{dale02} and the \spitz\ fluxes from \cite{bendo12b} when available, unless otherwise noted. Otherwise, \iras\ fluxes are used. Star formation rates (SFR) has been derived from {\it L$_{TIR}$} using the formula of \cite{kennicutt98} or from H$\alpha$ and H$\beta$ when no IR data is available. A dash indicates that no data is available. 

\noindent 
{\bf References for distances }: (1) \cite{bergvall06} ; (2) \cite{kennicutt03} ; (3) this work, calculated from the redshifts available in NED, the Hubble flow model from \cite{mould00} and assuming H$_0$ = 70 km s$^{-1}$ Mpc$^{-1}$ ; (4) \cite{pustilnik03} ; (5) \cite{aloisi07} ; (6) \cite{kim09} ; (7) \cite{schulte-ladbeck01} ; (8) \cite{moll07} ; (9) \cite{grocholski12} ; (10) \cite{tosi01} ; (11) \cite{karachentsev/Users/madden/Herschel/SPIRE/SAG2/Dwarf Galaxy Survey/DGS_Overview_paper/PASP version/aastex52/submit/resubmit/resubmit2/Figures/Madden_resubmit3_8april/PASP-350905_p1.pdf02} ; (12) \cite{karachentsev04} ; (13) \cite{karachentsev03} ; (14) \cite{deVaucouleurs91} ; (15) \cite{cannon03} ; (16) \cite{gieren06} ; (17) \cite{aloisi05} ; (18) \cite{sharina96} ; (19) \cite{lynds98} ; (20) \cite{bordalo09} ; (21) \cite{tully88} 

\noindent 
{\bf References for metallicities} : (1) \cite{guseva12} ; (2) \cite{kongcheng02} ; (3) \cite{izotov04} ; (4) \cite{kobulnicky99} ; (5) \cite{ugryumov03} ; (6) \cite{pustilnik03} ; (7) \cite{izotov07} ; (8) \cite{popescu00} ; (9) \cite{guseva03a} ; (10) \cite{izotov99} ; (11) \cite{guseva00} ; (12) \cite{magrini09} ; (13) \cite{lopezsanchez10} ; (14) \cite{izotov94} ; (15) \cite{izotov06} ; (16) \cite{izotov97} ; (17) \cite{izotov98} ; (18) \cite{kobulnicky97} ; (19) \cite{lee04} ; (20) \cite{saviane08} ; (21) \cite{mccall85} ; (22) \cite{skillman03} ; (23) \cite{lee06} ; (24) \cite{guseva07} ; (25) \cite{thuan95} ; (26) \cite{guseva03b} ; (27) \cite{masegosa94} ; (28) \cite{van_zee06} ; (29) \cite{van_zee96} ; (30) \cite{moles94} ; (31) \cite{pilyugin07} 

\noindent 
{\bf References for apparent magnitudes} : (1) \cite{deVaucouleurs91} ; (2) \cite{ugryumov03} ; (3) \cite{gildepaz05}; (4) \cite{vennik00} ; (5) \cite{ugryumov01} ; (6) \cite{maddox90} ; (7) SDSS DR4 ; (8) \cite{paturel03} ; (9) \cite{fricke01} ; (10) \cite{smoker00} 

\noindent 
{\bf References for HI masses} : (1) \cite{machattie13} assuming T$_{spin}$ of 91 K.  Assuming upper limit on T$_{spin}$ (320 K) results in upper limit of M$_{HI}$: $\leq$4.2 $\times$ 10$^9$ \msun ; (2) \cite{thuan04} ; (3) \cite{gordon81} ; (4) \cite{sauvage97} ; (5) \cite{pustilnik07} ; (6) \cite{machattie13} ; (7) \cite{chengalur06} ; (8) \cite{lelli12} ; (9) \cite{bettoni03} ; (10) \cite{huchtmeier88} ; (11) \cite{williams91} ; (12) \cite{huchtmeier05} ; (13) \cite{thuan81} ; (14) \cite{thuan99b} ; (15) \cite{walter08} ; (16) \cite{meurer98} ; (17) \cite{hunter11} ; (18) \cite{van_eymeren09} ; (19) \cite{lopez-sanchez12} ; (20) \cite{deblok06} ; (21) \cite{begum05} ; (22) \cite{ekta09} ; (23) \cite{pustilnik02} ; (24) \cite{davoust04} ; (25) \cite{paturel03} ; (26) \cite{ekta10} ; (27) \cite{moles94} ; (28) \cite{van_zee98} ; (29) \cite{cannon04}  

\noindent 
{$^a$ For these galaxies, no line intensities were available in the literature to determine the metallicities using the PT05 method. We use \cite{ugryumov03} where the [OIII]$\lambda$4363 line is used to obtain the electron temperature and the method from \cite{izotov94}. I06 is an updated version of the method of \cite{izotov94}. }

\noindent 
{$^b$ These galaxies are part of a group of galaxies. We list the value of the metallicity for the galaxy only (1$^{st}$ value quoted) and for the whole group (2$^{nd}$ value quoted), which is the mean of all of the metallicities in the group. For Mrk1089, the galaxy is region A-C from \cite{lopezsanchez10}. For UM311, the galaxy is region 3 following \cite{moles94}. } 

\noindent 
 {$^c$ For these galaxies, no line intensity uncertainties, from which the metallicities are determined, are quoted. We assume a conservative value of 0.1 which is the mean of the variation of the difference between this metallicity calibration, PT05, and the I06 method (see Appendix A).} 
 
\noindent 
{$^d$ corrected for Galactic extinction} 

\noindent
{$^e$ IRAS fluxes from NED}

\noindent
{$^f$ MIPS fluxes from \cite{engelbracht08} }

}
\newpage


The improvement in spatial resolution of \hers\ over \spitz\ can be immediately seen in Figure~\ref{ngc1705}, presenting the nearby (5 Mpc) galaxy, NGC~1705, in the DGS \citep{ohalloran10}. Two clusters that are blended at FIR wavelengths with \spitz\ are resolved with PACS on \hers, allowing for detailed studies of the cooler dust properties for the first time.

 \begin{figure}
     \centering
\includegraphics[width=15cm]{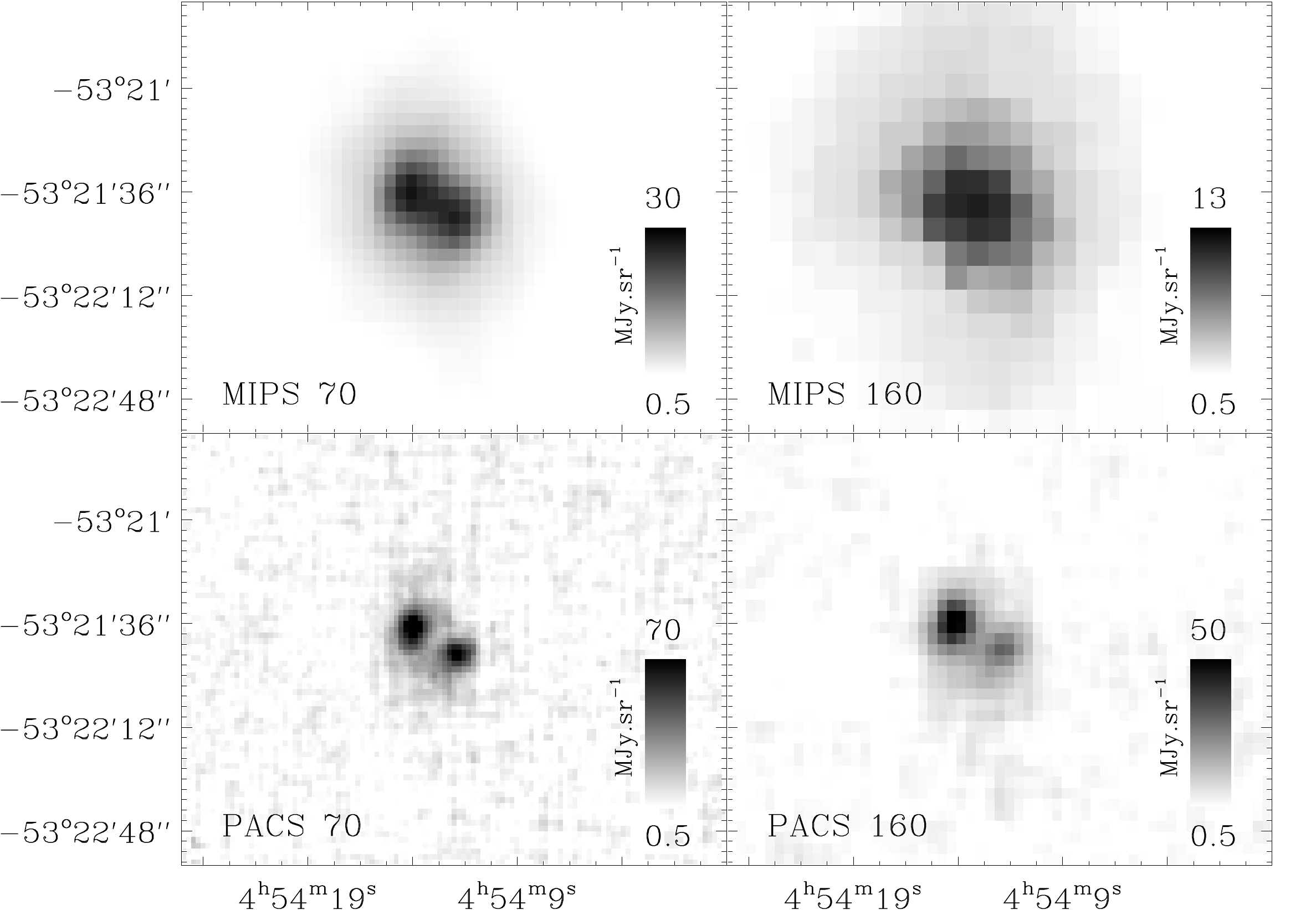}
\caption{ Improvement in spatial resolution of \hers\ in the FIR over \spitz\ is demonstrated on the DGS galaxy NGC1705 \citep{ohalloran10} in the comparison of MIPS (top row) with PACS (bottom row). The two clusters are well separated in both PACS images whereas they are blurred in MIPS images and indistinguishable at 160 \mic. The  FWHM of the PACS beams for the 70 and 160 \mic\ bands are 5.7" and 12" while for the MIPS bands the FWHM values are 18" and 38". \label{ngc1705}}
  \end{figure}

\subsection{PACS spectroscopic observations}
\label{spectro_obs} 
The PACS spectrometer footprint consists of 5 x 5 pixels of 9.4" each, with a total field of view of 47" x 47". Each spatial pixel covers 16 spectral elements. In only a few cases were all of the brightest 7 FIR fine structure lines observed (Table \ref{spectro_sample}): 158 \mic\ [CII], 63 and 145 \mic\ [OI], 88 \mic\ [OIII], 57 \mic\ [NIII] and the 122 and 205 \mic\ [NII] lines. In all but a few of the faintest galaxies, at least the [CII] and [OIII] were observed (Figure \ref{detection_histo}).  The spectral resolutions for the [CII], [OIII] and [OI] lines are: $\sim$ 1250 (240 km s$^{-1}$), $\sim$ 2400 (125 km s$^{-1}$),  and $\sim$ 3300 (90 km s$^{-1}$), respectively.
The chop/nod line observing mode was used for the compact sources which fit within the spectrometer footprint. For sources too extended for chopping (maximum chop of +/- 3 arcmin), we made raster maps using the unchopped grating scan mode, which replaced the decomissioned wavelength switching mode, used only for LMC/N11B \citep{lebouteiller12}. The data have been reduced to Level 1 within HIPE and then input into PACSman, an IDL package to handle line-fitting and map projection for PACS observations \citep{lebouteiller12}.  Maps are projected on a subpixel grid with with each final pixel size about 3" $\times$ 3", roughly 1/3 of the original spatial pixel. Details of the PACS data processing of the survey will be presented in \cite{cormier13}.

 The size of the spectroscopic maps vary depending on source and line. (Table~\ref{spectro_sample}).   
With the FWHM of 9.5" at 60-90 \mic\ and 11.5" at 150 \mic, the spatial resolution of \hers\ allows us to explore the morphology of the galaxies, and determine the physical properties traced by the FIR cooling lines for the first time {\it within} dwarf galaxies, not just around the brightest star forming regions. 
We achieved a dynamic range in flux values greater than or equal to 10 in [CII], [OI] 63 \mic, and [OIII] observations - in some cases as high as 100, in the brightest sources, thus, providing the sensitivity to extract physical information from the more diffuse ISM. In some Local Group galaxies, the S/N can be as high as 100 for some lines in the extended regions. For the more compact sources (often more distant), the median S/N is around: 13 for [CII], 18 for [OI] 63\mic, 38 for [OIII], 6 for [NII]122\mic, and 15 for [OI]145\mic.  As an example of the kind of sensitivity brought to these studies by PACS, \cite{madden97} detected low surface brightness extended [CII] emission in IC~10 at the level of 1x10$^{-15}$~W~m$^{-2}$ in a 55'' beam of FIFI on the KAO. If such emission were spread uniformly over the beam, the expected signal at the PACS spatial resolution would be $\sim$ 4x10$^{-17}$~W~m$^{-2}$. In comparison, the sensitivity of PACS after single spectroscopic integration (effective on-source time of 150s) for this line, in an equivalent KAO beam, is 2.9x10$^{-18}$~W~m$^{-2}$, and thus was rapidly detected with PACS with a S/N ranging from 10 to 100 across IC10.



\subsection{SPIRE imaging}
SPIRE \citep{griffin10} observed 250, 350 and 500 \mic\ bands simultaneously, with FWHM beam sizes of 18.1", 25.2" and 36.6", respectively.  The details of the data reduction are described in  \cite{remy13}.  Sixteen galaxies  including those with sizes larger than 1.5' were observed in the large scan-map mode, while the remaining galaxies were observed in small scan-map mode. Observations were performed as cross-scans, thus removing 1/f noise and at nominal speed (30" s$^{-1}$).   SPIRE data were processed mainly through HIPE, using the waveletDeglitcher. The residual baseline subtraction applies the median of the bolometer timelines for the whole observation, not only a single scan leg (BriGadE method, Smith et al, in preparation). Calibration uncertainties are estimated to be $\sim$7\% for all SPIRE bands.  At 250 \mic, 8 sources are detected and unresolved and 11 are not detected. At 500 \mic, 5 sources are unresolved and 19 are not detected (Figure~\ref{Histo_SAG2}). The 1 $\sigma$ sensitivities we achieved with SPIRE range from $\sim$ 0.4 MJy~$ sr^{-1}$ - 8.8 MJy~$sr^{-1}$, 0.3 - 4.8 MJy~$sr^{-1}$ and 0.1 - 1.8 MJy~$sr^{-1}$, for 250, 350 and 500 \mic, respectively (Table~\ref{rms_photom}) and final pixel sizes are 6", 8" and 12", respectively. Mapping the 11 larger galaxies takes about 1h at most per galaxy to reach the desired sensitivities with SPIRE while the small scan mode required about 10 minutes per galaxy.
 
\begin{deluxetable}{lcccccc} 
\tabletypesize{\scriptsize} 
\tablecolumns{7} 
\tablewidth{0pt} 
\tablecaption{\hers\ PACS and SPIRE 1$\sigma$ rms values (MJy sr$^{-1}$) of final maps. \label{rms_photom} }
\tablehead{ 
\colhead{Source}  & \colhead{70 \mic\ } & \colhead{100 \mic\ }& \colhead{160 \mic\ } & \colhead{250 \mic\ } & \colhead{350 \mic\ } & \colhead{500 \mic\ } 
        }
\startdata            
Haro11          &           5.29 &           7.18 &           2.55 &           0.87 &           0.52 &           0.18 \\
Haro2           &           4.95 &           4.87 &           2.06 &           0.87 &           0.51 &           0.20 \\
Haro3           &           7.60 &           7.54 &           5.57 &           1.01 &           0.68 &           0.25 \\
He2-10          &           8.46 &           8.27 &           4.10 &           1.04 &           0.62 &           0.30 \\
HS0017+1055     &           2.93 &           2.96 &           1.27 &           0.62 &           0.33 &           0.12 \\
HS0052+2536     &           2.89 &           2.72 &           1.39 &           0.76 &           0.36 &           0.15 \\
HS0822+3542     &           2.33 &           2.01 &           1.04 &            - &            - &            - \\
HS1222+3741     &           2.90 &           3.07 &           1.36 &            - &            - &            - \\
HS1236+3937     &           3.24 &           3.28 &           1.46 &           0.77 &           0.44 &           0.21 \\
HS1304+3529     &           3.24 &           4.08 &           1.55 &           0.94 &           0.49 &           0.16 \\
HS1319+3224     &           1.91 &           1.90 &           1.03 &            - &            - &            - \\
HS1330+3651     &           2.38 &           2.40 &           0.98 &            - &            - &            -\\
HS1442+4250     &           3.83 &           3.54 &           1.78 &           0.93 &           0.56 &           0.21 \\
HS2352+2733     &           2.30 &           2.11 &           0.88 &           0.72 &           0.47 &           0.13 \\
IZw18           &           1.94 &           1.80 &           0.98 &           0.48 &           0.24 &           0.10 \\
IC10            &           4.45 &           4.71 &           3.55 &           8.82 &           4.79 &           1.82 \\
IIZw40          &           5.23 &           5.09 &           3.15 &           1.67 &           0.81 &           0.36 \\
Mrk1089         &           7.88 &           7.77 &           6.02 &           1.14 &           0.64 &           0.29 \\
Mrk1450         &           2.67 &           2.67 &           1.22 &           0.88 &           0.77 &           0.37 \\
Mrk153          &           4.42 &           4.72 &           2.52 &           0.56 &           0.41 &           0.20 \\
Mrk209          &           7.08 &           7.28 &           3.79 &           0.62 &           0.38 &           0.13 \\
Mrk930          &           3.78 &           3.49 &           1.73 &           0.79 &           0.40 &           0.15 \\
NGC1140         &           3.77 &           3.68 &           1.77 &           0.87 &           0.50 &           0.19 \\
NGC1569         &           3.78 &           3.80 &           2.07 &           2.38 &           1.28 &           0.46 \\
NGC1705         &           4.08 &           3.96 &           1.79 &           0.86 &           0.54 &           0.25 \\
NGC2366         &           3.28 &           3.27 &           1.44 &           0.65 &           0.39 &           0.17 \\
NGC4214         &           3.79 &           3.66 &           1.72 &           0.91 &           0.56 &           0.25 \\
NGC4449         &           3.45 &           3.34 &           1.63 &           0.83 &           0.48 &           0.21 \\
NGC4861         &           3.20 &           3.06 &           1.40 &           0.80 &           0.49 &           0.22 \\
NGC5253         &           3.43 &           3.35 &           1.57 &           0.90 &           0.49 &           0.21 \\
NGC625          &           3.70 &           3.55 &           1.60 &           0.80 &           0.46 &           0.19 \\
NGC6822         &           4.21 &           4.05 &           1.99 &           3.06 &           1.71 &           0.64 \\
Pox186          &           3.69 &           3.34 &           1.40 &           0.67 &           0.39 &           0.18 \\
SBS0335-052     &           2.26 &           1.99 &           1.01 &           0.67 &           0.38 &           0.16 \\
SBS1159+545     &           2.65 &           1.96 &           0.98 &           0.58 &           0.33 &           0.16 \\
SBS1211+540     &           1.21 &           1.29 &           0.64 &           0.59 &           0.34 &           0.13 \\
SBS1249+493     &           3.45 &           3.34 &           1.45 &           0.80 &           0.49 &           0.17 \\
SBS1415+437     &           3.75 &           3.80 &           1.67 &            - &            - &            - \\
SBS1533+574     &           5.35 &           5.27 &           2.29 &           1.18 &           0.87 &           0.39 \\
Tol0618-402     &           1.50 &           1.50 &           0.90 &            - &            - &            - \\
Tol1214-277     &           1.53 &           1.44 &           0.73 &           0.64 &           0.38 &           0.18 \\
UGC4483         &           5.32 &           6.50 &           2.26 &           0.60 &           0.34 &           0.15 \\
UGCA20          &           3.65 &           3.80 &           1.55 &            - &           - &            - \\
UM133           &           3.74 &           3.64 &           1.86 &           0.75 &           0.44 &           0.18 \\
UM311           &           3.63 &           3.46 &           1.55 &           0.82 &           0.51 &           0.21 \\
UM448           &           7.74 &            NaN &           4.47 &           0.74 &           0.47 &           0.23 \\
UM461           &           3.81 &           3.67 &           1.79 &           0.79 &           0.56 &           0.20 \\
VIIZw403        &           7.22 &           7.46 &           3.48 &           0.55 &           0.31 &           0.15 \\
\enddata
\tablecomments
 {A dash indicates lack of data for this band.}
\end{deluxetable}
\clearpage

   
\subsection{SPIRE FTS observations}
 
The SPIRE Fourier Transform Spectrometer (FTS) covers 194 to 671 \mic\  
(447 to 1550 GHz), providing simultaneous coverage, for example, of CO  
rotational lines ranging from J=4-3 through J=13-12, as well as other  
molecules and atomic fines structure lines, such as [NII] 205 \mic\  
and [CI] lines at 492 and 809 GHz.
The spectrometer array can be operated in three spectral resolutions, low~($\delta\,\nu=25\mathrm{GHz}$), medium~($\delta\,\nu=7\mathrm{GHz}$), and, high~($\delta\,\nu=1.2\mathrm{GHz}$), and is split into the Spectrometer Long Wave array  
(SLW: 303 -671 \mic) and the Spectrometer Short Wave array (SSW:  
194-313 \mic). The SLW and SSW consist of 7 and 17 unvignetted  
bolometers arranged in a hexagon pattern covering a field of view of about 3' $ 
\times$ 3'. The beam ranges from 17" to 40" across the FTS wavelength  
range. Depending on the image sampling mode, the bolometer arrays are moved around the requested pointing within the FOV in 1-point~(sparse), 4-point~(intermediate), or, 16-point~(full) jiggles, resulting in a map with beam spacings of 50.5'', 25.3'', or, 12.7'', for the SLW, and, 32.5'', 16.3'', or, 8.1'', for the SSW.

  NGC~4214 and IC10 are observed with the FTS {\revised in high spectral resolution}, using the sparse  
sampling mode for a total of 2.7 and 3.7h, respectively. He2-10 and   
30 Doradus, the most massive star forming region in our neighboring  
LMC, are observed in intermediate sampling mode for 4.6 and 2.8h,  
respectively, {\revised and also with high spectral resolution}.
The SPIRE FTS data is first processed using the extended source  
calibration of the \hers\ Science Center pipeline which assumes the source is uniformly distributed within the FOV. Since this assumption is
not necessarily accurate, we correct for the source  
distribution using the photometry maps using the method introduced in Wu et al. (2013, in  
preparation). Calibration uncertainties in the SPIRE FTS range from   
about 5\% at the lower wavelength to $\sim$10\% toward the long  
wavelength end.
 
 \begin{deluxetable}{lcccccccccccccc} 
\tabletypesize{\tiny} 
\rotate
\tablecolumns{15} 
\tablewidth{-1pt} 

\tablecaption{The DGS spectroscopy: source coverage and 1$\sigma$ flux density rms values (MJy\,sr$^{-1}$) \label{spectro_sample}}
\tablehead
{ 
\colhead{Source}  & 
    \multicolumn{2}{c}{158 \mic\ [CII]} &  
    \multicolumn{2}{c}{63 \mic\ [OI]} &  
    \multicolumn{2}{c}{145 \mic\ [OI]}  &  
    \multicolumn{2}{c}{57 \mic\ [NIII]}  & 
    \multicolumn{2}{c}{88 \mic\ [OIII]}  &  
    \multicolumn{2}{c}{122 \mic\ [NII]} &  
    \multicolumn{2}{c}{205 \mic\ [NII]}  \\
 \cline{2-3} \cline{4-5} \cline{6-7} \cline{8-9} \cline{10-11} \cline{12-13} \cline{14-15}  
   \colhead{} &  
     \colhead{size(\arcsec)} &  \colhead{1 $\sigma$}     &  
     \colhead{size(\arcsec)} &  \colhead{1 $\sigma$}   &  
      \colhead{size(\arcsec)} &  \colhead{1 $\sigma$}    &  
      \colhead{size(\arcsec)} &  \colhead{1 $\sigma$}     &  
      \colhead{size(\arcsec)} &  \colhead{1 $\sigma$}     &   
      \colhead{size(\arcsec)} &  \colhead{1 $\sigma$}     &   
       \colhead{size(\arcsec)} & \colhead{1 $\sigma$}     
        }
\startdata            
Haro\,11                   &51$\times$51& 55    &51$\times$51& 152    &51$\times$51& 41    &51$\times$51& 178   &51$\times$51& 128    &51$\times$51& 34    &47$\times$47& 49 \\
Haro\,2                   &71$\times$71& 69    &47$\times$47&135    &-&-         &-& -     &47$\times$47& 150    &-&-         &-&-\\
Haro\,3                   &95$\times$95&60    &79$\times$79&185    &71$\times$71&54    &63$\times$63&300    &79$\times$79& 145    &47$\times$47& 54    & -     &\\
He\,2-10$^a$                   &51$\times$51&85    &53$\times$53&171    &47$\times$47&49    &47$\times$47& 435    &53$\times$53&137     &-&-        &-&-\\
HS\,0017+1055      &47$\times$47& 25    &-&-        &-&-        &-&-        &47$\times$47& 105    &-&-        &-&-\\
HS\,0052+2536      &47$\times$47&49    &47$\times$47&115    &-&-        &-&-        &47$\times$47&87    &-&-        &-&-\\
HS\,0822+3542    &47$\times$47&25    &-&-        &-&-        &-&-        &-&-        &-&-        &-&-\\
HS\,1222+3741    &47$\times$47&30   &-&-        &-&-        &-&-        &47$\times$47&73    &-&-        &-&-\\
HS\,1236+3937      &47$\times$47& 50    &-&-        &-&-        &-&-        &-&-        &-&-        &-&-\\
HS\,1304+3529      &47$\times$47& 48    &-&-        &-&-        &-&-        &47$\times$47& 126    &-&-        &-&-\\
HS\,1319+3224      &47$\times$47&48    &-&-        &-&-        &-&-        &47$\times$47&93    &-&-        &-&-\\
HS\,1330+3651      &47$\times$47&48    &47$\times$47&167    &-&-        &-&-        &47$\times$47&57    &-&-        &-&-\\
HS\,1442+4250      &47$\times$47& 44    &-&-        &-&-        &-&-        &-&-        &-&-        &-&-\\    
HS\,2352+2733      &47$\times$47&29    &-&-        &-&-        &-&-        &47$\times$47&72    &-&-        &-&-\\
II\,Zw\,40            &95$\times$95&57    &79$\times$79&186    &85$\times$85&60    &-&-        &71$\times$71&176    &47$\times$47&84    &-&-\\
I\,Zw\,18            &47$\times$47&27    &-&-        &-&-        &-&-        &-&-        &-&-        &-&-\\
IC\,10$^a$            &290$\times$150&109        &250$\times$150& 242-256        &200$\times$60&39-74    &-&-        &250$\times$120&171-233        &200$\times$60& 38-79    &-&-\\
LMC-30Dor$^a$        &350$\times$180& 124-1440        &350$\times$180& 244-2040        &190$\times$140& 89        &-&-        &310$\times$190&238-1439        &190$\times$100& 107-141        &-&-\\
LMC-N11A        &47$\times$47&134    &47$\times$47&356    &-&-        &-&-        &-&-        &-&-        &-&-\\
LMC-N11B        &127$\times$127&1287        &127$\times$127&1865        &127$\times$127&49-1612        &127$\times$127& 2511 
                &127$\times$127& 1736        &127$\times$127& 50-3082        &127$\times$127& 4250 \\
LMC-N11C        &260$\times$150& 96-135        &160$\times$100& 216-241        &-&-        &-&-        &260$\times$150& 202-240    &-&-        &-&-\\
LMC-N11I            &115$\times$47&141    &115$\times$47&229    &-&-        &-&-        &115$\times$47& 189    &-&-        &-&-\\
LMC-N159S        &81$\times$81&88    &81$\times$81&247    &-&-        &-&-        &-&-        &-&-        &-&-\\
LMC-N160        &76$\times$105&130    &75$\times$103&252    &60$\times$100&122-138    &-&-        &76$\times$105& 220    &-&-        &-&-\\
Mrk\,1089            &51$\times$51& 53    &53$\times$53& 109    &47$\times$47& 67    &-&-        &47$\times$47& 152    &51$\times$51& 45    &-&-\\
Mrk\,1450            &51$\times$51&49    &47$\times$47&108    &-&-        &-&-        &47$\times$47&127    &47$\times$47&28    &-&-\\
Mrk\,153            &47$\times$47&35    &47$\times$47&81    &-&-        &-&-        &47$\times$47&103    &-&-        &-&-\\
Mrk\,209             &47$\times$47&42    &47$\times$47&156    &-&-        &-&-        &47$\times$47&128    &-&-        &-&-\\
Mrk\,930                &51$\times$51&59    &47$\times$47&140    &47$\times$47&30    &-&-        &53$\times$53&109    &47$\times$47&29    &-&-\\
NGC\,1140        &95$\times$95&68    &47$\times$47&188    &47$\times$47&34    &-&-        &63$\times$63&184    &47$\times$47&27    &-&-\\
NGC\,1569        &95$\times$143& 70    &95$\times$79& 133    &-&-        &-&-        &119$\times$71& 112    &-&-        &-&-\\
NGC\,1705        &71$\times$71&61    &47$\times$47&169    &-&-        &-&-        &63$\times$63&149    &-&-        &-&-\\
NGC\,2366        &71$\times$71&71    &63$\times$63&229    &-&-        &-&-        &47$\times$47&108    &-&-        &-&-\\
NGC\,4214$^a$        &95$\times$95&108    &111$\times$111&150        &95$\times$95&39    &-&-        &111$\times$111&242        &95$\times$95&41    &95$\times$95&117\\
NGC\,4449        &200$\times$180&82-3035        &200$\times$180&140-6175        &90$\times$70&54    &-&-        &190$\times$120&121-140        &90$\times70$& 52    &-&-\\
NGC\,4861        &85$\times$215& 79   &85$\times$100& 141-283    &85$\times$100&37-71    &47$\times$47& 378    
                &85$\times$100& 137-242    &47$\times$47& 59    &47$\times$47&197\\
NGC\,5253        &94$\times$141& 69    &78$\times$78&264    &62$\times$47&55    &-&-        &63$\times$63&216    &47$\times$47&107    &-&-\\
NGC\,625            &85$\times$85&74    &47$\times$47&183    &-&-        &-&-        &63$\times$63&158    &-&-        &-&-\\
NGC\,6822-HV        &71$\times$71&87    &71$\times$71&168    &47$\times$47& 60    &-&-        &71$\times$71&126     &71$\times$71&63    &-&-\\
NGC\,6822-HIV    &77$\times$47&78    &81$\times$81&167    &-&-        &-&-        &77$\times$47&140    &-&-        &-&-\\
NGC\,6822-HX        &47$\times$47&102    &47$\times$47&180    &-&-        &-&-        &47$\times$47&139    &-&-        &-&-\\
Pox\,186              &47$\times$47& 22    &-&-        &-&-        &-&-        &47$\times$47& 92    &-&-        &-&-\\
SBS\,0335-052        &47$\times$47& 47    &47$\times$47& 29    &-&-        &-&-        &47$\times$47& 38    &-&-        &-&-\\
SBS\,1159+545     &47$\times$47&30     &47$\times$47&75    &-&-        &-&-        &47$\times$47&85    &-&-        &-&-\\
SBS\,1211+540    &47$\times$47&35    &-&-        &-&-        &-&-        &-&-        &-&-        &-&-\\
SBS\,1249+493    &47$\times$47&24    &-&-        &-&-        &-&-        &-&-        &-&-        &-&-\\
SBS\,1415+437    &47$\times$47&48    &47$\times$47& 73   &-&-        &-&-        &47$\times$47& 74    &-&-        &-&-\\
SBS\,1533+574    &47$\times$47&37    &47$\times$47& 72   &-&-        &-&-        &47$\times$47&76    &-&-        &-&-\\
SMC-N66            &149$\times$90& 134-179    &149$\times$90& 280-339    &-&-        &-&-        &149$\times$47& 231    &-&-        &-&-\\
Tol\,1214-277          &47$\times$47&51    &47$\times$47&71    &-&-        &-&-        &47$\times$47&78    &-&-        &-&-\\
UGC\,4483          &47$\times$47&33    &47$\times$47&87    &-&-        &-&-        &47$\times$47&64    &-&-        &-&-\\
UM\,133              &47$\times$47&51    &47$\times$47&112   &-&-        &-&-        &47$\times$47&115    &-&-        &-&-\\
UM\,311              &71$\times$71&67    &63$\times$63&216    &-&-        &-&-        &63$\times$63& 162    &47$\times$47& 38    &-&-\\
UM\,448            &51$\times$51& 61    &53$\times$53&308    &51$\times$51&38    &-&-        &53$\times$53&179    &47$\times$47&86    &-&-\\
UM\,461            &51$\times$51& 48    &47$\times$47& 166    &-&-        &-&-        &53$\times$53& 91    &47$\times$47& 20    &-&-\\
VII\,Zw\,403          &71$\times$71& 49    &63$\times$63& 164    &47$\times$47& 33    &-&-        &63$\times$63& 86    &47$\times$47& 29    &-&-\\
 \enddata 
 \tablecomments
 { {\revised The 1$\sigma$ flux density rms values are determined from the projected spectra and normalized by the area of a single spatial pixel (9.4" x 9.4"). The large-scale galaxy continuum was removed to reflect only the variations around the signal. For unresolved lines, expected line fluxes can be computed from flux density and the corresponding  instrumental resolution element (Section~\ref{spectro_obs}).}
 A dash indicates lack of data for this band. \\
 $^a$ SPIRE FTS data exists for these 4 DGS sources \\
  }
\end{deluxetable} 
\clearpage

 \subsection{Ancillary data - for a multiphase, multiwavelength approach}
 \label{ancillarysection}
 
The DGS galaxies have been observed with numerous ground-based and space-based observatories (Figure~\ref{detection_histo}). Having the full range of UV to radio observations available opens the door to some very powerful multiphase modeling of the dust and gas, and to relate this to a coherent picture of chemical evolution of galaxies. An important goal of the studies to come out of the DGS is to characterize the ISM as a function of metallicity and make the link to the ISM and star formation processes in truly primordial galaxies. With the vast multi-wavelength \hers\ data set, along with the complementary data that exists or is being collected, the physical characteristics of the nebular gas, the photodissociation regions, the neutral gas and stellar properties can be characterized.  


The first step in exploiting the \hers\ observations has been the construction of the detailed database of existing observations from UV to radio wavelengths.  In addition to the new PACS and SPIRE photometry and spectroscopy, the database includes the NIR to mm bands necessary to best characterize the dust properties using existing IRAS data, \spitz\ MIR to FIR observations, as well as the critical ground-based submm observations: 850 \mic, 870 \mic\ and 1.2 mm data from SCUBA/\jcmt\ or SCUBA-2/\jcmt, LABOCA/\apex, MAMBO/\iram\ and, complementing the longer submm to radio wavelengths, \planck\ data. Information on the atomic, ionized and molecular gas phases includes the \spitz\ IRS spectroscopy and is being complemented by data from the literature. A campaign of follow up observations to complete the molecular, atomic and dust submm/mm database is currently underway (\cite{cormier13}). To characterize the stellar activity, 2MASS and GALEX data have been collected for the database.

The ancillary data {\revised compiled for the DGS sample to date} include the following:
 \begin{itemize}
 \item GALEX maps in FUV and NUV (1528 and 2271\AA) observations are collected from the STScI MAST archive (http://galex.stsci.edu/GR6/). 
 These data give us a view into the star formation histories of the galaxies by capturing the direct UV emission from young stars to the most recent star formation activity \citep[e.g.][]{hunter10}. The dust absorption properties in the UV provide additional constraints on the SED modeling of the dust emission properties in the MIR to FIR. GALEX observations exist for all but 9 of the DGS sample.
\item 2MASS observations have been compiled from IRSA (http://irsa.ipac.caltech.edu/) and exist for 34 galaxies in the DGS. The J, H and K bands from 2MASS provide direct access to the emission from the oldest stellar population.
 
\item \spitz\ has observed all 50 targets of the DGS  with various combinations of IRAC bands at 3.6, 4.5, 5.8 and 8.0 \mic\ (beam sizes : 1.7'', 1.7'', 1.9'', 2.0'') and MIPS bands at 24, 70 and 160 \mic\ (beam sizes : 6'', 18'', 38''). The IRAC and MIPS   micron data consist of a combination of archival data and the cycle 5 program: Dust Evolution in Low-Metallicity Environments (P.I. F. Galliano; ID: 50550). The MIPS data are reduced and compiled in \cite{bendo12} and recently publicly released.  \spitz\  IRAC bands cover the emission from the MIR-emitting small grains and the PAHs, the most important species in galaxies for gas heating via the photoelectric effect \citep{bakes94}. The MIR and FIR range of the dust SEDs and the evolved stellar population are well constrained with the IRAC and MIPS bands. Furthermore MIPS shares two bands with PACS at 70 and 160 \mic. This is important to compare the two instruments and assure that they correspond and to understand cases of disagreement \citep{remy13}. \spitz\ IRS spectroscopy, covering 5 to 35 \mic, exists for all but 5 of the DGS galaxies and have been recently reprocessed (R\'emy-Ruyer et al. 2013b, in preparation).  The IRS spectroscopy covers the PAHs bands as well as many ionized lines which are relatively bright in dwarf galaxies and are thus important to constrain the properties of the ionizing sources and the nebular properties \citep[e.g.][]{cormier12, lebouteiller12}.  For example, MIR nebular lines, such as [NeII], [NeIII], [SIV], and [SIII], accessible with the \spitz\ IRS, have relatively high critical densities ($>$ 1000 cm$^{-3}$), complementing the PACS low density tracers, [NII], [NIII], and [OIII] lines (Table~\ref{FIR_lines}). Also, the 33 \mic\ [SiII] line, which requires an ionization energy of 8.1 eV, complements the PACS neutral PDR gas tracers such as [CII] and [OI].

\item The sensitivity of IRAS at 12, 25, 60 and 100 \mic\ (beam sizes: 10.8' $\times$ 4.5' to 3.0' $\times$ 5.0') provides data for about one half of the DGS sample (data from the IRSA site). The 12 \mic\ observations provide an additional valuable MIR constraint useful for modeling the global SEDs of the DGS galaxies.

\item  The DGS galaxies have also been detected by WISE at 3.4, 4.6, 12 and 22 \mic\ with spatial resolutions of 6.1", 6.4", 6.5", and 12.0", respectively, for all but one galaxy (data from the IRSA site). The 22 \mic\ observations, especially, will give the necessary constraint on the MIR wavelength range of the SED when neither IRAS nor Spitzer 24 \mic\ are available. 

\item Submm observations from ground-based telescopes are crucial to determine the presence and characterization of the submm excess. Usually this excess may begin to appear around 450-500 \mic\ compelling  observations longward of 500 \mic. About 30\% of the DGS sample have already been observed in the submm range with either the MAMBO 1.2 mm (\iram\ - 30m telescope), LABOCA (\apex),  SCUBA(\jcmt) pr SCUBA-2 (\jcmt). New observations for DGS sources are currently underway with SCUBA-2  as well as with LABOCA  on {\it APEX} to augment the submm observations to characterize and study the submm excess ).

\item \planck\ has observed the sky with bands from 30 to 857 GHz. Due to the relatively large \planck\ beams (4.5' to 33') and limited sensitivity, only a handful of galaxies are included in the Planck Early Release Catalog. Further \planck\ data processing in the next data-release in 2013 is expected to produce more results for the DGS targets.
\end{itemize}

A summary of the available ancillary data currently included in the DGS database, can be found in Table \ref{ancillary}. Figure~\ref{detection_histo} gives a view of range of existing \hers\ and ancillary for the DGS sample, and how many galaxies were detected for the different data sets. 

\begin{deluxetable}{lccccccccc} 
\tabletypesize{\scriptsize} 
\rotate
\tablecolumns{10} 
\tablewidth{0pt}
\tablecaption{Summary of available \hers\ photometry and ancillary data for the DGS galaxies \label{ancillary}
} 
\tablehead{   
\colhead{Source} & \colhead{Herschel}  & \colhead{ GALEX$^1$} & \colhead{2MASS$^2$} & \colhead{IRAS$^2$} &  \colhead{Spitzer $^3$} & \colhead{WISE$^2$} & \colhead{ground-based $^4$} & \colhead{Planck$^5$ bands:GHz} & \colhead{} \\
\colhead{} & \colhead{FIR/submm}  & \colhead{NUV/FUV} & \colhead{NIR} & \colhead{IR} &  \colhead{MIR/FIR} & \colhead{NIR/MIR} & \colhead{submm} & \colhead{submm} & \colhead{} 
}
\startdata            
Haro 11   	  	  	& PACS - SPIRE		&GALEX	&JHK	& IRAS	&IRAC - MIPS - IRS$^f$		&WISE	&LABOCA 	&- &\\
Haro 2  	  	  	&PACS - SPIRE	 	&GALEX   &JHK	& IRAS	&IRAC - MIPS - IRS$^g$ 		&WISE	&SCUBA 			&- &\\
Haro 3 	  	  	&PACS - SPIRE 		&GALEX	&JHK	& IRAS	&IRAC - MIPS - IRS$^f$		&WISE	&- 			&857 - 545 &\\
He 2-10 			&PACS - SPIRE 		&-	&JHK	& IRAS	&IRAC - MIPS - IRS$^f$ 			&WISE      &SCUBA MAMBO 		&- &\\
HS 0017+1055  	&PACS - SPIRE		&-	&JHK	& -	&IRAC - MIPS24 - IRS$^f$			&WISE &- 			&- &\\
HS 0052+2536  	&PACS - SPIRE		&GALEX	&JHK	& IRAS	&IRAC - MIPS24 - IRS$^f$ 	&WISE	& LABOCA			&- &\\
HS 0822+3542 	&PACS 				&GALEX	&JHK	& -	&IRAC - MIPS - IRS$^f$			&WISE	&- 			&- &\\
HS 1222+3741	  	&PACS 				&GALEX	&JHK	& -	&IRAC - MIPS70 - IRS$^f$		&WISE	&- 			&- &\\
HS 1236+3937  	&PACS - SPIRE 		&GALEX	&-	& -	&IRAC - IRS$^f$ 				&-	&- 			&- &\\
HS 1304+3529  	&PACS - SPIRE		&-	&JHK	& -	&IRAC - MIPS24 - IRS$^f$ 		&WISE	&- 			&- &\\
HS 1319+3224  	&PACS				&-	&-	& -	&IRAC - IRS$^f$				&WISE	&- 			&- &\\
HS 1330+3651  	&PACS				&-	&-	& -	&IRAC - IRS$^f$ 				&WISE	&- 			&- &\\
HS 1442+4250  	&PACS - SPIRE		&GALEX	&-	& -	&IRAC4.5/8 - MIPS - IRS$^f$ 		&WISE	&- 			&- &\\	
HS 2352+2733  	&PACS - SPIRE		&GALEX	&-	& -	&MIPS24 - IRS$^f$				&WISE	&- 			&- &\\
I Zw 18  			&PACS - SPIRE 		&GALEX	&JHK	& -	&IRAC - MIPS - IRS$^{f,g}$	&WISE			&- 			&- &\\
II Zw 40  			&PACS - SPIRE	 	&GALEX	&JHK	& IRAS	&IRAC - MIPS - IRS$^{f,h}$	&WISE			&SCUBA - MAMBO 		&- &\\
IC 10  	 		&PACS - SPIRE		&-	&JHK	& IRAS	&IRAC - MIPS - IRS$^{f,g,h}$	&WISE			&SCUBA		&353 - 217 - 143 &\\
Mrk 1089 			&PACS - SPIRE 		&GALEX	&JHK	& IRAS	&IRAC - MIPS24/70 - IRS$^{f,g}$	&WISE		&LABOCA 	&- &\\
Mrk 1450  		&PACS - SPIRE 		&-	&JHK	& IRAS	&IRAC - MIPS - IRS$^f$ 		&WISE		&- 			&- &\\
Mrk 153  			&PACS - SPIRE 		&GALEX	&JHK	& IRAS	&IRAC - MIPS - IRS$^{f,h}$ 	&WISE			&- 			&- &\\
Mrk 209  			&PACS - SPIRE 		&GALEX	&JHK	& -	&IRAC4.5/8 - MIPS - IRS$^f$	&WISE		&- 			&- &\\
Mrk 930  	  		&PACS - SPIRE 		&GALEX	&JHK	& IRAS	&IRAC - MIPS - IRS$^f$	&WISE			& LABOCA			&- &\\
NGC 1140  		&PACS - SPIRE		&GALEX	&JHK	& IRAS	&IRAC - MIPS - IRS$^f$	&WISE			&SCUBA - MAMBO 		&- &\\
NGC 1569  		&PACS - SPIRE		&GALEX	&JHK	& IRAS	&IRAC - MIPS - IRS$^{f,g}$ 	&WISE			&SCUBA - MAMBO 		&- &\\
NGC 1705  		&PACS - SPIRE		&GALEX	&JHK	& IRAS	&IRAC - MIPS - IRS$^g$  		&WISE		&LABOCA	&- &\\
NGC 2366  		&PACS - SPIRE		&GALEX &JHK & IRAS  &IRAC - MIPS - IRS$^{f,g}$  
&WISE			& SCUBA2			&- &\\
NGC 4214  		&PACS - SPIRE		&GALEX	&JHK	& IRAS	&IRAC - MIPS - IRS$^{f,g}$  	&WISE			&SCUBA		&857 - 545 - 353 &\\
NGC 4449  		&PACS - SPIRE		&GALEX	&JHK	& IRAS	&IRAC - MIPS - IRS$^{g}$ 		&WISE		&SCUBA - MAMBO		&857 - 545 - 353 &\\
NGC 4861  		&PACS - SPIRE		&GALEX	&JHK	& IRAS	&IRAC - MIPS - IRS$^{f,g}$  	&WISE			& SCUBA2 			&- &\\
NGC 5253  		&PACS - SPIRE		&GALEX	&JHK	& IRAS	&IRAC - MIPS - IRS$^{g,h}$  		&WISE		&SCUBA - LABOCA		&857 - 545 - 353 &\\
NGC 625  		&PACS - SPIRE		&GALEX	&JHK	& IRAS	&MIPS - IRS$^{f}$ 			&WISE		&LABOCA 	&857 - 545 - 353 &\\
NGC 6822  		&PACS - SPIRE		&GALEX	&JHK	& IRAS	&IRAC - MIPS - IRS$^{f,g}$ 	&WISE			&LABOCA	&545 - 353 &\\
Pox 186  			&PACS - SPIRE 		&-	&-	& -	&IRAC - MIPS24 - IRS$^{c,f}$	&WISE		&- 			&- &\\
SBS 0335-052 		&PACS - SPIRE		&GALEX	&JHK	& -	&IRAC - MIPS - IRS$^{f}$		&WISE		&- 			&- &\\
SBS 1159+545 	&PACS - SPIRE 		&-	&-	& -	&IRAC - MIPS24 - IRS$^{c,f}$ 	&WISE		&- 			&- &\\
SBS 1211+540	  	&PACS - SPIRE		&GALEX	&-	& -	&IRAC - MIPS24 - IRS$^{f}$	&WISE		&- 			&- &\\
SBS 1249+493 	&PACS - SPIRE		&GALEX	&-	& -	&IRAC - MIPS24 - IRS$^{f}$ 	&WISE		&- 			&- &\\
SBS 1415+437	   	&PACS 				&GALEX	&-	& -	&IRAC4.5/8 - MIPS - IRS$^{f}$	&WISE		&- 			&- &\\
SBS 1533+574  	&PACS - SPIRE		&GALEX	&JHK	& IRAS	&IRAC - MIPS70 - IRS$^{f}$	&WISE		&- 			&- &\\
Tol 0618-402  		&PACS 				&GALEX	&JHK	& -	&IRAC - MIPS - IRS$^f$ 		&WISE		&- 			&- &\\
Tol 1214-277  		&PACS - SPIRE 		&GALEX	&-	& -	&IRAC - MIPS - IRS$^{f,h}$		&WISE		&- 			&- &\\
UGC 4483  		&PACS - SPIRE		&GALEX	&JHK	& IRAS	&IRAC - MIPS - IRS$^f$ 		&WISE		&- 			&- &\\          
UGCA 20  	 	&PACS				&GALEX	&-	& -	&IRAC - MIPS24 - IRS$^f$ 	&WISE		&- 			&- &\\
UM 133  			&PACS - SPIRE		&GALEX	&-	& -	&IRAC - MIPS24 - IRS$^g$ 	&WISE		&-			&- &\\
UM 311  			&PACS - SPIRE		&GALEX 	&-	& -	&IRAC4.5/8 - MIPS - IRS$^f$ 	&WISE		&LABOCA 	&- &\\
UM 448  			&PACS70/160 - SPIRE 	&GALEX		&JHK	& IRAS	&IRAC - MIPS - IRS$^f$ 		&WISE		&SCUBA - LABOCA 			&- &\\
UM 461  			&PACS - SPIRE 		&GALEX	&JHK	& IRAS	&IRAC - MIPS - IRS$^f$		&WISE		&- 			&- &\\
VII Zw 403  		&PACS - SPIRE		&GALEX	&JHK	& IRAS	&IRAC - MIPS - IRS$^{f,g}$	&WISE			&-  			&- &\\
\enddata

{
$^a$ Data from STScI MAST archive (http://galex.stsci.edu/GR6/). 

$^b$ Data collected from the NASA/IPACS Infrared Science Archive (IRSA) 

$^c$ : Data from \cite{bendo12}

$^d$ : Data collected from the literature: SCUBA or SCUBA2/\jcmt: 450 and/or 850 \mic;  Laboca/\apex: 870 \mic; MAMBO/\iram: 1.2 mm  \citep[e.g.][]{bolatto00, james02, lisenfeld02, bottner03, galliano03, kiuchi04, galliano05, galametz09, galametz11} or recent follow-up observations \citep{remy13}.

$^e$ : Data collected from the \planck\ database and the Early Release Source Catalog

$^f$ : \spitz\ IRS 1D spectrum available for at least one of the modules : ShortLow, ShortHigh, LongLow, LongHigh. Data extracted using the database tool, CASSIS \citep{lebouteiller11}

$^g$ : \spitz\ IRS maps available but for the largest galaxies not full galaxy coverage; IRS maps extracted for the whole galaxy and/or part of it only, using the database tool, CASSIS \citep{lebouteiller11}

$^h$: while MIPS data are technically available for all MIPS bands, the data from one or two of the bands may have limited usefulness (limited coverage in the case of IC~10; confusion issues in the 160 micron images for Mrk~153 and Tol~1214-277; saturated emission at 24 microns in NGC~5253; problems with background subtraction in the 160 micron image of II~Zw40
}
 \end{deluxetable} 
\clearpage  

 
\section{A taste of science results}
\label{science_section}
Here we give some examples of science that can be carried out with the DGS, using the full photometric and spectroscopic \hers\  data along with the available ancillary data at a variety of wavelengths. 

\subsection{Distribution of the ISM components in NGC~4214}
\label{section 5_1}
An example of the detailed FIR and submm dust distribution that \hers\ reveals is shown in Figure~\ref{ngc4214_combined} for NGC~4214, a nearby (D=2.9 Mpc) star-forming dwarf galaxy with metallicity 12 + log(O/H) =8.2 \citep{kobulnicky96}. The PACS and SPIRE images from the DGS are the highest spatial resolution data at these wavelengths. The distribution of the dust traced by the 3 FIR PACS bands, where the peak of the total luminosity exists, resembles, at first glance, that of the longer wavelength SPIRE bands, which are mostly tracing the overall colder dust component. {\revised There are two particularly prominent star-forming regions seen at \spitz\ and \hers\ images: the more northern of these 2 prominent peaks is called central (C), while the peak just to the south-east of this is called SE.} The quiescent extended "neck" region to the north-west is called NW. The C and SE sites contain clusters of widely varying ages and characteristics \citep{ubeda07a, fanelli97} are very prominent at all PACS bands and at least up to 350 \mic\ and are sitting in a more extended, diffuse dust component. These 2 \hers\ peaks are seen prominently in H$\alpha$ (Figure \ref{ngc4214_combined}): the massive central HII region harboring a range of clusters overall $\sim$ 3 to 4 Myr  \citep{mackenty00} and as well a super star cluster; the SE site being the second most prominent H$\alpha$ region configured in a series of knots of some of the youngest clusters in the galaxy, $\sim$ 2 Myr \citep{ubeda07b}.  The NW region is characterized by a more quiescent cluster of H$\alpha$ knots, also apparent in \hers\ observations.  The contrast between the bright star forming regions and the extended low surface brightness emission decreases toward longer wavelength, where, for example, at 500 \mic\ the emission is rather dominated by the  more extended cooler dust emission. 
  
 \begin{figure}
    \centering
     \vspace{-1cm}
\includegraphics[width=16cm, height=18cm]{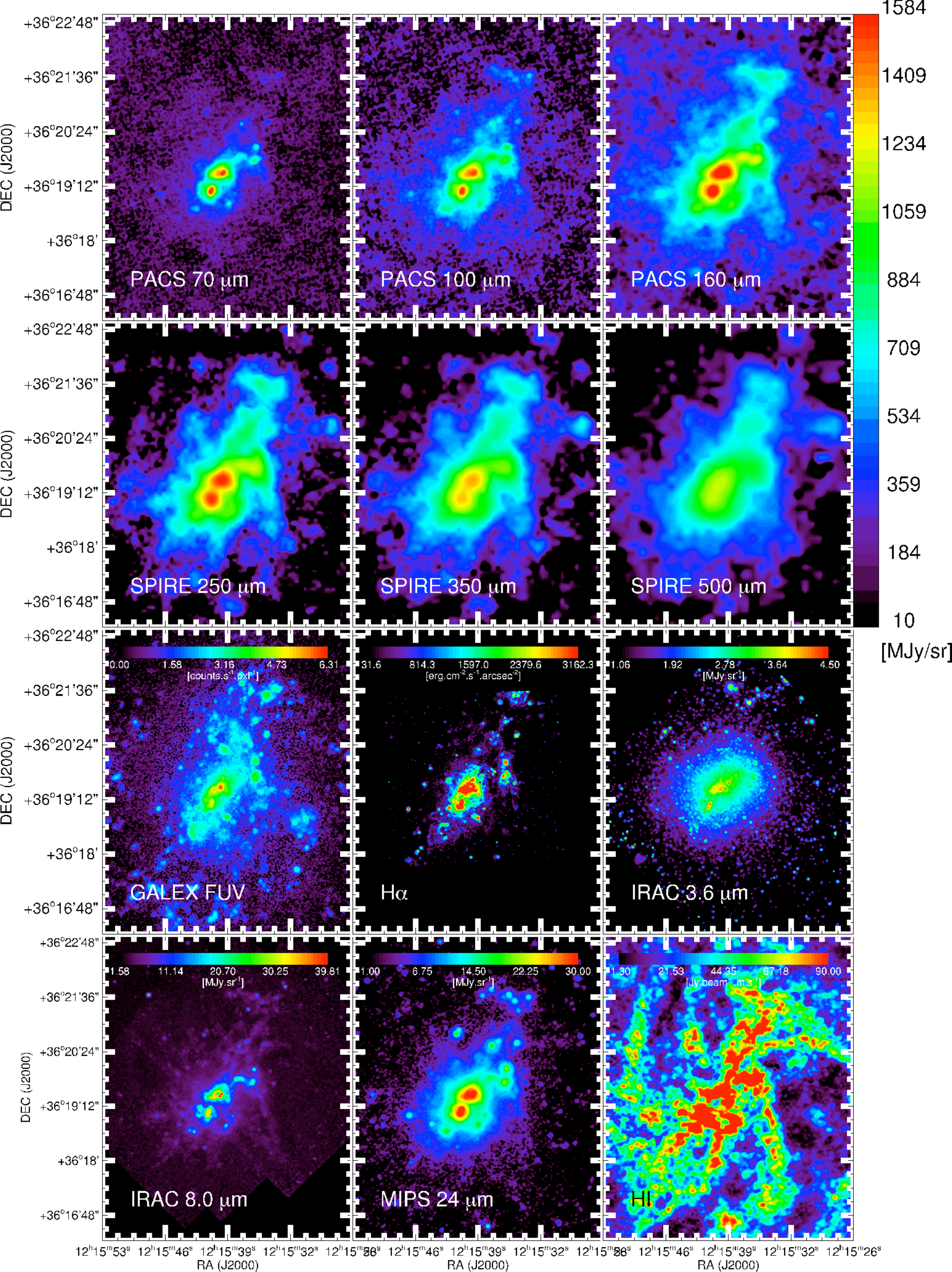}
\vspace{-0.6cm}
\caption{\small \hers\ and ancillary images for the DGS galaxy, NGC4214. The PACS maps have been reduced with Scanamorphos and the SPIRE maps with a customized version of the official pipeline (see \cite{remy13}).  Note the 2 star forming regions especially prominent at PACS FIR wavelengths: {\revised NGC4214-C is the more northern of the 2 brightest star forming regions, and NGC4214-SE is just south-east of this.} NGC4214-C harbors a super star cluster. The rather quiescent extended "neck" region to the north-west is called NW.}
\label{ngc4214_combined}
  \end{figure}

At other wavelengths, there are some obvious similarities to the \hers\ data. In Figure \ref{ngc4214_combined}, we include maps from the rich ancillary dataset available for NGC4214: GALEX tracing the youngest stellar components, H$\alpha$ showing the distribution of the ionized gas, IRAC 3.6 \mic\ shows the evolved stellar component, IRAC 8 \mic\  the PAHs and/or hot dust component, MIPS 24 \mic\ \citep{bendo12} tracing the warm dust component and 21 cm HI, from the THINGS survey \citep{walter08}. Note how extended the HI emission is, while the star formation activity is primarily concentrated toward the central 2 star forming regions (NGC4214-C and NGC4214-SE ; see \ref{ngc4214_combined}).

\begin{figure*}
\centering
\vspace{-1.0 cm}
 \includegraphics[width=5.0cm]{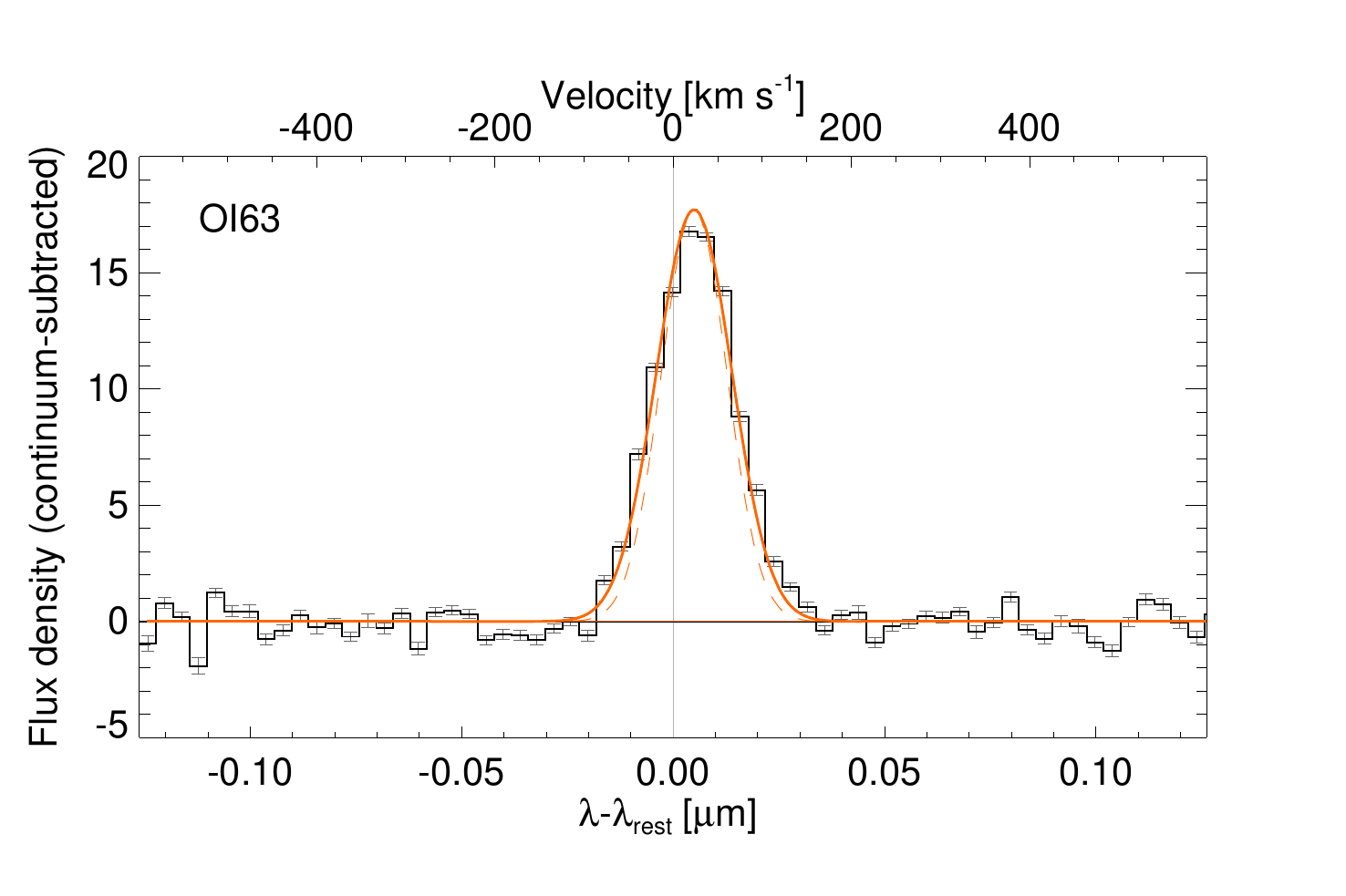}
 \includegraphics[width=5.0cm]{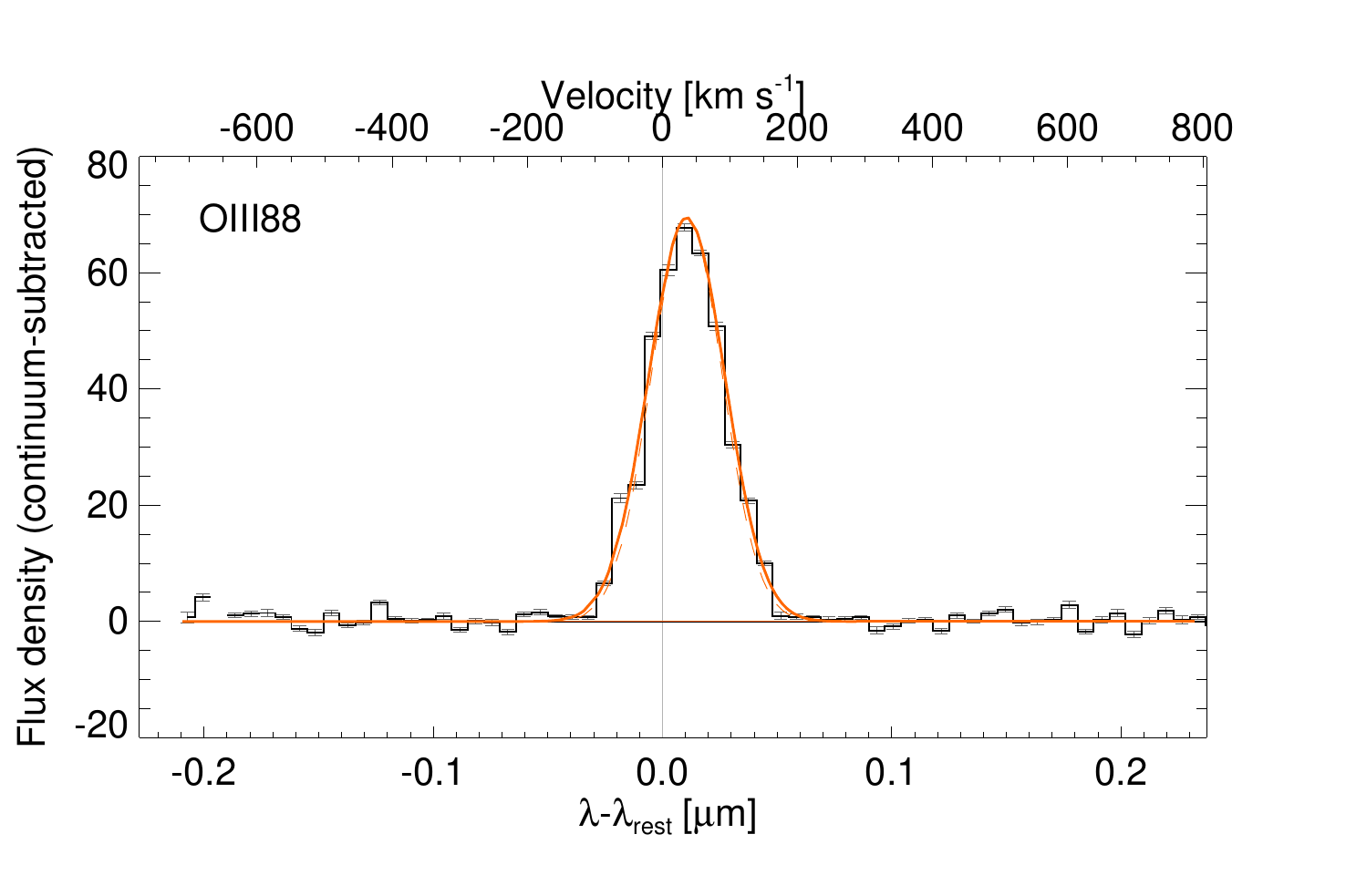}
 \includegraphics[width=5.0cm]{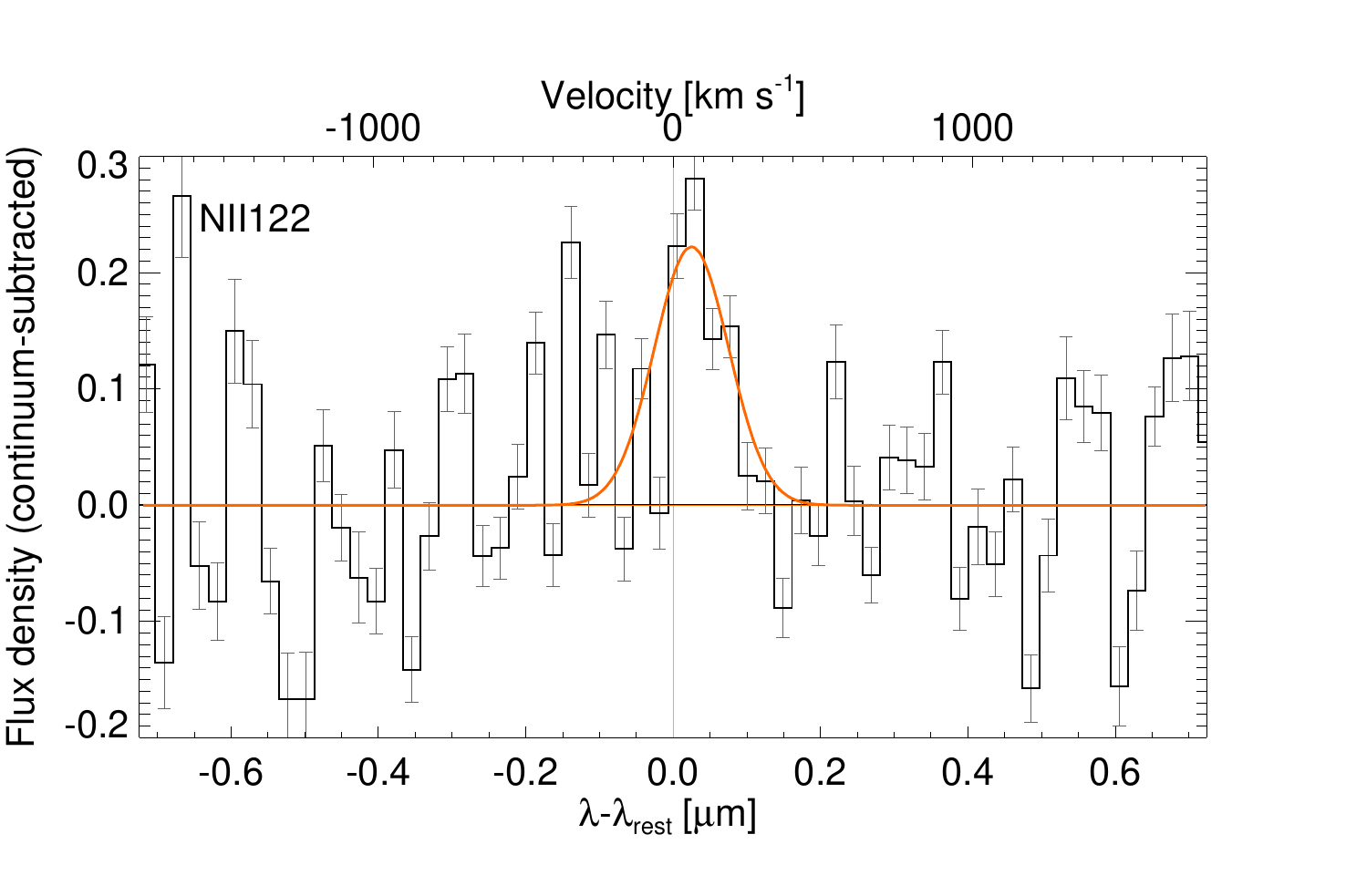}
 \includegraphics[width=5.0cm]{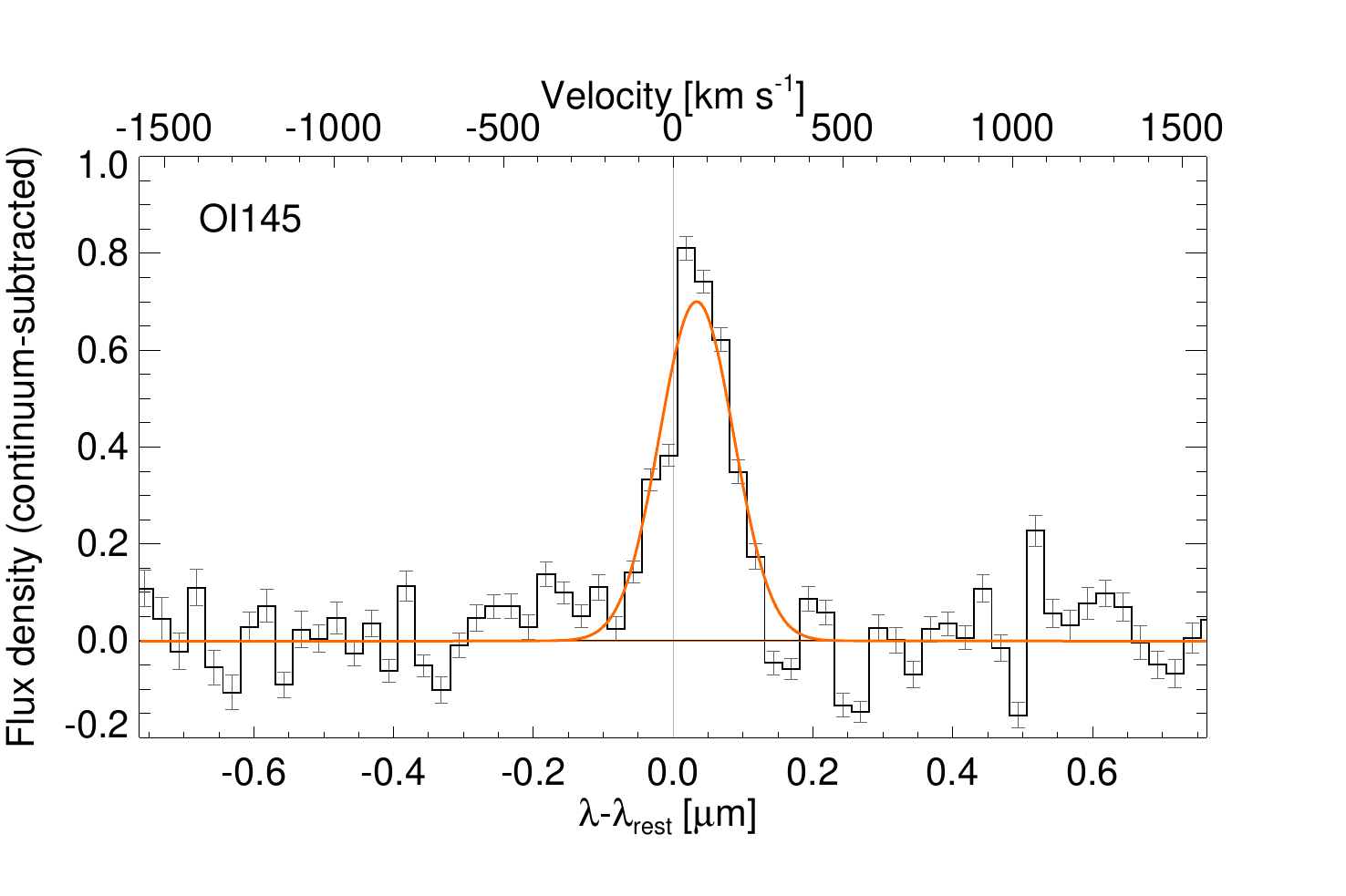}
 \includegraphics[width=5.0cm]{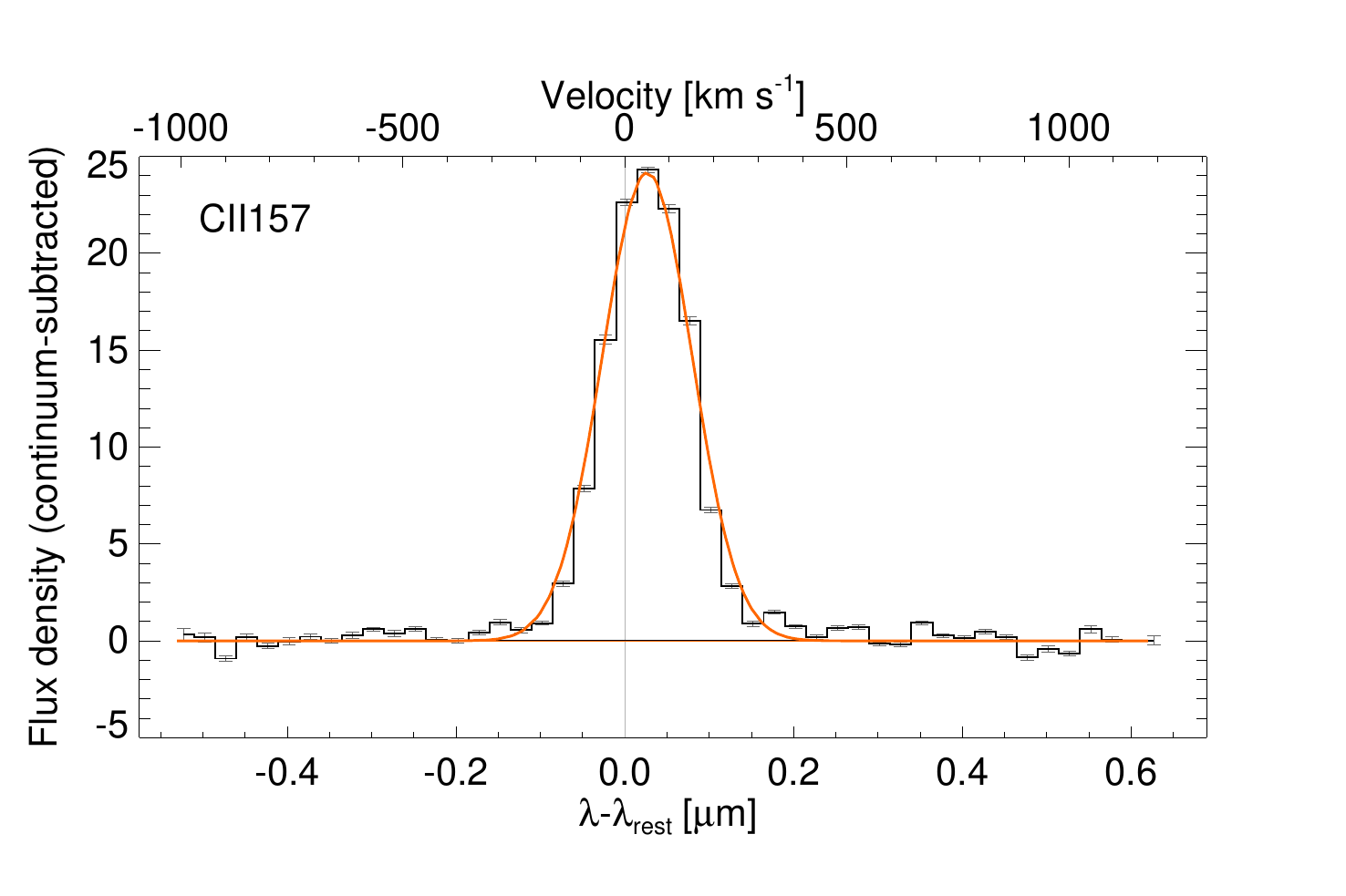}
 \includegraphics[width=5.0cm]{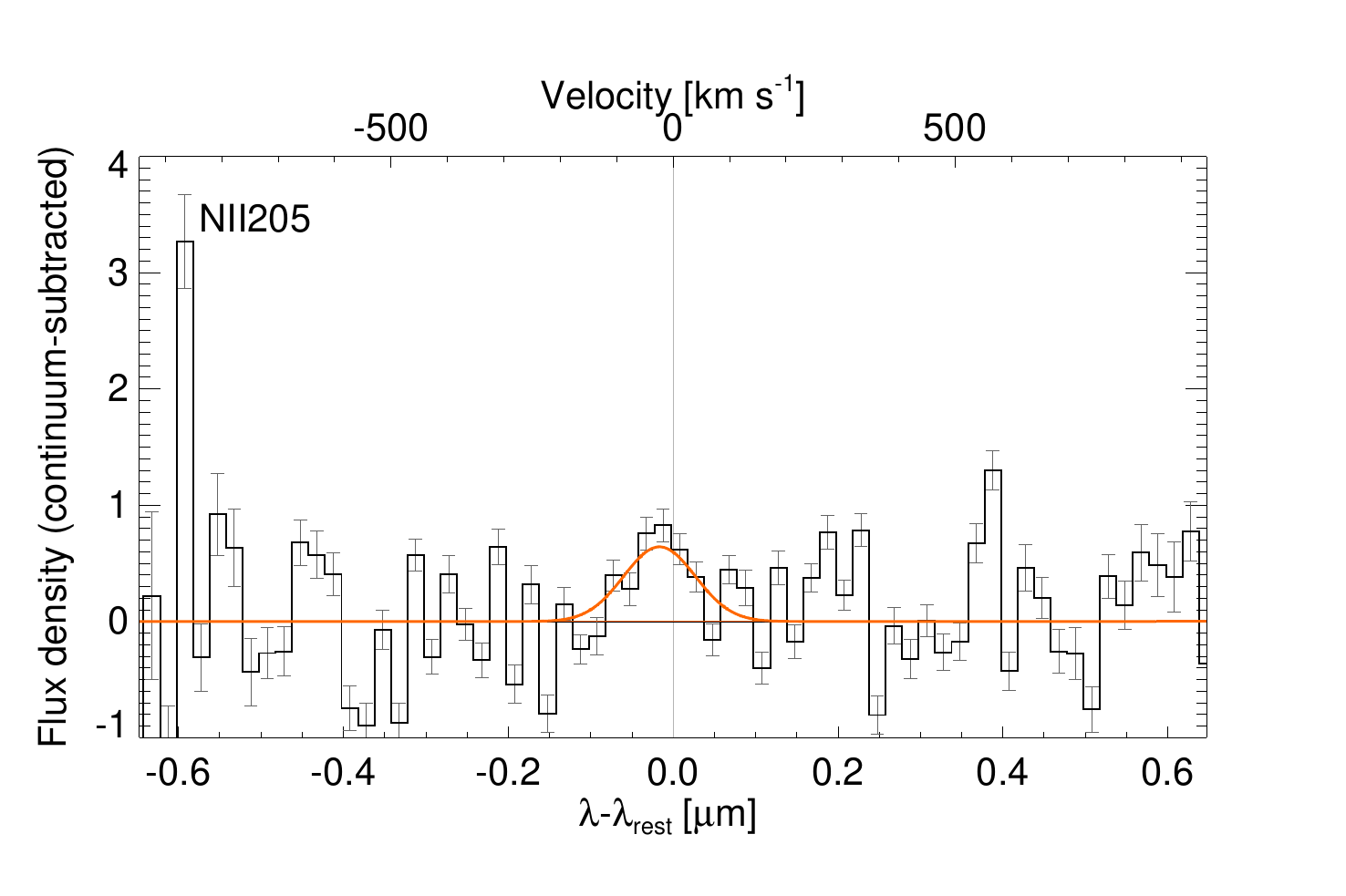}

 \vspace{-0.8 cm}
 \caption{ \small Example of the PACS spectral line quality observed near the central star forming region in NGC~4214. \revised {For these cases the lines are not resolved and the spectra have been binned to 1/5 of the instrument spectral resolution (see section \ref{spectro_obs}). }}
\label{spectra_ngc4214}
\end{figure*}
\clearpage

\begin{figure*}
\centering
  \includegraphics[width=13cm]{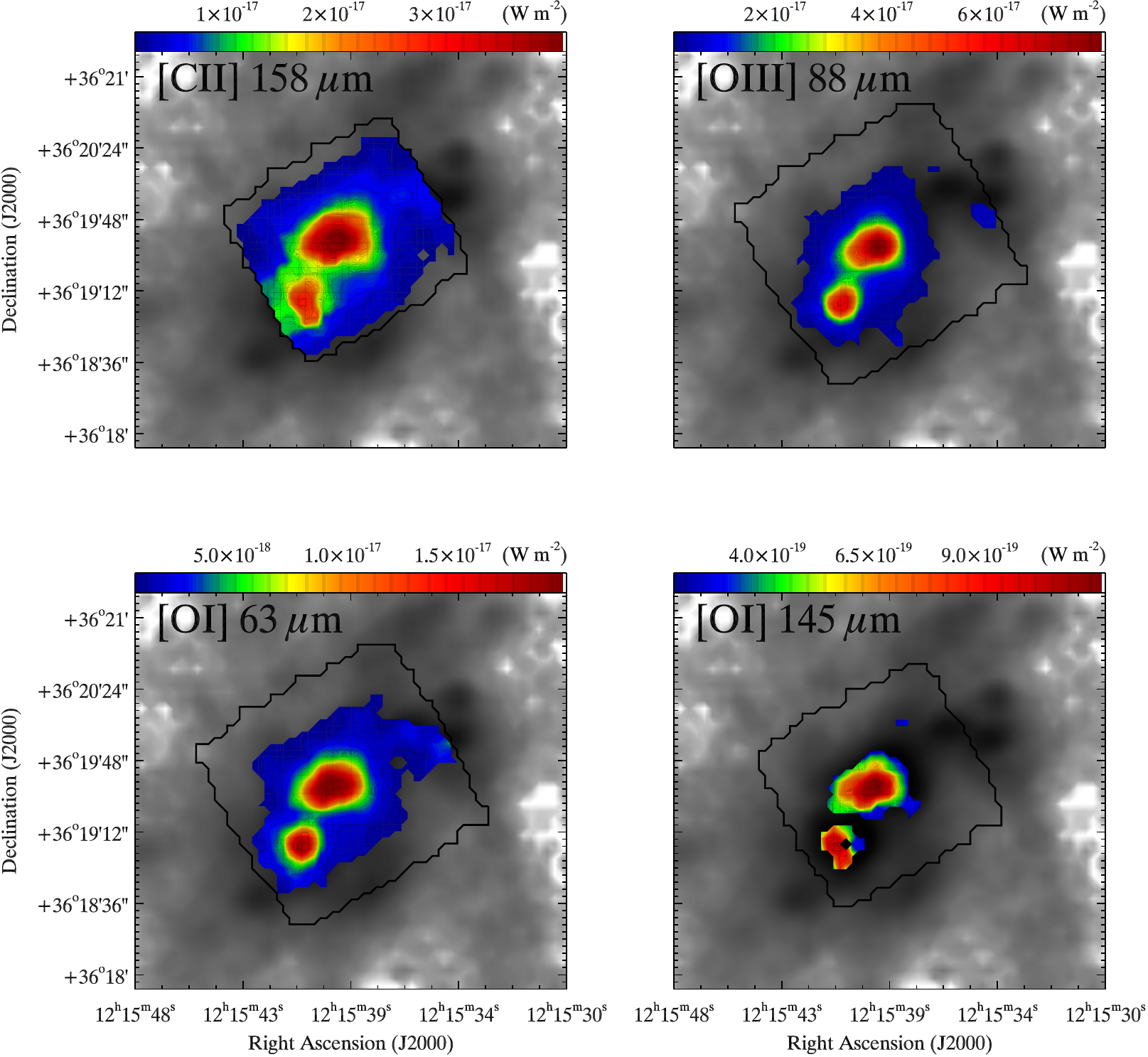}
\caption{ \small PACS spectroscopic maps on the 160 \mic\ photometry map of NGC~4214. The [CII] peaks are very broad, covering the star forming regions (NGC-4214-C and NGC4214-SE) and the associated extended PDRs. The [OIII] line emission is much more peaked toward region C, peaking on the location of the super star cluster, the ionizing source, whereas, the [OI] emission peaks toward C and SE, but the peak at C is shifted from the [OIII] peak \citep{cormier10}. The extension to the NW of region C (NGC4214-NW) is illuminated primarily in the PDR tracers, [CII] and [OI]. }
\label{maps_ngc4214}
\end{figure*}
\clearpage

\subsection{FIR dust and gas properties within NGC~4214}

Four PACS lines are very bright throughout NGC~4214:  [OIII] 88\mic, [CII] 158 \mic\ and the [OI] 63 and 145 \mic\ lines (Figure \ref{spectra_ngc4214}). The [OIII] 88\mic\ line is the brightest everywhere, not the [CII] 158 \mic\ line, which is most often the brightest line throughout normal metallicity galaxies. The very porous ISM in NGC4214, typical of {\revised low metallicity} star forming galaxies, is very evident in the extent of the wide-scale distribution of [OIII] 88 \mic, requiring the presence of 35 eV photons, consistent with the deficit of PAHs found on galaxy-wide scales, presumably destroyed by the permeating unattenuated UV photons in dwarf galaxies \cite[e.g.][]{engelbracht05, madden06}. The porosity of NGC~4214 and other dwarf galaxies is also shown by studies concluding the importance of the escaping ionizing radiation on the dust SED \citep{hermelo13, cormier12}. 
The [NII] 122 and 205 \mic\ lines are also detected toward the peak star forming regions, but are 1 or 2 orders of magnitudes fainter than the other lines. All of the lines peak toward the 2 main star forming regions (Figure \ref{maps_ngc4214}): the [OIII] peak coincides with the location of the super star cluster in region C, while the [CII] peak is shifted off of the super star cluster and coincides with the peak of the neutral PDR tracers, the 2 [OI] lines. The S/N of the [CII] map is remarkable - more than 10 throughout the region mapped as illustrated in Figure~\ref{linemaps_ngc4214}, allowing for detailed spatial studies of the heating and cooling of the near and beyond the low metallicity star forming regions.  

To illustrate the kind of spectral and spatial studies that can be carried out on the gas and dust properties, we present the combined gas and dust SEDs of the 3 main HII regions, created from  \spitz\ IRAC, MIPS, IRS and \hers\ observations (Figure \ref{seds_ngc4214}). The data treatment and dust modelling details are described in \cite{remy13}. The shapes of the dust SEDs are representative of typical compact HII regions.  These SEDs include the MIR and FIR lines from the PACS spectroscopic maps to present the full gas plus dust SEDs. Note the relatively wide diversity of behavior of the high excitation vs. low excitation lines and the variation in the line to FIR continuum levels. For example, the NW region shows a much more quiescent dust SED as well as low levels of high excitation ionic lines, which are in contrast to the those of the 2 bright star forming regions at the origin of the peaks of the warm dust emission (C and SE).  Notice that toward the central and south-eastern star forming regions, the PDR tracers, [CII] 158 \mic\  and [OI] 63 \mic\ lines, as well {\revised as the ionized gas tracer, [OIII] 88 \mic, dominate} the infrared line emission and account for 2 to 4 \% of the total {\it L$_{TIR}$} \citep{cormier10}.  The rich MIR wavelength range shows prominent high excitation ionic lines, such as [NeII] 12.8 \mic, [NeIII] 15.5 \mic, [OIV] 25.9 \mic and  [SIV] 10.5 \mic, arising from the rather dense HII regions as well as the PAH bands, emitting from 3 to 17 \mic. \cite{cormier12} demonstrate the level of complexity possible now in modeling $\sim 20$ MIR and FIR fine structure lines self-consistently with the dust continuum.

\begin{figure*}
\centering
\vspace{-1.0 cm}
\includegraphics[width=10 cm]{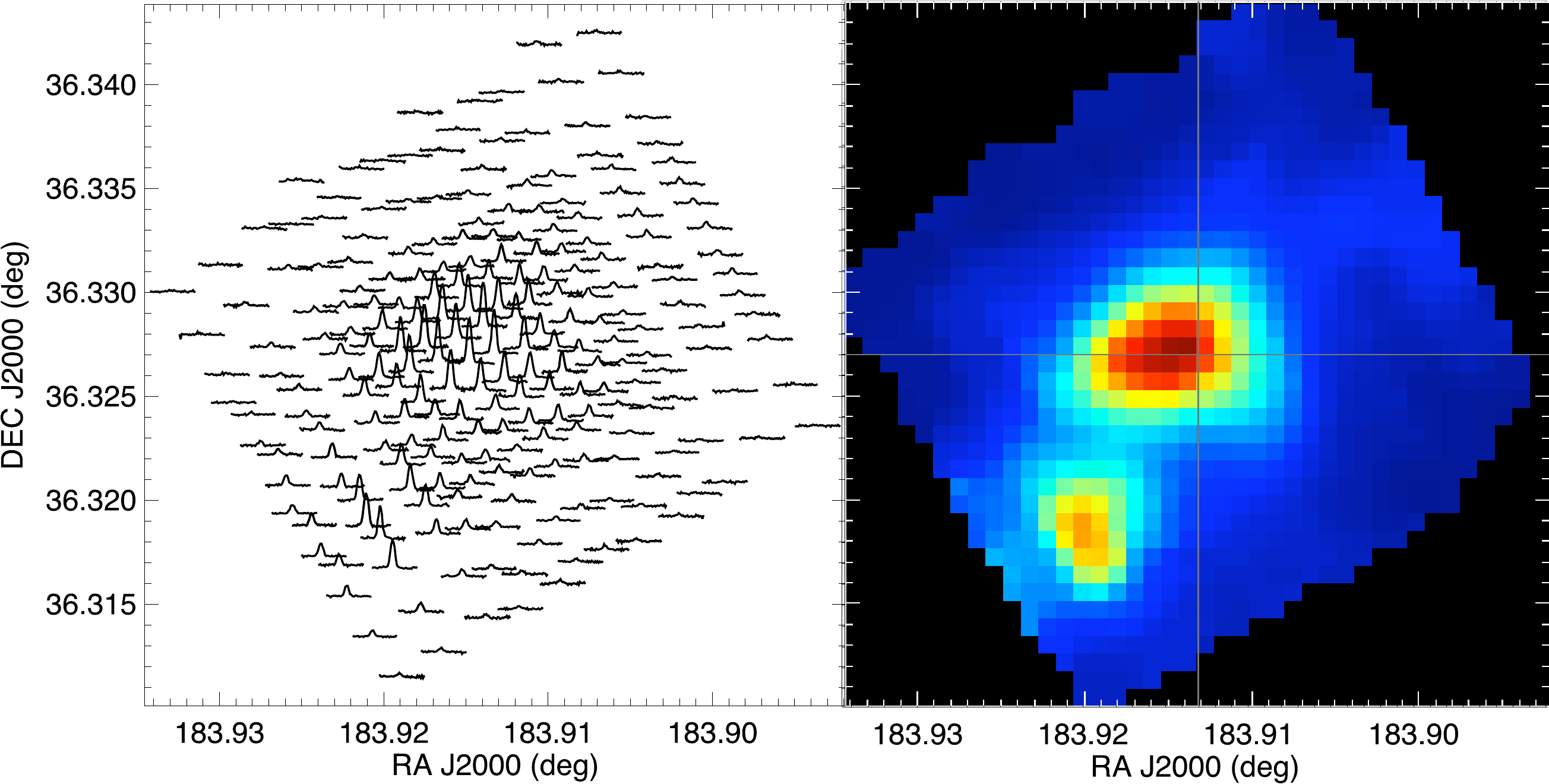}
\vspace{-0.5 cm}
\caption{ \small ({\it left}): Distribution of the PACS [CII] spectra. Even toward the edges of the region mapped (2' $\times$ 2'), the S/N of the [CII] line is at least 10. ({\it right}) Image of the [CII] line map, showing the peak emission toward regions C and SE. }
\label{linemaps_ngc4214}
\end{figure*}
  
\begin{figure*}
\centering
\includegraphics[width=7.0cm]{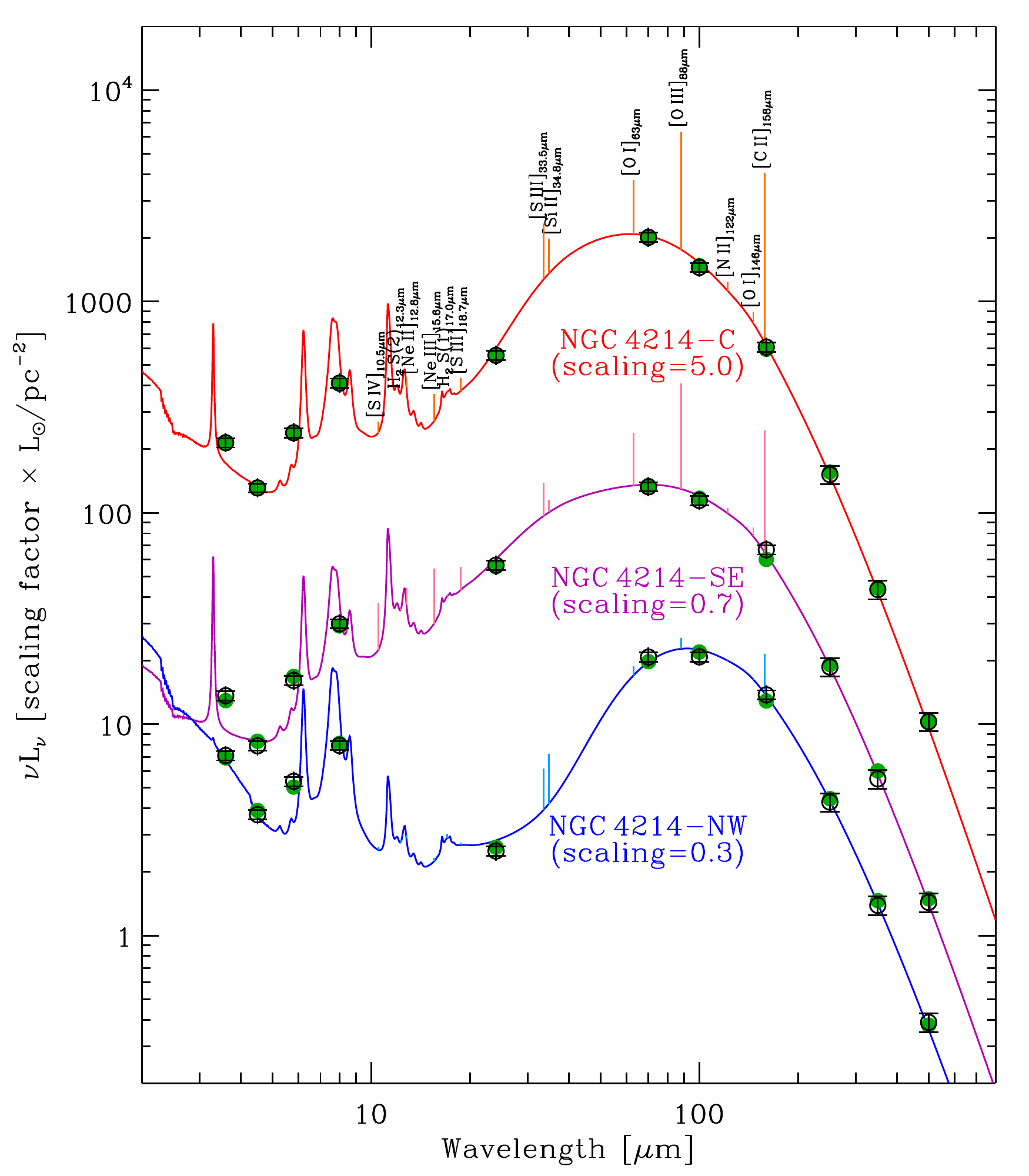}
\vspace{-0.6 cm}
\caption{\small SEDs of the 2 main star forming regions, in the center (NGC4214-C) and to the south-east (NGC4214-SE) along with the SED of a more quiescent region to the north-west (NGC4214-NW) for comparison. For clarity, the luminosities have been multiplied by the scaling factor indicated below each SED. The black circles with error bars are the integrated \spitz\ and \hers\  broadband observations. The solid lines are the fitted SED models \citep{galliano11}, fit to the broadband observations. The green dots are the synthetic photometry of the model. The vertical spectra on top of the SED represent the observed integrated line luminosities from \spitz\ IRS and \hers\ PACS \citep{cormier10}.}
\label{seds_ngc4214}
\end{figure*}
\clearpage

  \subsection{Submillimetre excess}

In Figure~\ref{SED} we present examples of modelled SEDs for the new \hers\ data as well as with the ancillary dataset available for VIIZw403 and NGC~1569. From these modeled SEDs we will have access to new parameters for the galaxies, such as the total dust mass, total luminosity, new SFR and dust composition and dust size distribution \citep{remy13}. Comparison of the observed SEDs with modeled SEDs in dwarf galaxies often points to an as-yet inexplicable excess (Section~\ref{sciencegoals}) which begins to become apparent near $\sim$500 \mic, the longest wavelength observed by \hers, or sometimes becomes more evident at longer submm wavelengths, in which case it may be detectable with ground-based submm telescopes, such as {\it JCMT}(SCUBA2), {\it APEX}(LABOCA),  {\it IRAM}(GISMO) or {\it ALMA}, for example. VIIZw403 already shows a submm excess beginning at $\sim$ 500 \mic\ (Figure~\ref{SED}). 
In NGC~1569, on the other hand, there is no hint of submm excess throughout the \hers\ SPIRE bands, including $\sim$ 500 \mic. However, a small excess beginning at $\sim$ SCUBA 850 \mic\ observations is present, growing at mm wavelengths \citep{galliano03}. In this case, the submm excess would not be detectable without wavelengths beyond \hers. The \hers\ data are crucial to properly constrain the submm dust emission to be able to accurately extrapolate and interpret the SED to wavelengths $>$ 500 \mic. Without the \hers\ data, the Rayleigh-Jeans side of the dust SED is poorly constrained, making it more difficult to properly assess whether the submillimetre excess is present. {\revised Accurate quantification of the submm excess at SPIRE wavelengths must take into account careful consideration of the most recent calibration and beam sizes in the aperture photometry.}

Note that these are global SEDs presented here, and spatial SED modelling throughout well-resolved galaxies, may uncover regions where the presence of this submm excess may be apparent even at \hers\ wavelengths, as has been detected with SPIRE within the LMC, where the presence of this excess is anti-correlated with the dust surface density \citep{galliano11}.

 \begin{figure*}
\includegraphics[width=8.0cm]{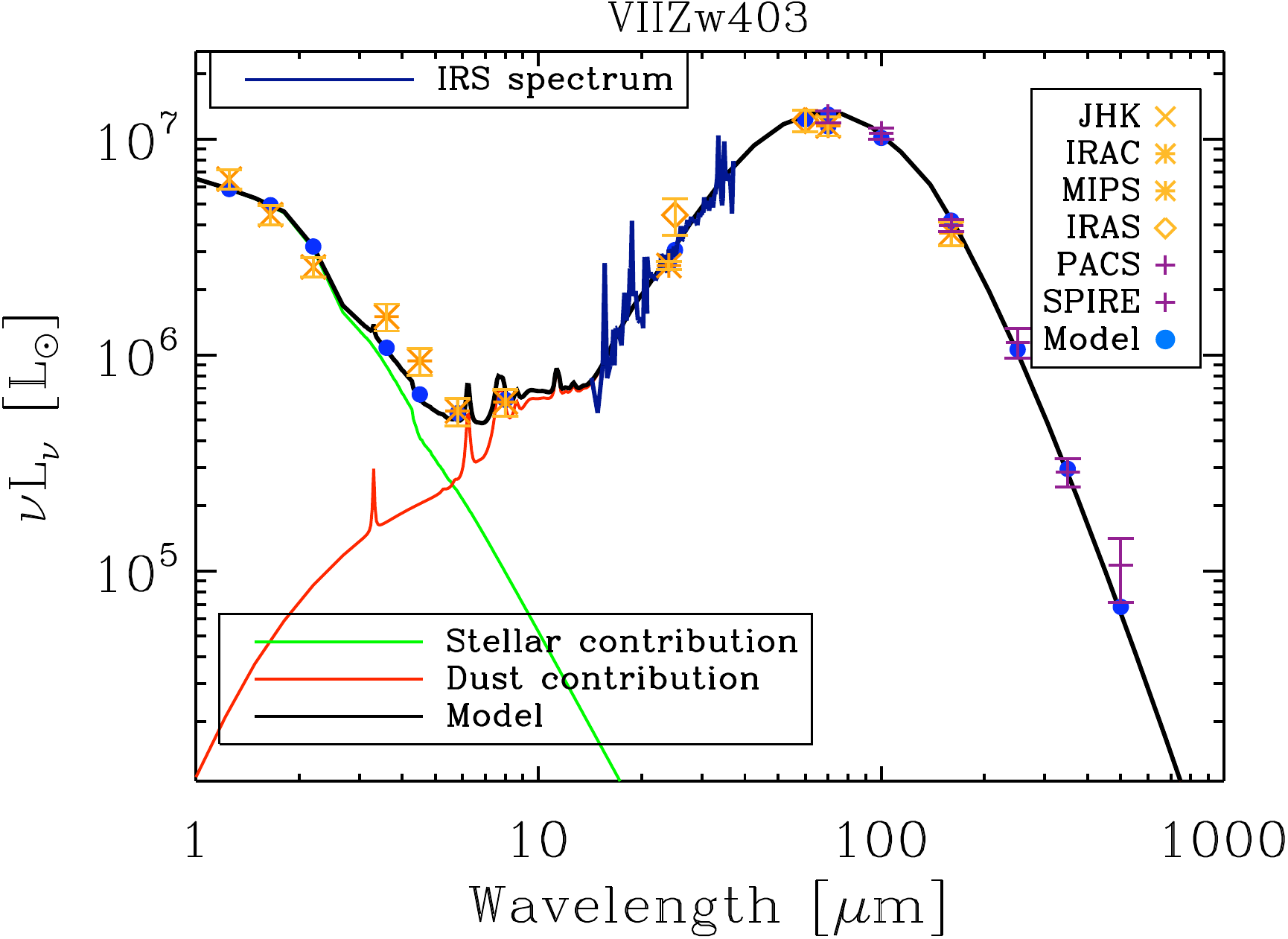}
\includegraphics[width=8.0cm]{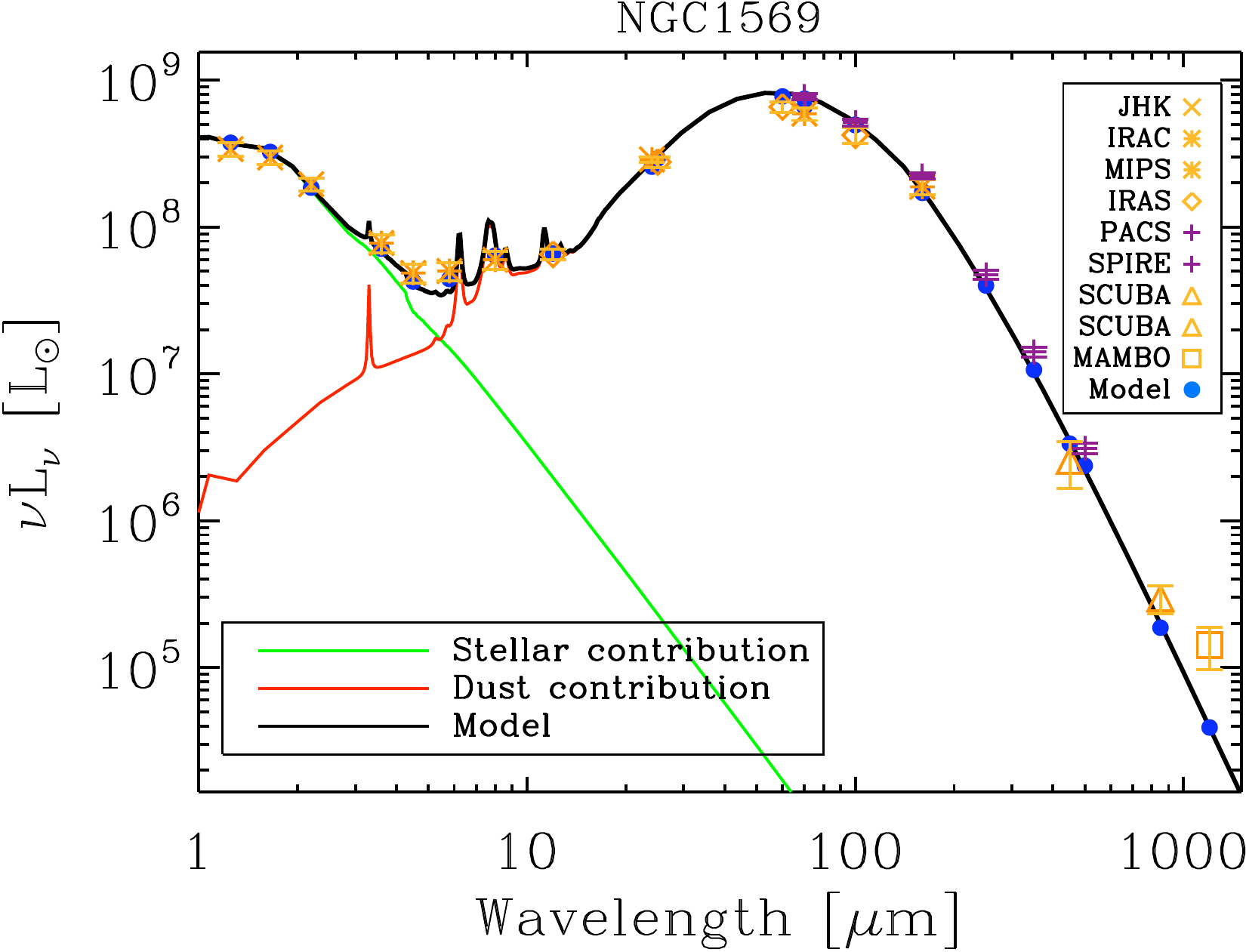}
\caption{\small Examples of galaxy-wide SEDs for two DGS galaxies : VIIZw403 {\it (left)}, and NGC~1569 {\it ( right)}. Both SEDs have been modelled as in \cite{galliano11} and include the new \hers\ data (purple) as well as any available ancillary data (orange). The different symbols refer to the different instruments that have been used. For VIIZw403, the IRS spectrum is displayed in dark blue. While limited IRS spectra do exist for NGC1569, we do not include them in this SED fit, due to lack of full galaxy coverage. The total modelled SED in black is the sum of the contributions displayed here: stellar (green) and dust (red). The model points in the different considered bands are the filled blue circles. The dust SED has been corrected for CO line contamination as well as free-free emission. Notice the submm excess that becomes apparent in VIIZw403 at the SPIRE 500 \mic\ band while some galaxies, such as NGC~1569, for example, the submm excess may only begin at longer wavelengths, such as 850/870 \mic\ or longer \citep{remy13}, {galliano03, galametz09}.
}
\label{SED}
\end{figure*}
\clearpage

 \section{DGS Products}
\label{products_section}
The final products for the PACS and SPIRE photometry data will become public 
on the HeDaM database (http://hedam.oamp.fr/; \citep{roehlly12})
 by the beginning of 2014. All of the \hers\ maps for the DGS galaxies will be available in FITS format as well as Jpg format.  
A catalog of the new \hers\ fluxes will also be available in \cite{remy13}.
Since the PACS and SPIRE spectroscopic data is still in the process of being finalized, this data is forseen to become available in mid-2014. This data will be delivered in the form of final total spectra and total flux values.  

As an illustration of the wealth of data the DGS brings, in Appendix A we have assembled an Atlas of 3 color images of the DGS composed of \spitz\ MIPS 24 \mic, tracing the hot dust component, \hers\ PACS 100 \mic, close to the peak of the dust SED, and \hers\ 250 \mic, tracing the cooler dust.
 

\section{Summary}
\label{summary_section}
We present an overview of the Dwarf Galaxy Survey which has used 230h of \hers\ observations to study the FIR and submm gas and dust properties in 50 star forming dwarf galaxies of the local Universe with 6 PACS and SPIRE bands at 70, 100, 160, 250, 350, 500 \mic\ and the PACS and SPIRE spectrometers. The PACS FIR spectroscopy gives us maps of [CII]~158 \mic, [OI]~63 and 145~\mic, [OIII]~88 \mic, [NIII]~57 \mic\ and [NII]~122 and 205 \mic\ The SPIRE FTS covers 194 to 671 \mic\ and includes high-J CO lines (J=4-3 to J=13-12), [NII] 205 mu and [CI] lines at 370 and 609 \mic.  The targets span a wide range of metallicity values, as low as 1/50 Z$_{\sun}$, and cover 4 orders of magnitude of star formation rates. The full DGS database goes beyond the FIR and submm observations of \hers, and includes a rich array of ancillary observations, including \spitz\ MIR photometry and spectroscopy and FIR photometry, {\it 2MASS} NIR data, {\it WISE} MIR data, {\it GALEX} UV and FUV observations, {\it IRAS}, and ground-based submm data, for example. We present some science goals the full survey will bring and show some examples to illustrate the potential of this survey for futures studies of the distribution of the gas and dust properties within galaxies as well as on global scales. We anticipate that advances on the evolution of the dust and gas as a function of metallicity will bring us closer to understanding the enrichment of the Universe over cosmic times.

\clearpage

\appendix
 
 \section{Determination of Metallicities}
 \label{appendixA}
The metallicity of a galaxy is an intrinsic parameter tracing the evolution of the system:  the ISM becoming more enriched with elements as a function of age. Calibrating this important parameter from observations and models has lead to discrepancies that have  been studied extensively (see \cite{kewley08} for a review on this subject). Here we compare the results in determining the metallicity values of the DGS sample using the method of \cite{izotov06} (I06), who determine the electron temperature, T$_{e}$, from the [OIII]$\lambda$4363 line with the method of \cite{pilyugin05} (PT05) which uses the R$_{23}$ ratio : R$_{23}$ = ([OII]$\lambda$3727+[OIII]$\lambda\lambda$4959,5007)/H$\beta$. Degeneracies in the upper and lower branches of this calibration are removed using the [NII]$\lambda6584$/[OII]$\lambda$3727 ratio \citep{kewley08}.  In Table~\ref{tablea} we provide the metallicity values of the DGS targets determined for the I06 method. Uncertainty values have been determined by propagating the errors in the observed lines from the literature. Table~\ref{sample} includes the metallicity values of the PT05 method, which has been adopted for the DGS survey.
 
 Figure~\ref{metallicitycomp} shows the variation between the I06 and PT05 methods of metallicity calibration. We find a mean deviation of $\pm$0.1 dex, as shown by the dashed lines.
 
\begin{figure*}
\centering
  \includegraphics[width=20cm, angle=90]{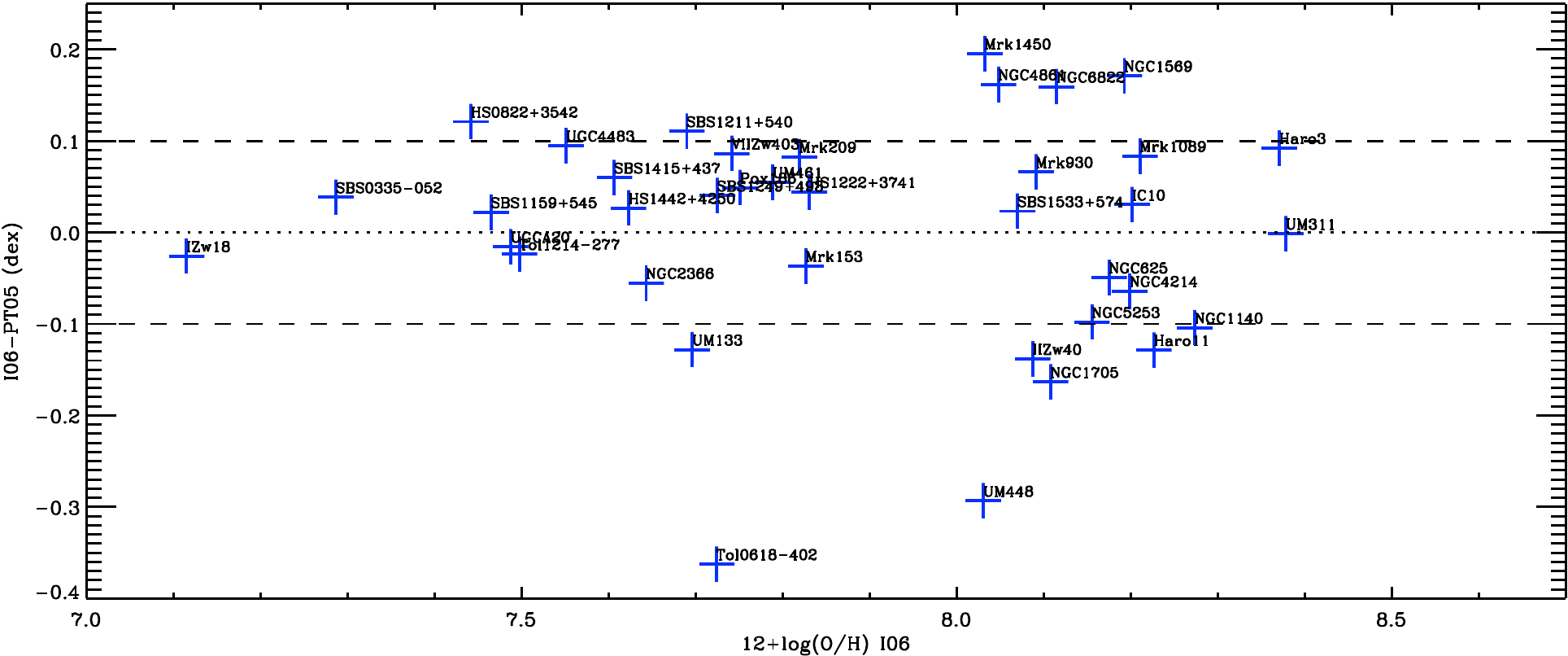}
\caption{ \small Comparison of I06 and PT05 metallicity calibrations for the DGS galaxies. The dashed lines delineate the mean variation of the sample, $\pm$ 0.1 dex. }
\label{metallicitycomp}
\end{figure*}

\begin{deluxetable}{lcc} 
\tabletypesize{\scriptsize} 
 \tablecolumns{3} 
 \tablewidth{0pt} 
\tablecaption{Metallicities for the DGS galaxies determined using the I06 calibration \label{tablea} }
\tablehead{ 
\colhead{Source}  & \colhead{12+log(O/H) (I06)} & \colhead{Ref} 
 }
\startdata            
 Haro11        &    8.23  $\pm$    0.03  &    1 \\
Haro2$^a$        &     -  &    - \\
Haro3         &    8.37  $\pm$    0.02  &    3 \\
He2-10$^a$        &     -  &    - \\
HS0017+1055$^b$   &    7.63  $\pm$    0.10  &  5 \\
HS0052+2536$^b$   &    8.07  $\pm$    0.20  &   5 \\
HS0822+3542   &    7.44  $\pm$    0.06  &     6 \\
HS1222+3741   &    7.83  $\pm$    0.03  &     7 \\
HS1236+3937$^c$   &    7.47  $\pm$    0.1   &     8 \\
HS1304+3529$^c$   &    7.66  $\pm$    0.1   &      8 \\
HS1319+3224$^c$   &    7.59  $\pm$    0.1   &     8 \\
HS1330+3651$^c$   &    7.66  $\pm$    0.1   &     8 \\
HS1442+4250   &    7.62  $\pm$    0.02  &      9 \\
HS2352+2733$^b$   &    8.40  $\pm$    0.20  &   5 \\
IZw18         &    7.12  $\pm$    0.04  &   10 \\
IIZw40        &    8.09  $\pm$    0.02  &  11 \\
IC10          &    8.20  $\pm$    0.06  &    12 \\
Mrk1089$^d$      &    8.39/8.21  $\pm$    0.12  &  13 \\
Mrk1450       &    8.03  $\pm$    0.01  &  14 \\
Mrk153        &    7.83  $\pm$    0.05  &    15 \\
Mrk209        &    7.82  $\pm$    0.01  &   16 \\
Mrk930        &    8.09  $\pm$    0.03  &    17 \\
NGC1140       &    8.27  $\pm$    0.08  &     3 \\
NGC1569       &    8.19  $\pm$    0.03  &    18 \\
NGC1705       &    8.11  $\pm$    0.13  &    19 \\
NGC2366       &    7.64  $\pm$    0.03  &    20 \\
NGC4214       &    8.20  $\pm$    0.03  &     4 \\
NGC4449$^a$       &     -  & - \\
NGC4861       &    8.05  $\pm$    0.01  &   16 \\
NGC5253       &    8.16  $\pm$    0.03  & 4 \\
NGC625        &    8.18  $\pm$    0.03  &   22 \\
NGC6822       &    8.12  $\pm$    0.07  &   23 \\
Pox186        &    7.75  $\pm$    0.01  &    24 \\
SBS0335-052   &    7.29  $\pm$    0.01  &   17 \\
SBS1159+545   &    7.46  $\pm$    0.02  &  14 \\
SBS1211+540   &    7.69  $\pm$    0.01  &   14 \\
SBS1249+493   &    7.73  $\pm$    0.01  &    25 \\
SBS1415+437   &    7.61  $\pm$    0.01  &  26 \\
SBS1533+574   &    8.07  $\pm$    0.02  &   16 \\
Tol0618-402   &    7.72  $\pm$    0.02  &    27 \\
Tol1214-277   &    7.50  $\pm$    0.01  &    4 \\
UGC4483       &    7.55  $\pm$    0.03  &   28 \\
UGCA20        &    7.49  $\pm$    0.04  &    29 \\
UM133         &    7.70  $\pm$    0.01  &    4 \\
UM311$^d$         &    8.38/8.36  $\pm$    0.05  &  17,30,31 \\
UM448         &    8.03  $\pm$    0.04  &  17 \\
UM461         &    7.79  $\pm$    0.03  &    15 \\
VIIZw403      &    7.74  $\pm$    0.01  &    16 \\

\enddata
\scriptsize{
{\bf Notes : } \\
$^a$ : For these galaxies, the [OIII]$\lambda$4363 line was not detected and a direct determination of the metallicity with the I06 method was therefore not possible.

$^b$ : For these galaxies, we quote \cite{ugryumov03} who use the [OIII]$\lambda$4363 line to determine the electron temperature and the method from \cite{izotov94}. I06 is an updated version of the method of \cite{izotov94}.

$^c$ : For these galaxies, no line intensity uncertainties, from which the metallicities are determined, are quoted. We assume a conservative value of 0.1 which is the mean of the variation of the difference between PT05, and the I06 method.

$^d$ : These galaxies are members of groups of galaxies. We quote the value of the metallicity for the individual galaxy (left value) and that for the whole group (right value), which is the mean of all of the metallicities in the group. For Mrk1089, the galaxy is region A-C from \cite{lopezsanchez10}. For UM311, the galaxy is region 3 following \cite{moles94}.

{\bf References for metallicities} : (1) \cite{guseva12} ; 
(3) \cite{izotov04} ; (4) \cite{kobulnicky99} ; (5) \cite{ugryumov03} ; (6) \cite{pustilnik03} ; (7) \cite{izotov07} ; (8) \cite{popescu00} ; (9) \cite{guseva03a} ; (10) \cite{izotov99} ; (11) \cite{guseva00} ; (12) \cite{magrini09} ; (13) \cite{lopezsanchez10} ; (14) \cite{izotov94} ; (15) \cite{izotov06} ; (16) \cite{izotov97} ; (17) \cite{izotov98} ; (18) \cite{kobulnicky97} ; (19) \cite{lee04} ; (20) \cite{saviane08} ; 
(22) \cite{skillman03} ; (23) \cite{lee06} ; (24) \cite{guseva07} ; (25) \cite{thuan95} ; (26) \cite{guseva03b} ; (27) \cite{masegosa94} ; (28) \cite{van_zee06} ; (29) \cite{van_zee96} ; (30) \cite{moles94} ; (31) \cite{pilyugin07} \\
 }
 \end{deluxetable}
\clearpage


\section{DGS {\it Spitzer-Herschel} Atlas}
As an illustration of the wealth of data the DGS brings,we have assembled an atlas of 3-color images of the DGS galaxies composed of \spitz\  MIPS 24 \mic\ (blue), tracing the hot dust component, PACS 100 \mic\ (green), close to the peak of the dust SED, and SPIRE 250 \mic\ (red), tracing the cooler dust. Note that the images here are produced using their original resolution. The beams for the different bands are indicated in the lower right of each image. When a band is not present in the 3-color image, the source was not observed at that wavelength.
 


\begin{figure*}
\centering
\vspace{-1.5cm}
\includegraphics[width=18.0cm]{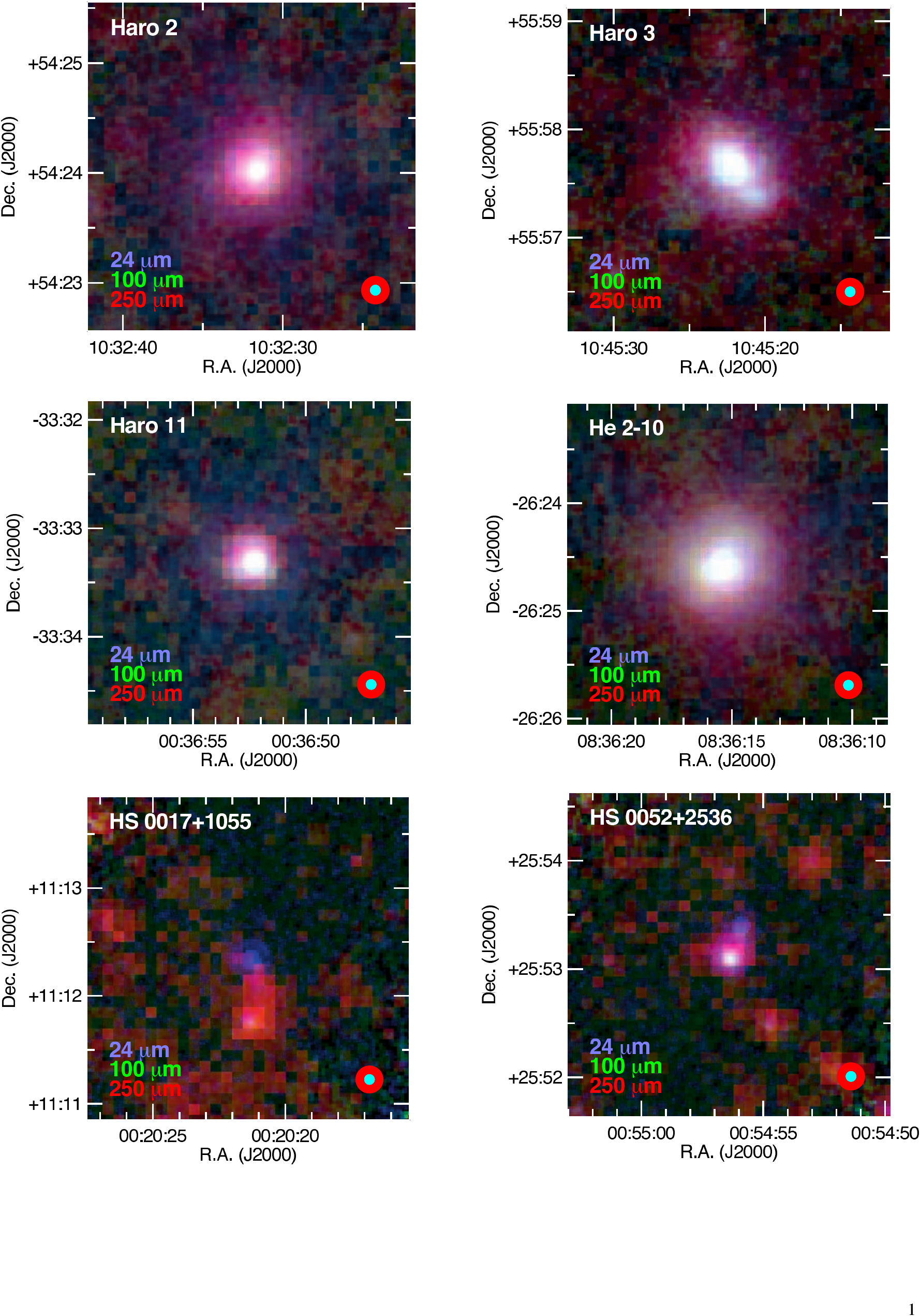}
\caption{Appendix A - DGS {\it Spitzer-Herschel} Atlas}
\end{figure*}

\begin{figure*}
\centering
\vspace{-1.5cm}
\includegraphics[width=18.0cm]{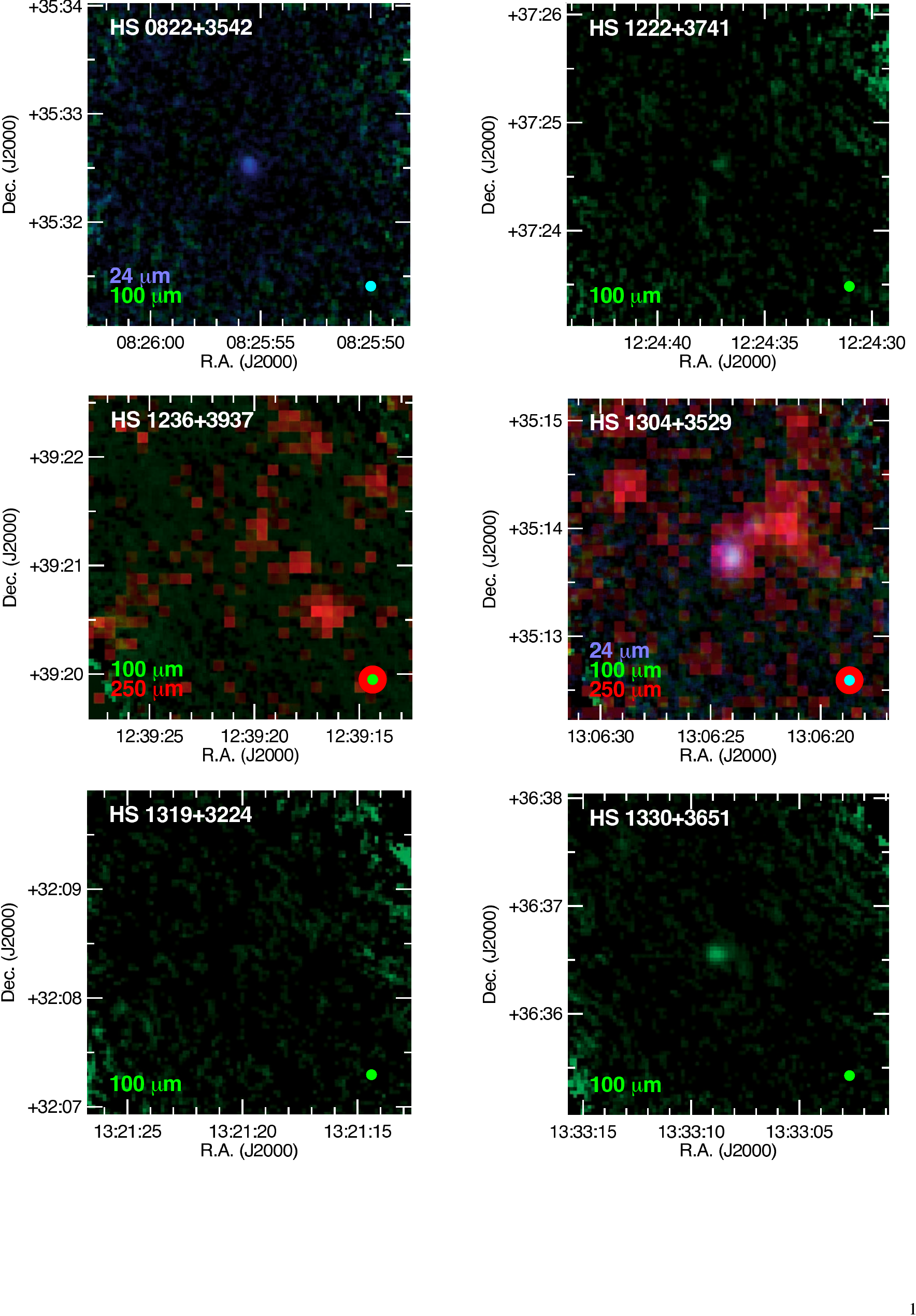}
\caption{Appendix A - DGS {\it Spitzer-Herschel} Atlas - continued}
\end{figure*}

\begin{figure*}
\centering
\vspace{-1.5cm}
\includegraphics[width=18.0cm]{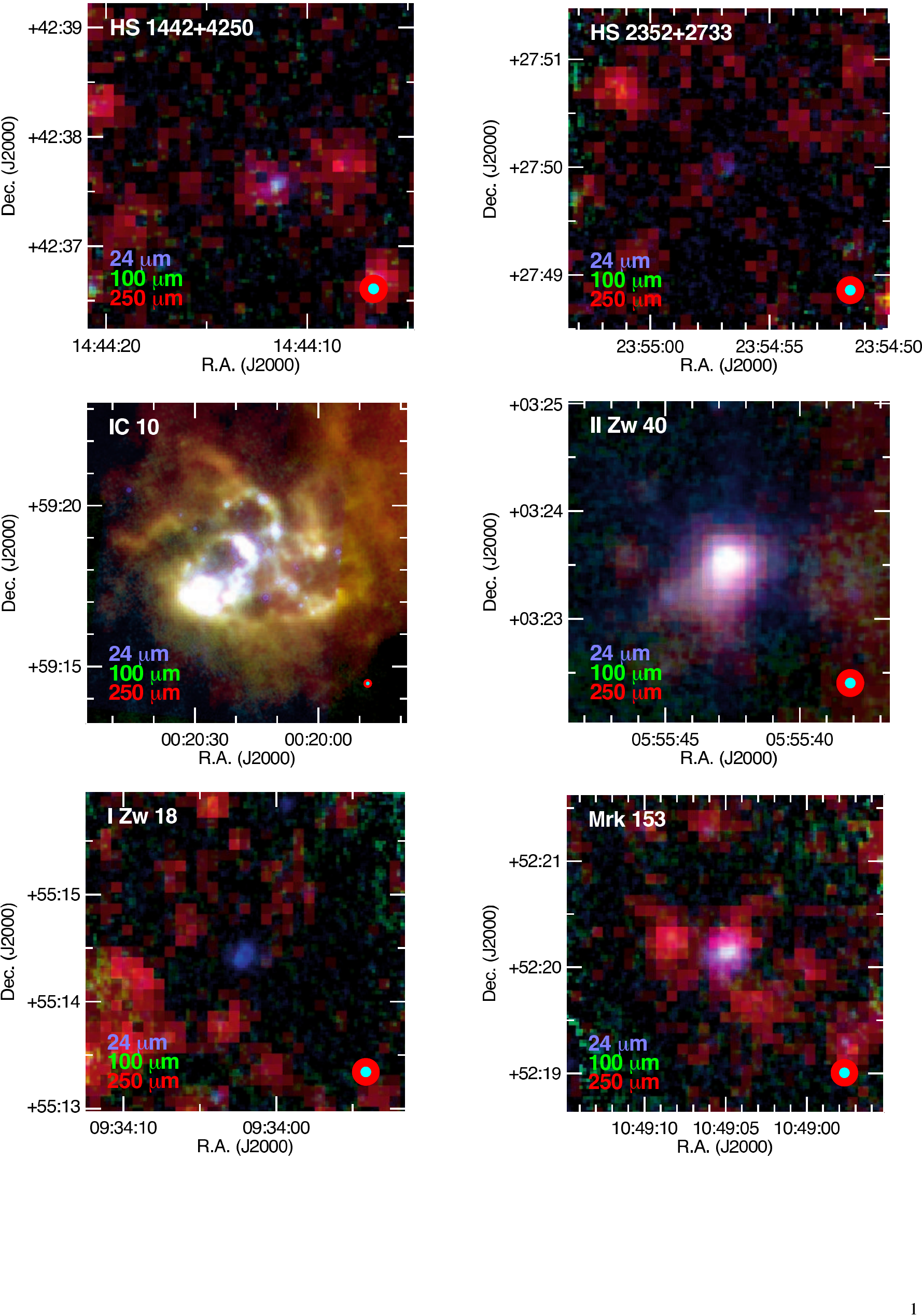}
\caption{Appendix A - DGS {\it Spitzer-Herschel} Atlas - continued}
\end{figure*}

\begin{figure*}
\centering
\vspace{-1.5cm}
\includegraphics[width=18.0cm]{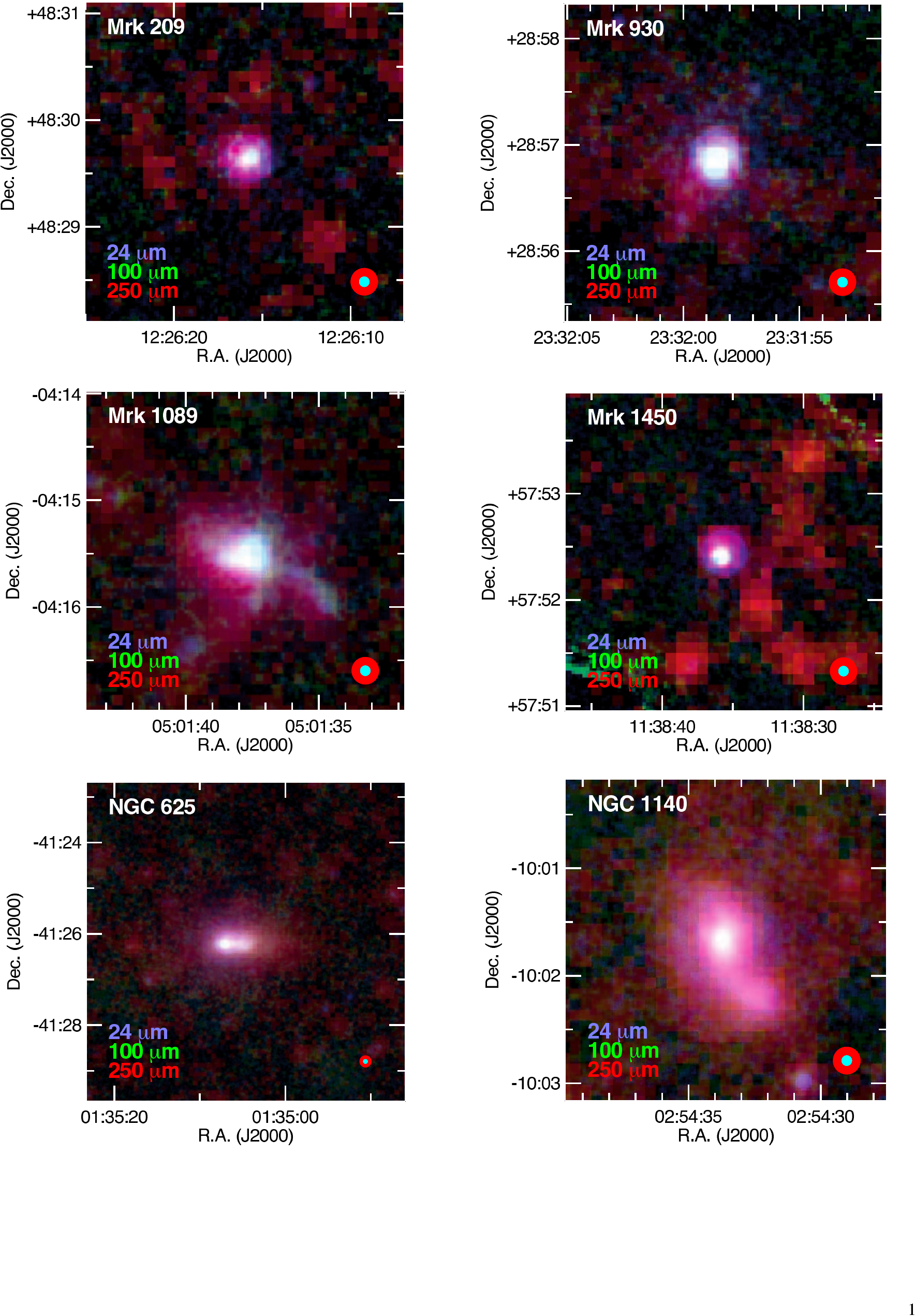}
\caption{Appendix A - DGS {\it Spitzer-Herschel} Atlas - continued}
\end{figure*}

\begin{figure*}
\centering
\vspace{-1.5cm}
\includegraphics[width=18.0cm]{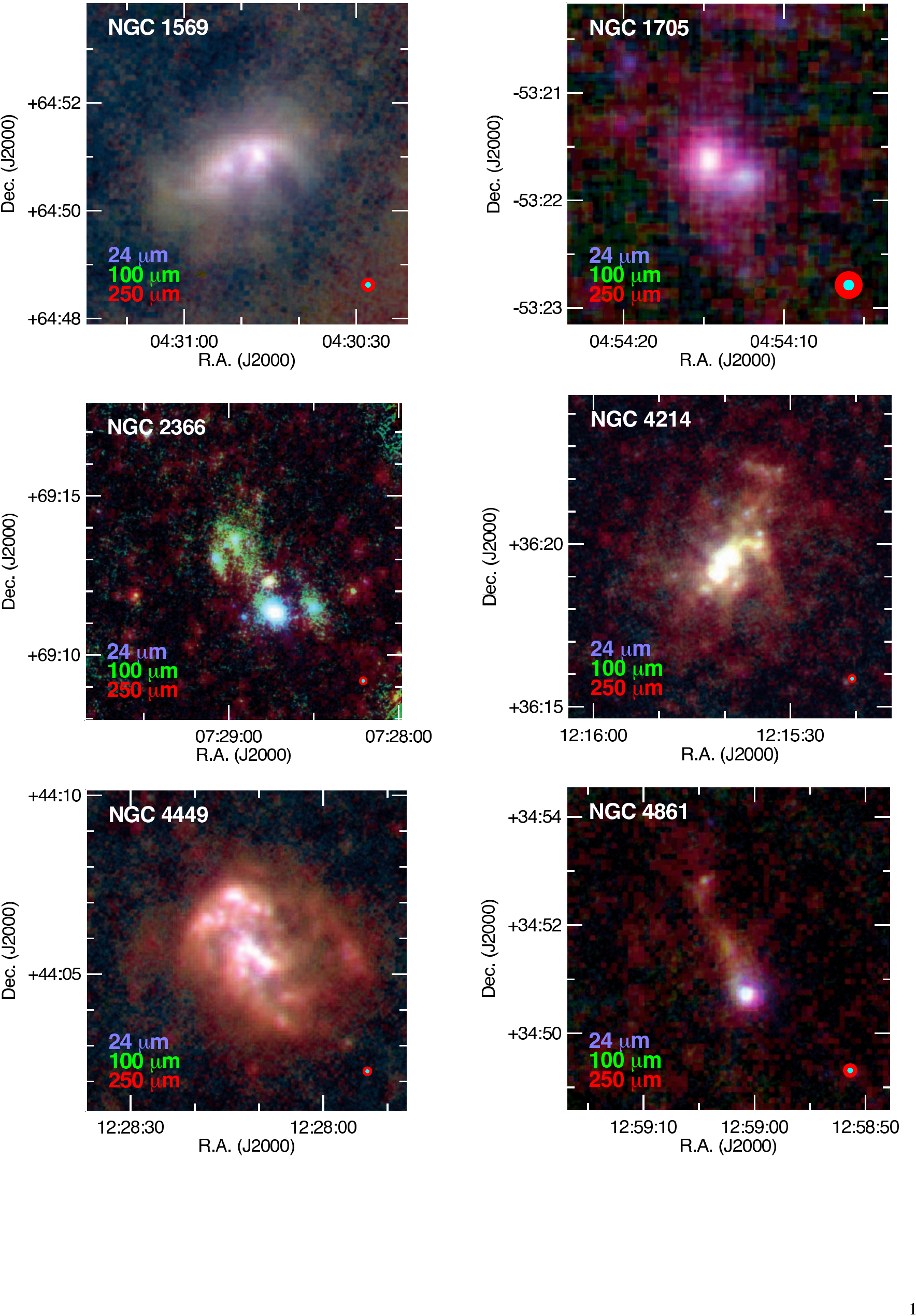}
\caption{Appendix A - DGS {\it Spitzer-Herschel} Atlas - continued}
\end{figure*}

\begin{figure*}
\centering
\vspace{-1.5cm}
\includegraphics[width=18.0cm]{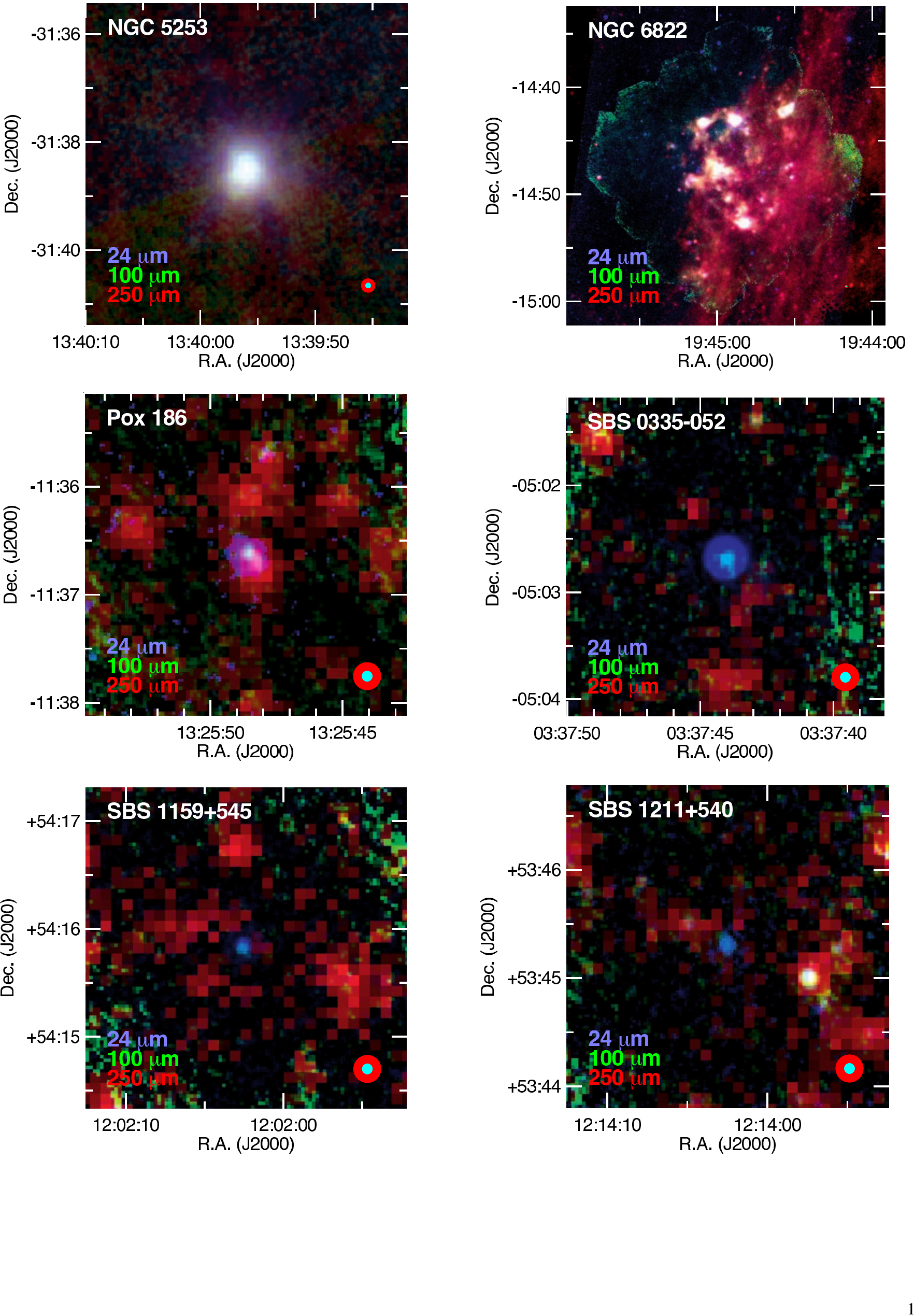}
\caption{Appendix A - DGS {\it Spitzer-Herschel} Atlas - continued}
\end{figure*}

\begin{figure*}
\centering
\vspace{-1.5cm}
\includegraphics[width=18.0cm]{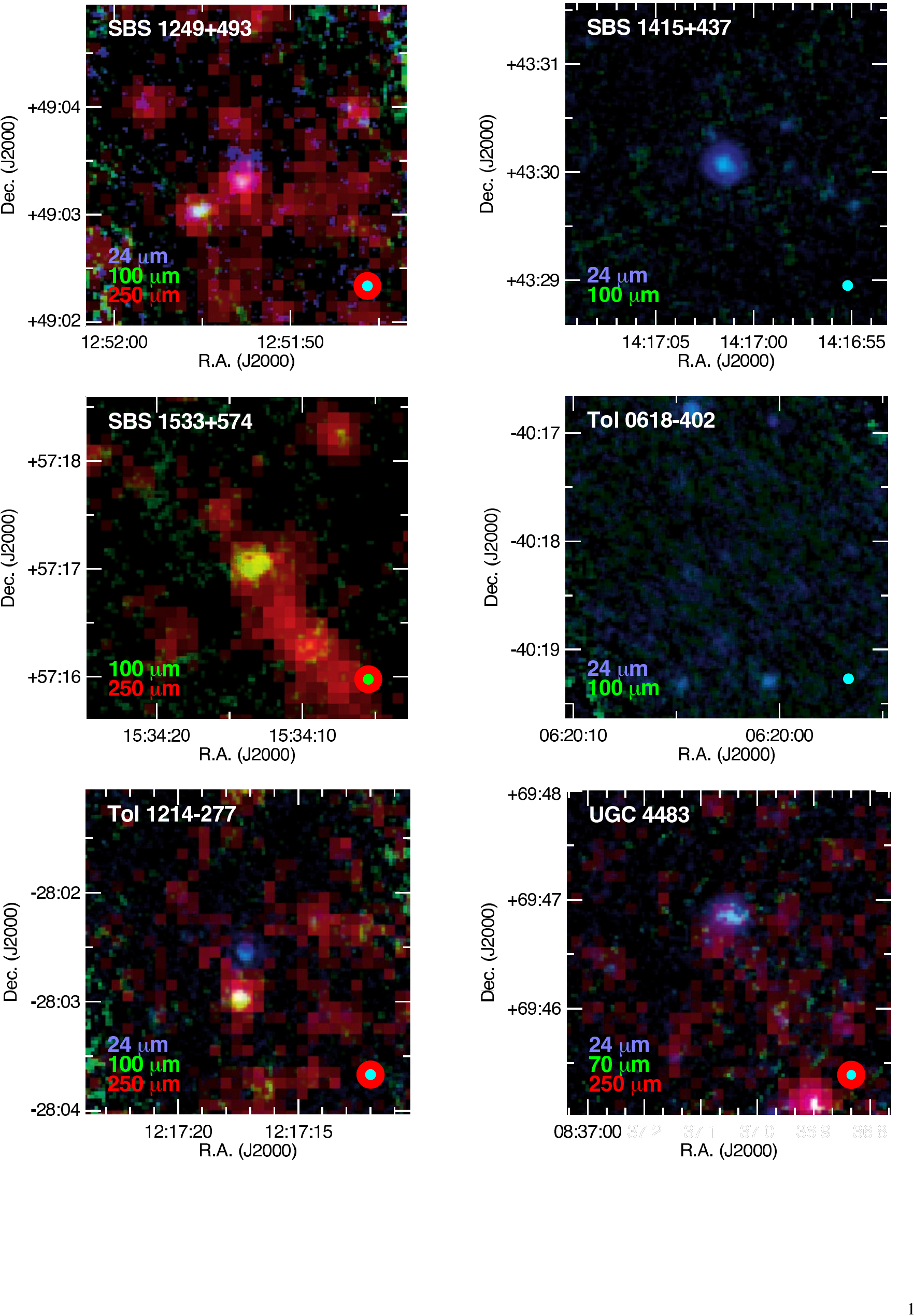}
\caption{Appendix A - DGS {\it Spitzer-Herschel} Atlas - continued}
\end{figure*}

\begin{figure*}
\centering
\vspace{-1.5cm}
\includegraphics[width=18.0cm]{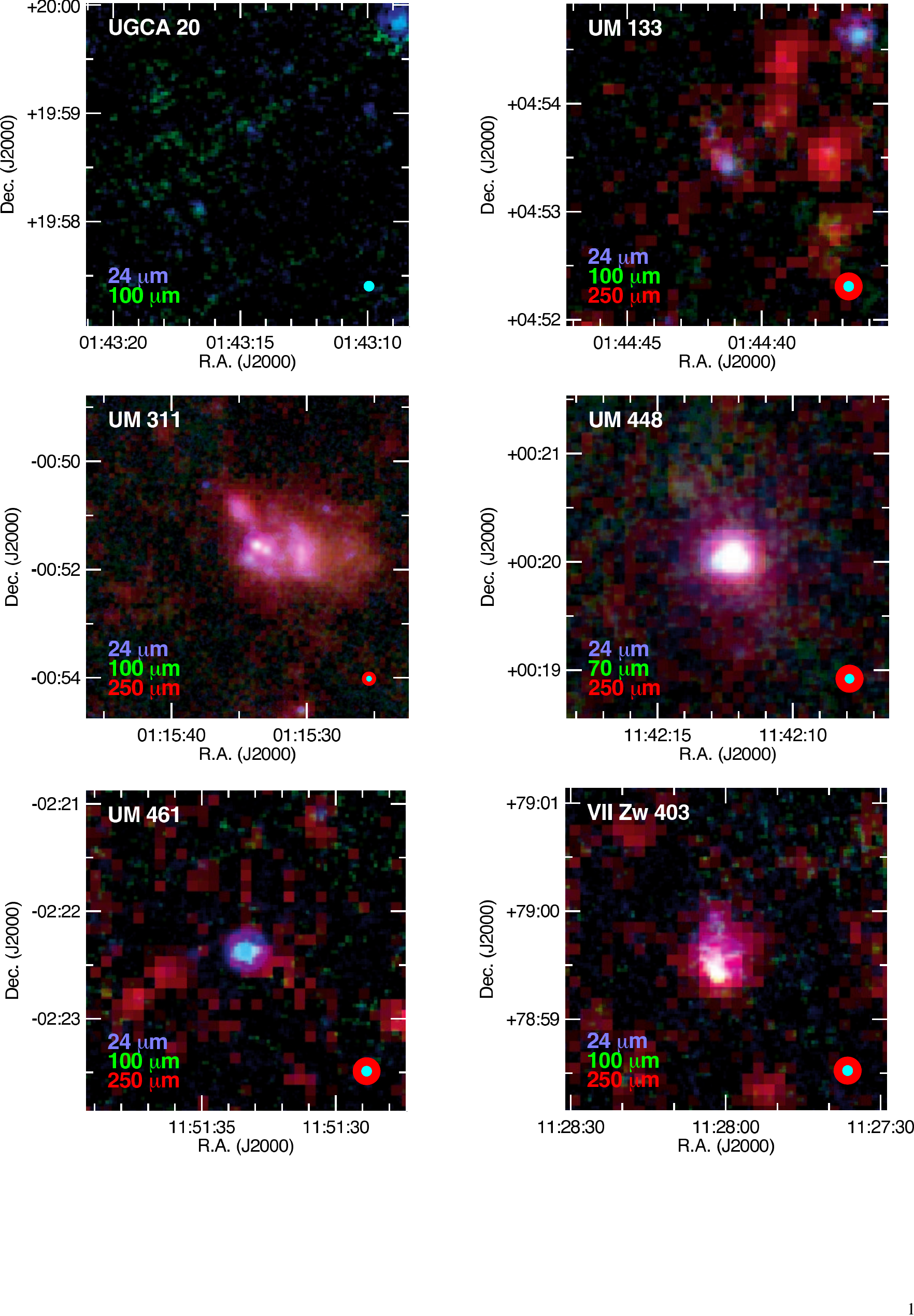}
\caption{Appendix A - DGS {\it Spitzer-Herschel} Atlas - continued}
\end{figure*}
\clearpage

\begin{acknowledgements}
We thank the anonymous referee for useful suggestions which help to improve the clarity of the manuscript. PACS has been developed by MPE (Germany); UVIE (Austria); KU Leuven, CSL, IMEC (Belgium); CEA, LAM (France); MPIA (Germany); INAF-IFSI/OAA/OAP/OAT, LENS, SISSA (Italy); IAC (Spain). This development has been supported by BMVIT (Austria), ESA-PRODEX (Belgium), CEA/CNES (France), DLR (Germany), ASI/INAF (Italy), and CICYT/MCYT (Spain). SPIRE has been developed by Cardiff University (UK); Univ. Lethbridge (Canada); NAOC (China); CEA, LAM (France); IFSI, Univ. Padua (Italy); IAC (Spain); SNSB (Sweden); Imperial College London, RAL, UCL-MSSL, UKATC, Univ. Sussex (UK) and Caltech, JPL, NHSC, Univ. Colorado (USA). This development has been supported by CSA (Canada); NAOC (China); CEA, CNES, CNRS (France); ASI (Italy); MCINN (Spain); Stockholm Observatory (Sweden); STFC (UK); and NASA (USA).

SPIRE has been developed by a consortium of institutes led by Cardiff Univ. (UK) and including: Univ. Lethbridge (Canada); NAOC (China); CEA, LAM (France); IFSI, Univ. Padua (Italy); IAC (Spain); Stockholm Observatory (Sweden); Imperial College London, RAL, UCL-MSSL, UKATC, Univ. Sussex (UK); and Caltech, JPL, NHSC, Univ. Colorado (USA). This development has been supported by national funding agencies: CSA (Canada); NAOC (China); CEA, CNES, CNRS (France); ASI (Italy); MCINN (Spain); SNSB (Sweden); STFC, UKSA (UK); and NASA (USA).

The research leading to these results has received funding from the European Community's Seventh Framework Programme (/FP7/2007-2013/) under grant agreement No 22951

\end{acknowledgements}


\newpage
\bibliographystyle{apj}
\bibliography{references}
 
\end{document}